\documentclass[aps,prd,nofootinbib,twocolumn,showpacs,superscriptaddress,letterpaper]{revtex4-2}
\pdfoutput=1 

\usepackage[T1]{fontenc}

\usepackage{bm,amsmath,amsfonts,amssymb,wasysym}
\usepackage{graphicx}
\usepackage{breakurl}
\usepackage{array}
\usepackage{slashed}
\usepackage{xspace}
\usepackage{hhline}
\usepackage[italic]{hepnames}
\usepackage{multirow}
\usepackage{color}
\usepackage{placeins}
\usepackage{adjustbox}
\usepackage{rotating} 
\usepackage{tabularx} 
\usepackage[vlines]{tabularht}
\usepackage[alsoload=hep]{siunitx}
\usepackage{physics}
\usepackage{gensymb}
\usepackage{hyphenat}

\usepackage{adjustbox}
\usepackage{rotating}
\usepackage{booktabs}

\usepackage[colorlinks,hyperindex,breaklinks]{hyperref}
\hypersetup{linkcolor=blue,citecolor=blue,filecolor=black,urlcolor=blue}

\usepackage[capitalise]{cleveref}
\usepackage[displaymath]{lineno}

\def\bulnu{\ensuremath{B \to X_u \, \ell\, \bar \nu_{\ell}}\xspace}
\def\bclnu{\ensuremath{B \to X_c \, \ell\, \bar \nu_{\ell}}\xspace}

\def\lumi{\ensuremath{62.8 \, \mathrm{fb}^{-1}}\xspace}
\def\NBB{\ensuremath{\left(68.2 \pm 0.9 \right) \times 10^{6}}\xspace}

\def\bdlnu{\ensuremath{B \to D \, \ell\, \bar \nu_{\ell}}\xspace}
\def\bdslnu{\ensuremath{B \to D^* \, \ell\,\bar \nu_{\ell}}\xspace}
\def\bdbsblnu{\ensuremath{B \to D^{(*)} \, \ell\, \bar \nu_{\ell}}\xspace}
\def\bddslnu{\ensuremath{B \to D^{**} \, \ell\, \bar \nu_{\ell}}\xspace}

\def\GF{\ensuremath{G_{F}}\xspace}

\def\AbsVcb{\ensuremath{\left| V_{cb} \right|}\xspace}

\def\b2{Belle~II\xspace}

\def\Bbar    {\kern 0.18em\overline{\kern -0.18em B}\xspace}

\newcommand{\qsquared}{\ensuremath{q^2}\xspace}

\newcommand{\qsquaredn}[1]{\ensuremath{q^{2#1}\xspace}}

\newcommand{\qsquarednreco}[1]{\ensuremath{\qsquaredn{#1}_\mathrm{reco}}\xspace}
\newcommand{\qsquaredncalib}[1]{\ensuremath{\qsquaredn{#1}_\mathrm{calib}}\xspace}

\newcommand{\qsqmoment}[1]{\ensuremath{\langle \qsquaredn{#1} \rangle}\xspace}

\newcommand{\qsqmomentcalib}[1]{\ensuremath{\langle \qsquaredn{#1}_\mathrm{calib} \rangle}\xspace}
\newcommand{\qsqmomenttrue}[1]{\ensuremath{\langle \qsquaredn{#1}_\mathrm{gen,sel} \rangle}\xspace}
\newcommand{\qsqmomentgen}[1]{\ensuremath{\langle \qsquaredn{#1}_\mathrm{gen} \rangle}\xspace}
 
\newcommand{\qsqxmoment}[1]{\ensuremath{\langle q^{#1} \rangle}\xspace}

\newcommand{\qsquarednevent}[3]{\ensuremath{\qsquaredn{#1}_{\mathrm{#2},#3}}}

\newcommand{\btag}{\ensuremath{B_{\rm{tag}}}\xspace}
\newcommand{\bsig}{\ensuremath{B_{\rm{sig}}}\xspace}
\newcommand{\FEIprob}{\ensuremath{\mathcal{P}_\mathrm{FEI}}\xspace}

\newcommand{\mbc}{\ensuremath{M_{\mathrm{bc}}}\xspace}
\newcommand{\DeltaE}{\ensuremath{\Delta E}\xspace}

\newcommand{\ccalib}{\ensuremath{\mathcal{C}_\mathrm{calib}}\xspace}
\newcommand{\ctrue}{\ensuremath{\mathcal{C}_\mathrm{gen}}\xspace}

\begin{document}

\title{\Large{Measurement of Lepton Mass Squared Moments in \bclnu Decays with the \b2 Experiment}}
\date{\today}


\author{\large{The \b2 Collaboration}}

\noaffiliation

\begin{abstract}
We present measurements of the first to fourth moments of the lepton mass squared $q^2$  of \mbox{\bclnu} decays for $\ell = e, \mu$ and with $X_c$ a hadronic system containing a charm quark.
These results use a sample of electron-positron collisions at the $\Upsilon(4S)$ resonance corresponding to \lumi of integrated luminosity and collected by the \b2 experiment in 2019 and 2020.
To identify the $X_c$ system and reconstruct $q^2$, one of the $B$ mesons from an $\Upsilon(4S) \to B \Bbar$ decay is fully reconstructed in a hadronic decay mode using a multivariate $B$ tagging algorithm.
We report raw and central moments for $ \qsquared > \SI{1.5}{GeV^2/\clight^4}$ up to $\qsquared > \SI{8.5}{GeV^2/\clight^4}$, probing up to 77\% of the accessible \mbox{\bclnu} phase space. This is the first measurement of moments in the experimentally challenging range of $[ 1.5, 2.5 ] \, \mathrm{GeV^2/\clight^4}$.
\end{abstract}


\maketitle


\section{Introduction}\label{sec:introduction}

Existing measurements of $|V_{cb}|$ use either exclusive final states with \bdslnu and \bdlnu providing the most precise values or inclusive final states. 
In inclusive determinations of $|V_{cb}|$, the total decay rate can be expressed as an expansion of a small number of non-perturbative matrix elements with the heavy-quark expansion (HQE). 
Using HQE, the total semileptonic rate can be expanded in powers of $\Lambda_{\mathrm{QCD}}/m_b$, the ratio of the QCD scale parameter and the bottom-quark mass and perturbative corrections proportional to the strong coupling constant $\alpha_s$ can also be systematically incorporated~\cite{Jezabek:1988iv,Aquila:2005hq,Pak:2008cp,Melnikov:2008qs,Becher:2007tk,Mannel:2014xza,Alberti:2013kxa,Mannel:2010wj}. 

The current world averages~\cite{Amhis:2019ckw} for \AbsVcb determined from inclusive and exclusive approaches are
\begin{linenomath}
\begin{align}
  \left| V_{cb}^\mathrm{incl.} \right| &=  (42.19 \pm 0.78) \times 10^{-3}\quad \mathrm{and} \\
  \left| V_{cb}^\mathrm{excl.} \right| &=  (39.25 \pm 0.56) \times 10^{-3},
\end{align}
\end{linenomath}
respectively. 
The uncertainties are the sum of experimental and theoretical uncertainties; the world averages differ by about three standard deviations. 
The 2\% relative uncertainty in the world average for the inclusive approach is largely due to the theory uncertainty associated with the truncation of HQE and perturbative expansion~\cite{Benson:2003kp,Gambino:2011cq}.
To further reduce this uncertainty, higher order non-perturbative matrix elements must be determined from measured spectral moments. 
This is complicated by the proliferation of HQE parameters at higher orders in the expansion. 
At $\mathcal{O}(1/m_{b}^4)$ in the HQE thirteen non-perturbative matrix elements contribute to the total rate and the spectral energy and mass moments. 

Reference~\cite{Fael:2018vsp} outlines a novel and alternative approach to determine $|V_{cb}|$ from inclusive decays avoiding this proliferation of terms.
Exploiting reparameterization invariance, the authors reduce the number of parameters necessary to calculate the total rate at $\mathcal{O}(1/m_{b}^4)$ to only eight.
Unfortunately, spectral moments of lepton-energy and hadron-mass spectra violate reparameterization invariance. However, reparameterization invariance is retained in the spectral moments of the lepton mass squared $q^2 \equiv \left( p_\ell + p_\nu\right)^2 = (p_B - p_X)^2$ where $p_i$ is the four-momentum of the particle $i$. 

We present measurements of the spectral moments of the lepton mass squared \qsqmoment{n} with \mbox{$n=1$--4} for $q^2 > \SI{1.5}{GeV^2/\clight^4}$ up to $\SI{8.5}{GeV^2/\clight^4}$.
The simultaneous analysis of these moments can determine the non-perturbative matrix elements as their contributions vary with the $q^2$ threshold~\cite{Fael:2018vsp}; moments with a lower \qsquared threshold retain more information about the inclusive \bclnu process.
Charge conjugation is implied throughout this paper, and $\mathcal{B}(\bclnu)$ is defined as the average of the branching fraction with $B^0$ and $B^+$ and $\ell = e, \mu$.

We present raw and central moments, with the latter having the benefit of smaller correlations between $q^2$ thresholds and the orders of moments.
The first measurement of the first \qsquared moment was reported in Ref.~\cite{Csorna:2004kp} with an implicit lower requirement on the lepton energy of $\SI{1}{GeV}$. 
This requirement renders the measured moment unsuitable for the analysis outlined in Ref.~\cite{Fael:2018vsp}.
To avoid this problem, we restrict our measurement to the region $ \qsquared > \SI{1.5}{GeV^2/\clight^4}$: events passing this selection have leptons that can be reliably identified.

A measurement of the \qsquared moments, similar to the one presented in this paper, using the full Belle data set was recently reported by the Belle collaboration~\cite{Belle:2021idw} for $\qsquared > \SI{3.0}{GeV^2/\clight^4}$, covering 58\% of the accessible \bclnu phase space. 
We report measurements of the raw and central \qsquared moments with comparable precision and include the experimentally challenging \qsquared region $[ 1.5, 2.5 ] \, \mathrm{GeV^2/\clight^4}$, covering up to 77\% of the accessible \bclnu phase-space. 

The remainder of this paper is organized as follows: \cref{sec:data_set_sim_samples} describes the data set used in this analysis, the \b2 detector, and the simulation of $e^+ e^-$ collision events.
\cref{sec:analysis_strategy} introduces the tag-side and the inclusive reconstruction of semileptonic $B$ decays.
\cref{sec:leptonic_inv_mass_moments} describes the background subtraction, calibration and calculation of the lepton mass squared moments. 
\cref{sec:systematics} discusses the systematic uncertainties affecting the measurement.
\cref{sec:results} presents the main findings and \cref{sec:summary} our conclusions.


\section{Belle~II Detector, Data Set and Simulated Samples}
\label{sec:data_set_sim_samples}

\subsection{SuperKEKB and the Belle~II Detector}

We analyze data collected in 2019 and 2020 by the \b2 detector~\cite{Abe:2010gxa} at the SuperKEKB $e^+ e^-$ accelerator complex~\cite{Akai:2018mbz}.
At SuperKEKB, $\SI{7}{GeV}$ electrons collide with $\SI{4}{GeV}$ positrons giving a centre-of-mass (CM) energy of $\sqrt{s}=\SI{10.58}{GeV}$, corresponding to the mass of the $\Upsilon(4S)$ resonance.
This results in a boost of $\beta \gamma = 0.28$ of the CM frame relative to the laboratory frame.
The integrated luminosity of \lumi~\cite{Abudin_n_2020} of the data corresponds to \NBB $B$ pairs. 
We use $\SI{9.2}{fb^{-1}}$ of data recorded $\SI{60}{MeV}$ below the $\Upsilon(4S)$ resonance to constrain contributions from $e^+ e^- \to q \bar q$ continuum processes.
 
The \b2 detector is a substantial upgrade of the Belle detector~\cite{Abashian:2000cg} with improved reconstruction of charged and neutral particles and particle identification performance. 
The detector consists of several subdetectors arranged in a cylindrical structure around the $e^+ e^-$ interaction point (IP). 
The IP is enclosed by a beryllium beam pipe with an inner radius of \SI{1}{\centi \metre}.
The part of the detector closest to the IP is the pixel detector (PXD), consisting of two layers of depleted p-channel field-effect-transistor pixel-sensor modules~\cite{Kemmer:1986vh}.
The first layer comprises sixteen modules arranged in eight ladders.
The second layer was only partially installed for data taking and consists of four modules.
The PXD is surrounded by four layers of double-sided silicon strip modules: the silicon vertex detector (SVD).
The first SVD layer is arranged parallel to the beam axis, while the forward sections of the second to fourth layers are tilted with respect to the beam axis in order to reduce the overall material budget and the number of sensors.
Both silicon tracking detectors are enclosed by the central drift chamber~(CDC), which is filled with a He (50\%) and $\text{C}_2 \text{H}_6$ (50\%) gas mixture.
The CDC contains 56576 sense and field wires oriented along the beam direction or tilted and arranged into 56 radial layers.
By combining the information from axial and stereo wires the full three-dimensional trajectory of a charged particle is reconstructed and its specific ionization $\dd{E}/\dd{x}$ is measured.
Outside the CDC, a time-of-propagation detector~(TOP) and an aerogel ring-imaging Cherenkov detector~(ARICH) cover the barrel and forward endcap regions of the detector, respectively.
The TOP reconstructs spatial and temporal coordinates of the ring of Cherenkov light cones emitted from charged particles passing through quartz radiator bars.
The information from both the TOP and ARICH and the CDC are combined together to identify charged particles.
The electromagnetic calorimeter~(ECL) consists of a \SI{3}{\metre} long barrel section with an inner radius of \SI{1.25}{\metre} and annular endcaps.
In total 8736 CsI(Tl) crystals arranged in a pointing geometry allow for precise energy and timing measurements of neutral and charged particles.
The ECL is located outside the TOP and inside the remaining volume of a superconducting solenoid with a field strength of \SI{1.5}{\tesla}.
The $K^0_\mathrm{L}$ and muon detector (KLM) is located outside of the coil.
It consists of an alternating structure of $\SI{4.7}{cm}$ thick iron plates and active detector elements.
The iron plates are used as the magnetic flux return yoke for the solenoid and absorber material to range out charged hadrons.
The detector elements are glass-electrode resistive plate chambers and plastic scintillators in the barrel and endcap region, repsectively.

We define the $z$ axis of the laboratory frame as the central axis of the solenoid with the positive direction in the direction of the electron beam.
The polar angle $\theta$ and the longitudinal and transverse directions are defined with respect to the $z$ axis. 
Variables with asterisk superscripts are measured in the CM frame; variables without asterisks are measured in the laboratory frame.

\subsection{Reconstruction}

Charged particle tracks are reconstructed combining information from the PXD, SVD, and CDC~\cite{Bertacchi:2020eez}.
The reconstruction of energy depositions from neutral and charged particles in the ECL (ECL clusters) uses shower shapes and timing information~\cite{Kou:2018nap}.
Tracks are identified as electron or muon candidates by combining information from several subdetectors into a single lepton identification likelihood $\mathcal{L}_{\ell}$ (PID).
Muons are identified reliably by extrapolating tracks to the KLM.
The main features used for the construction of the likelihood are the longitudinal penetration depth and the transverse scattering of the extrapolated track in the KLM.
For electrons, the likelihood is constructed from information from the ECL, CDC, TOP, and ARICH.
The most important discriminant is the ratio of the reconstructed energy in the ECL to the estimated track momentum, which should be close to unity for electrons.
The identification of charged pions, kaons, and protons is based on likelihood information from the CDC, TOP, and ARICH.
Their likelihoods are denoted as $\mathcal{L}_{\pi}$, $\mathcal{L}_{K}$, and $\mathcal{L}_\mathrm{p}$.
Hadrons with momenta less than $\SI{700}{MeV/\clight}$ are primarily identified using $\dd{E}/\dd{x}$ measurements from the CDC. 
Hadrons with momenta larger than $ \SI{700}{MeV/\clight}$ are primarily identified using the TOP and ARICH measurements.
Photon candidates are identified using the ECL shower shape of clusters not matched to a track.
We require each photon candidate have a transverse energy greater than \SI{30}{MeV} when reconstructed in the barrel or \SI{20}{MeV} when reconstructed in either endcap.
A loose selection on a multivariate shower-shape classifier that uses multiple Zernike moments~\cite{ZERNIKE1934689} is imposed.
A more detailed overview of the \b2 PID algorithms and the photon reconstruction algorithms can be found in Ref.~\cite{Kou:2018nap}.

\subsection{Simulation}

Monte Carlo (MC) samples are used to determine reconstruction efficiencies and acceptance effects as well as to estimate background contamination.
MC samples of $B$ decays are simulated using the \texttt{EvtGen} generator~\cite{EvtGen}.
The simulation of $e^+ e^- \to q \bar q$ continuum processes is carried out with \texttt{KKMC}~\cite{Jadach:1999vf} and \texttt{PYTHIA8}~\cite{Sjostrand:2014zea}.
Electromagnetic final-state radiation (FSR) is simulated using \texttt{PHOTOS}~\cite{Photos} for all charged final-state particles.
Interactions of particles with the detector are simulated using \texttt{GEANT4}~\cite{Agostinelli:2002hh}.

The simulation is corrected using data-driven weights to account for differences in identification and reconstruction efficiencies. The PID for electrons is corrected as a function of the laboratory-frame momentum and polar angle and charge of the electron candidate using samples of $e^+ e^- \to e^+ e^- (\gamma)$ and $e^+ e^- \to  e^+ e^- e^+ e^-$ events and events with $J/\psi \to e^+ e^-$ decays.
The PID for muons is corrected using samples of $e^+ e^- \to \mu^+ \mu^- \gamma$ and $e^+ e^- \to e^+ e^- \mu^+ \mu^-$, and  events with $J/\psi \to \mu^+ \mu^-$ decays.
The average multiplicative corrections are $0.95$ and $0.89$ for electron and muon candidates, respectively.
The rates of misidentifying charged hadrons as charged leptons are corrected using samples of $K_S^0 \to \pi^+ \pi^-$, $D^{*\, +} \to D^0 \pi^+$, and $e^+ e^- \to \tau^+ \tau^-$, with average multiplicative misidentification-rate corrections of $1.50$ and  $0.98$ for electron and muon candidates, respectively.

All recorded $e^+ e^-$ collision data and simulated events are reconstructed and analyzed with the open-source \texttt{basf2} framework~\cite{basf2}.

\subsection{Simulation of \bclnu}

The analysis relies on accurate modeling of \bclnu decays.
Inclusive semileptonic \bclnu decays are dominantly \bdlnu\ and \bdslnu decays.
The \bdlnu\ decays are modeled using the BGL parameterization~\cite{Boyd:1994tt} with form-factor parameter values and uncertainties from the fit in Ref.~\cite{Glattauer:2015teq}.
For \bdslnu\, decays, the BGL implementation proposed in Refs.~\cite{Grinstein:2017nlq,Bigi:2017njr} with form-factor parameter values and uncertainties from a fit to the measurement of Ref.~\cite{Waheed:2018djm} is used.
Both branching fractions are normalized to the average branching fraction of Ref.~\cite{Amhis:2019ckw} assuming isospin symmetry.

Semileptonic \mbox{\bddslnu} decays with $D^{**} = D_0^*, D_1^\prime, D_1, D_2^* $ are modeled using heavy-quark-symmetry-based form-factors proposed in Ref.~\cite{Bernlochner:2016bci} and with $D^{**}$ masses and widths from Ref.~\cite{pdg:2020}.

For the \bddslnu branching fractions, we adopt the values of Ref.~\cite{Amhis:2019ckw} to account for missing isospin-conjugated and other established decay modes observed in studies of $B$ decays into fully hadronic final states.
This follows the prescription outlined in Ref.~\cite{Bernlochner:2016bci}. All existing exclusive \bddslnu measurements only use $D^{**\,0} \to D^{(*)+} \, \pi^-$ decay modes. To correct for the missing isospin modes we multiply the branching fractions with a multiplicative factor of $3/2$.

In the average in Ref.~\cite{Amhis:2019ckw}, all measurements of $B \to D_2^* \, \ell \bar \nu_\ell$  are relative to $\overline D_2^{*} \to D^{*\, -} \pi^+$. To account for $\overline D_2^{*} \to D^{-} \pi^+$ contributions, we apply a multiplicative factor of $1.54 \pm 0.15 $ calculated from the branching fractions of Ref.~\cite{pdg:2020}.

The world average for $B \to D_1^\prime \, \ell \bar \nu_\ell$ in Ref.~\cite{Amhis:2019ckw} combines measurements that only marginally agree with each other (the probability of the combination is below 0.01\%).
We exclude the measurement of  Ref.~\cite{Liventsev:2007rb} that is in conflict with the measured branching fractions of Refs.~\cite{Aubert:2008ea,Abdallah:2005cx}.
That measurement also conflicts with the expectation that $\mathcal{B}(B \to D_1^\prime \, \ell \bar \nu_\ell)$ is comparable to $\mathcal{B}(B \to D^\ast_0 \, \ell \bar \nu_\ell)$~\cite{Leibovich:1997em, Bigi:2007qp}.
By excluding Ref.~\cite{Liventsev:2007rb} we obtain
\begin{linenomath}
\begin{align}
 \mathcal{B}(B^+ \to \overline D_1^{\prime\, 0}(\to D^{*\,-} \pi^+) \, \ell \nu_\ell) & = \left(0.28 \pm 0.06 \right) \times 10^{-2} \, .
\end{align}
\end{linenomath}

The world average for $\mathcal{B}(B \to D_1 \, \ell \bar \nu_\ell)$ does not include contributions from $D_1 \to D \pi \pi$.
To account for these, we use a multiplicative factor $0.43 \pm 0.11$ calculated from the branching fractions of $D_1 \to D^{*-} \pi^+$ and $D_1 \to \bar D^{0} \pi^+ \pi^-$ from Ref.~\cite{Aaij:2011rj}. 
The contribution of $D_1 \to D \pi \pi$ decays is subtracted from the $B \to D \pi \pi \ell \bar \nu_\ell$ branching fraction measured in Ref.~\cite{Lees:2015eya}.
The three-hadron final states must be corrected for missing isospin-conjugated modes.
Following Ref.~\cite{Lees:2015eya}, we use an average isospin correction multiplicative factor of 
\begin{linenomath}
\begin{align}
f_{\pi\pi} =  \frac{ \mathcal{B}( \overline D^{**} \to \overline D^{(*)\, 0} \pi^+ \pi^-) }{ \mathcal{B}( \overline D^{**} \to \overline D^{(*)} \pi \pi) } = \frac{1}{2} \pm \frac{1}{6} \, ,
\end{align}
\end{linenomath}
whose uncertainty covers the isospin hypotheses for different resonant final states ($f_0(500) \to \pi \pi$ and $\rho \to \pi \pi$ result in $f_{\pi \pi} = 2/3$ and $1/3$, respectively) and non-resonant three-body decays ($f_{\pi\pi} = 3/7$).

Further, it is assumed that the resulting branching fractions saturate the branching fractions of orbitally excited states:
\begin{linenomath}
\begin{align}
 \mathcal{B}(\overline D_2^{*} \to \overline D \pi) + \mathcal{B}(\overline D_2^{*} \to \overline D^* \pi) = 1 \, , \nonumber \\
 \mathcal{B}(\overline D_1 \to  \overline D^{*} \pi) + \mathcal{B}(\overline D_1 \to \overline D \pi\pi)  = 1 \, , \nonumber \\
  \mathcal{B}(\overline D_1^\prime \to \overline D^{*} \pi) = 1 \, , \quad \text{and} \quad
  \mathcal{B}(\overline D^*_0 \to  \overline D \pi)  = 1 \,  \, .
\end{align}
\end{linenomath}

For the $B \to D^{(*)} \, \pi \, \pi \, \ell \, \nu_\ell$ contributions not covered by decays into $D_1 \to D \pi \pi$, we use values measured in Ref.~\cite{Lees:2015eya}. We neglect the small contribution from $B \to D_s^{(*)} \, K \, \ell \, \nu_\ell$ decays.

After this, there is still a difference between the sum of all exclusive modes and the inclusive \bclnu branching fraction of Ref.~\cite{pdg:2020}.
In the following, this missing component contributing to the total branching fraction is referred to as the `gap'.
We fill this gap with equal parts of $B \to D \, \eta \, \ell \, \nu_\ell$ and $B \to D^{*} \, \eta \, \ell \, \nu_\ell$ decays and assign an uncertainty of 100\% to its branching fraction.
These decays are simulated with final-state momenta uniformly distributed in the available phase space or an alternative model involving a broad resonance for the hadronic $X_c$ final state.

\begin{figure}[ht!]
  \centering
  \includegraphics[width=0.47\textwidth]{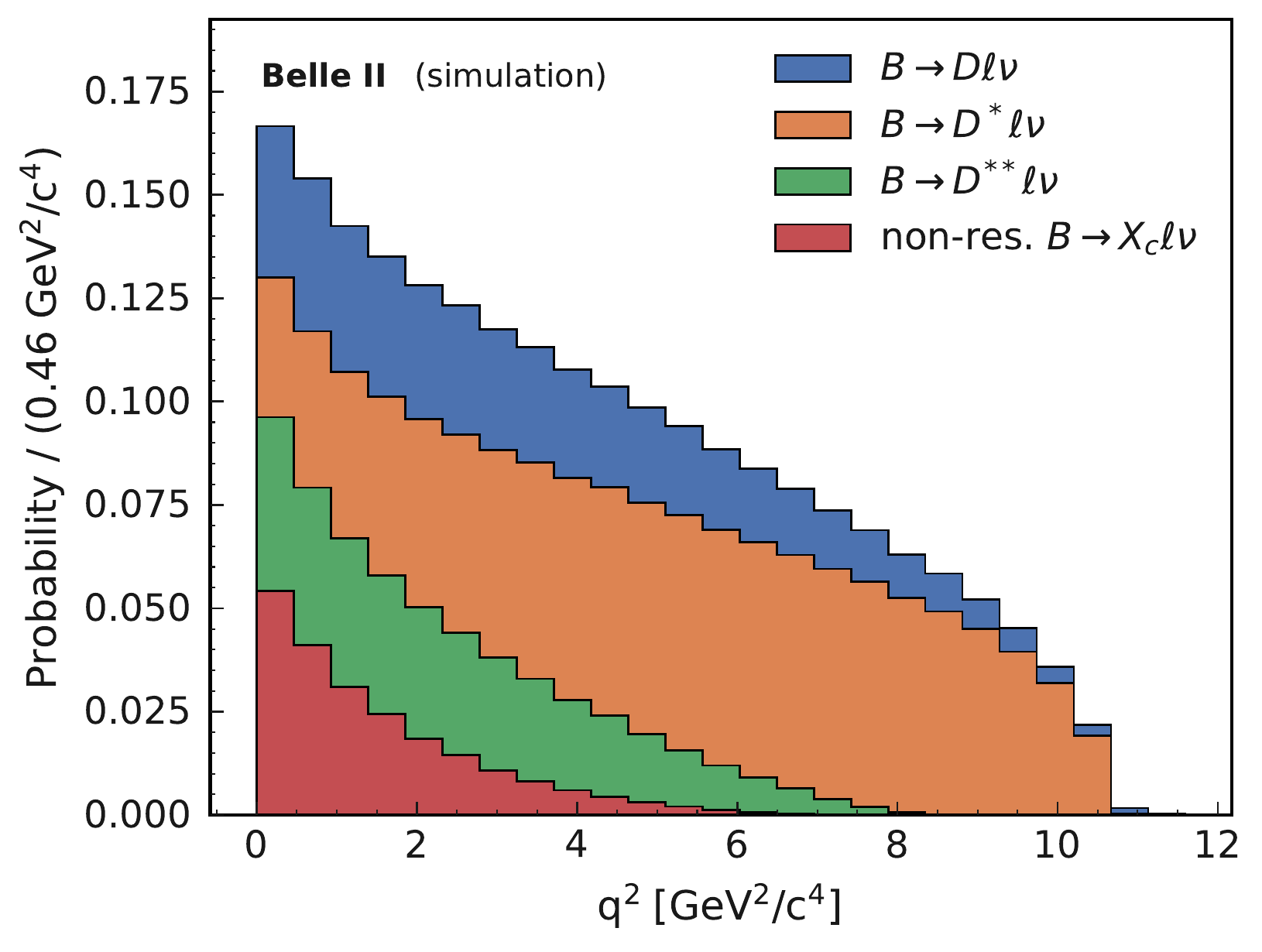}
  \caption{The \qsquared spectrum for different $X_c$ final states without reconstruction effects.}
  \label{fig:generator_q2_spectrum}
\end{figure}

\begin{table}[b]
  \caption{
      Branching fractions used in the simulation of \bclnu.
  }
  \vspace{0.2cm}
  \label{tab:bfs}
  \begin{tabular}{lrr}
 
    \toprule
    Decay & $\mathcal{B}(B^+)$ & $\mathcal{B}(B^0)$ \\
    
    \midrule
    $B \to D \, \ell \, \nu_\ell$ & $\left(2.4 \pm 0.1\right) \times 10^{-2} $ & $\left(2.2 \pm 0.1\right)\times 10^{-2} $ \\
    $B \to D^* \, \ell \, \nu_\ell$ & $\left(5.5 \pm 0.1 \right)\times 10^{-2} $ &$\left( 5.1 \pm 0.1 \right)\times 10^{-2} $ \\
  
    \midrule
    $B \to D_1 \, \ell \, \nu_\ell$ & $\left(6.6 \pm 1.1 \right) \times 10^{-3}$ & $\left(6.2 \pm 1.0\right) \times 10^{-3}$ \\
    $B \to D_2^* \, \ell \, \nu_\ell$ & $\left(2.9 \pm 0.3\right) \times 10^{-3}$ & $\left(2.7 \pm 0.3\right) \times 10^{-3}$  \\
    $B \to D_0^* \, \ell \, \nu_\ell$ & $\left(4.2\pm 0.8\right) \times 10^{-3}$ & $\left(3.9 \pm 0.7\right) \times 10^{-3}$ \\
    $B \to D_1^\prime \, \ell \, \nu_\ell$ & $\left(4.2 \pm 0.9\right) \times 10^{-3}$ & $\left(3.9 \pm 0.8 \right) \times 10^{-3}$  \\
   
    \midrule
    $B \to D \pi \pi \, \ell \, \nu_\ell$ & $\left(0.6 \pm 0.9 \right) \times 10^{-3}$ & $\left(0.6 \pm 0.9 \right) \times 10^{-3}$ \\
    $B \to D^* \pi \pi \, \ell \, \nu_\ell$ & $\left(2.2 \pm 1.0 \right) \times 10^{-3}$ & $\left(2.0 \pm 1.0 \right) \times 10^{-3}$ \\
    $B \to D \eta \, \ell \, \nu_\ell$ & $\left(4.0 \pm 4.0 \right) \times 10^{-3}$ & $\left(4.0 \pm 4.0 \right) \times 10^{-3}$ \\
    $B \to D^{*} \eta \, \ell \, \nu_\ell$ & $\left(4.0 \pm 4.0 \right) \times 10^{-3}$ & $\left(4.0 \pm 4.0 \right) \times 10^{-3}$ \\
    \midrule
    \bclnu & $\left(10.8 \pm 0.4\right) \times 10^{-2} $ & $\left(10.1 \pm 0.4\right) \times 10^{-2} $ \\
   
    \bottomrule
  \end{tabular}
\end{table}

\Cref{fig:generator_q2_spectrum} shows the resulting \qsquared spectrum evaluated without reconstruction effects for the different $X_c$ final states and \cref{tab:bfs} summarizes the semileptonic branching fractions.
 At high $q^2$ contributions from $B \to D^* \, \ell \, \nu_\ell$ dominate, whereas at low $q^2$ $B \to D^{**} \ell \, \nu_\ell$ and non-resonant $X_c$ ($B \to D^{(*)} \, \pi \, \pi \, \ell \, \nu_\ell$ and gap processes) have sizeable contributions.


\section{Inclusive Reconstruction of  \bclnu Decays and Event Selection}\label{sec:analysis_strategy}

\subsection{Tag-side Reconstruction}

We reconstruct $\Upsilon(4S) \to B \bar B$ events with the Full Event Interpretation (FEI) algorithm~\cite{Keck:2018lcd}.  
The algorithm reconstructs one of the $B$ mesons of the $B\bar B$ pair in fully hadronic decays.
In the following, the tag-side $B$ candidate reconstructed by the FEI is denoted as \btag.
The FEI uses a hierarchical bottom-up approach starting with the selection of charged and neutral final-state particles ($e^-$, $\mu^-$, $\pi^-$,  $K^-$, $p$, $\gamma$) from tracks, and ECL clusters, combining them into intermediate particles ($J/\psi, \pi^0, K_S^0, D, D_s, D^*, D_s^*, \Lambda, \Lambda_c, \Sigma^+$), and finally forming \btag candidates.
At each stage, the FEI uses an optimized implementation of gradient-boosted decision trees ~\cite{Keck2017} to estimate the signal probability \FEIprob of each candidate in a distinct decay chain to be correctly reconstructed.
For each candidate, the decision trees combines the signal probability of previous stages with additional kinematic and vertex-fit information.
More than 100 decay channels are reconstructed resulting in $\mathcal{O}(10,000)$ decay chains.

We select events that have at least three charged particles and three ECL clusters to suppress \btag candidates from continuum processes. 
The total visible energy of the event in  the CM frame must be greater than \SI{4}{GeV} and the total energy in the ECL is required to be between 2 and \SI{7}{GeV}.
To reduce continuum background, events must have $R_2 < 0.4$, with $R_2$ the ratio of the second and zeroth Fox-Wolfram moments~\cite{PhysRevLett.41.1581}.
We suppress continuum events by requiring $\cos(\theta_\mathrm{T}) < 0.7$, where $\theta_\mathrm{T}$ is the angle between the thrust axis of the decay products of the \btag and the thrust axis of the rest of the event~\cite{Bevan:2014iga}.
\btag candidates are selected by requiring $\FEIprob>0.01$.
The reconstruction efficiency with this requirement is approximatively $0.26$\% and $0.35\%$ for neutral and charged \btag candidates, respectively. More details on the FEI performance with \b2 data can be found in Ref.~\cite{Abudinen:2020dla}. 

We require \btag candidates to have beam-constrained mass values satisfying
\begin{equation}
  \mbc = \sqrt{\frac{s}{4} - \left| \mathbf{p}_{\btag}^\ast \right|^2} > \SI{5.27}{GeV/\clight^2},
\end{equation}
where $\mathbf{p}_{\btag}^\ast$  is the three-momentum of the \btag candidate. 
The energy difference
\begin{equation}
  \DeltaE = E^\ast_{\btag} - \frac{\sqrt{s}}{2}
\end{equation}
must be within $[-0.15, 0.1] \, \mathrm{GeV}$,
where $E^\ast_{\btag}$ is the energy of the \btag. 
All tracks and ECL clusters not used in the reconstruction of \btag candidate are used to define and reconstruct the signal side.
At this stage we allow for multiple \btag candidates in each event.


\subsection{Signal-side Reconstruction} \label{subsec:signal_side_reconstruction}
Semileptonic $B$ decays are identified by selecting electron and muon candidates with laboratory frame momenta greater than $\SI{0.5}{GeV/\clight}$.
These tracks are required to originate from the IP by requiring $d_{r} < \SI{1}{cm}$ and $\left|d_{z}\right| < \SI{2}{cm}$.
Here, $d_{r}$ and $d_{z}$ are the distances of closest approach to the IP transverse to and along the $z$ axis, respectively. 
Each lepton candidate is required to have a polar angle within the CDC acceptance $[17\degree, 150\degree]$, and at least one hit in the CDC. 

\begin{figure}[ht!]
  \includegraphics[width=0.47\textwidth]{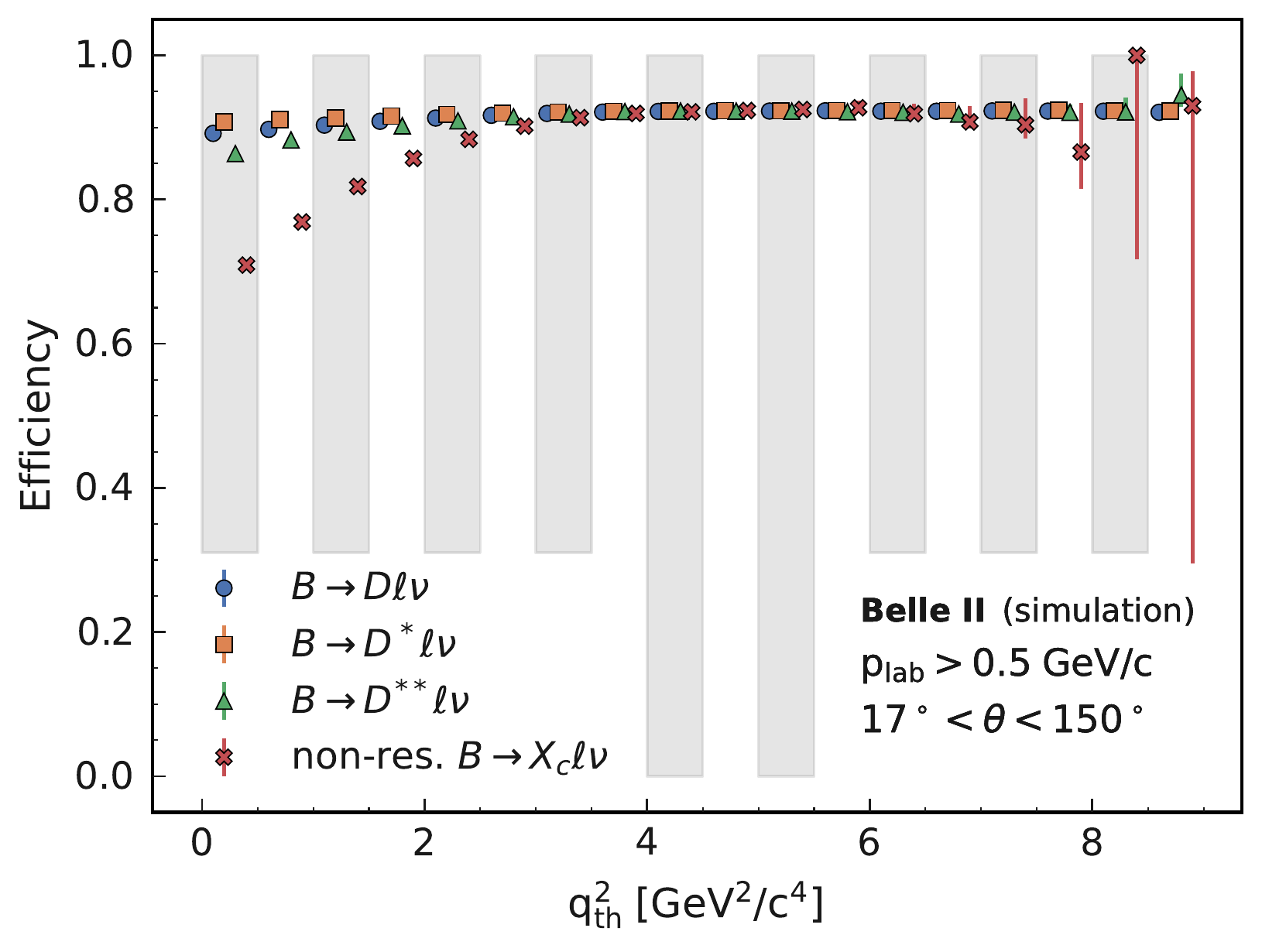} 
  \caption{Selection efficiencies as functions of $q^2$ threshold $q^2_{\mathrm{th}}$. The points for different $X_c$ final states and the same lower \qsquared threshold are shifted horizontally and the grey and white-bands visually group the same $q^2$ threshold.}
  \label{fig:xc_generator_efficiency_plab_theta}
\end{figure}

The momentum and polar angle selection affects the selection efficiency as a function of $q^2$, which is illustrated in \cref{fig:xc_generator_efficiency_plab_theta}.
At low $q^2$ thresholds, the efficiency depends on the final states.
A lower selection efficiency is observed for the $D^{**}$ and non-resonant contributions, introducing a dependence of the moments on modeling of \bclnu. To minimize extrapolation of the moments to unmeasured phase-space regions, we require $q^2 >1.5 \, \mathrm{GeV^2/\clight^4}$.

Lepton candidates are selected using $\mathcal{P}_\ell =  \mathcal{L}_{\ell}/(\mathcal{L}_{e}+\mathcal{L}_{\mu}+\mathcal{L}_{\pi}+\mathcal{L}_{K}+\mathcal{L}_\mathrm{p}+\mathcal{L}_\mathrm{d})$ and we require $\mathcal{P}_\ell > 0.9$ for both electrons and muons. 
To account for the energy of electrons lost to bremsstrahlung photons, the four-momenta of such photons are added to the four-momenta of electrons.
Bremsstrahlung photons are identified using the electron track, extrapolating its PXD and SVD hits and the estimated track intersections with the beam pipe and inner wall of the CDC to the ECL to search for clusters.
ECL clusters with energies between 2\% and 100\% of the electron energy and without any other track association are identified as potential bremsstrahlung photons. 
All clusters that lie within three times the expected resolutions in polar and azimuthal angles are used to correct the electron candidate.
These clusters are then removed from consideration for the remainder of the analysis.
For charged \btag candidates, we require the signal-side lepton have a charge opposite to that of the \btag. 

Particles with transverse momenta less than $\SI{275}{MeV/\clight}$ have radii of
curvature in the magnetic field sufficiently small that they loop within the CDC volume and may be
reconstructed as multiple tracks. 
To identify such tracks, we compare the proximity and the magnitude of the momenta of all low-momentum tracks.
When there are potential duplicates, we select the track with the smallest value of $(5 \times d_{r})^2 + |d_{z}|^2$. 
The size of the scaling factor on $d_{r}$ is optimized to minimize track duplicates. 

After reconstructing the \btag and signal-side lepton candidate, the $X_c$ system is identified as the remaining charged particles and photons.
The four-momentum for a charged particle is calculated from the reconstructed track momentum and the assigned mass hypothesis based on the largest identification probability. 
As we do not explicitly reconstruct  charmed states, we denote the reconstructed system in the following as $X$ and its four-momentum $p_X$ and mass $M_X$. 
A signal-side candidate is rejected if the $X$ system does not contain at least one charged particle and the absolute event charge is $>1$. 

The lepton mass squared is reconstructed as 
\begin{linenomath}
\begin{align}
  \qsquared = (p^\ast_{\bsig} - p^\ast_{X})^2 \, ,
\end{align}
\end{linenomath}
with $p^\ast_{\bsig}=(\sqrt{s}/2, - \mathbf{p}_{\btag}^\ast)$. 
The missing four-momentum in the event is reconstructed as 
\begin{linenomath}
\begin{align}
 p_\mathrm{miss} = p_{e^+ \, e^-} - p_{\btag} - p_{X} - p_{\ell} \, ,
\end{align}
\end{linenomath}
where $p_{e^+ \, e^-}$ is the four-momentum of the colliding electron-positron pair. 
We require $E_{\mathrm{miss}} > \SI{0.5}{GeV}$ and $|\mathbf{p}_{\mathrm{miss}}| > \SI{0.5}{GeV/\clight}$ to improve the resolution on the mass of the hadronic system.
The average multiplicity of  $\btag \ell$~candidates is $1.5$ per event. In each event, we retain only the one with the highest lepton momentum. 
When multiple $\btag \ell$~candidates share the same lepton, one is chosen randomly.

\begin{figure}[t]
  \centering
  \includegraphics[width=0.45\textwidth]{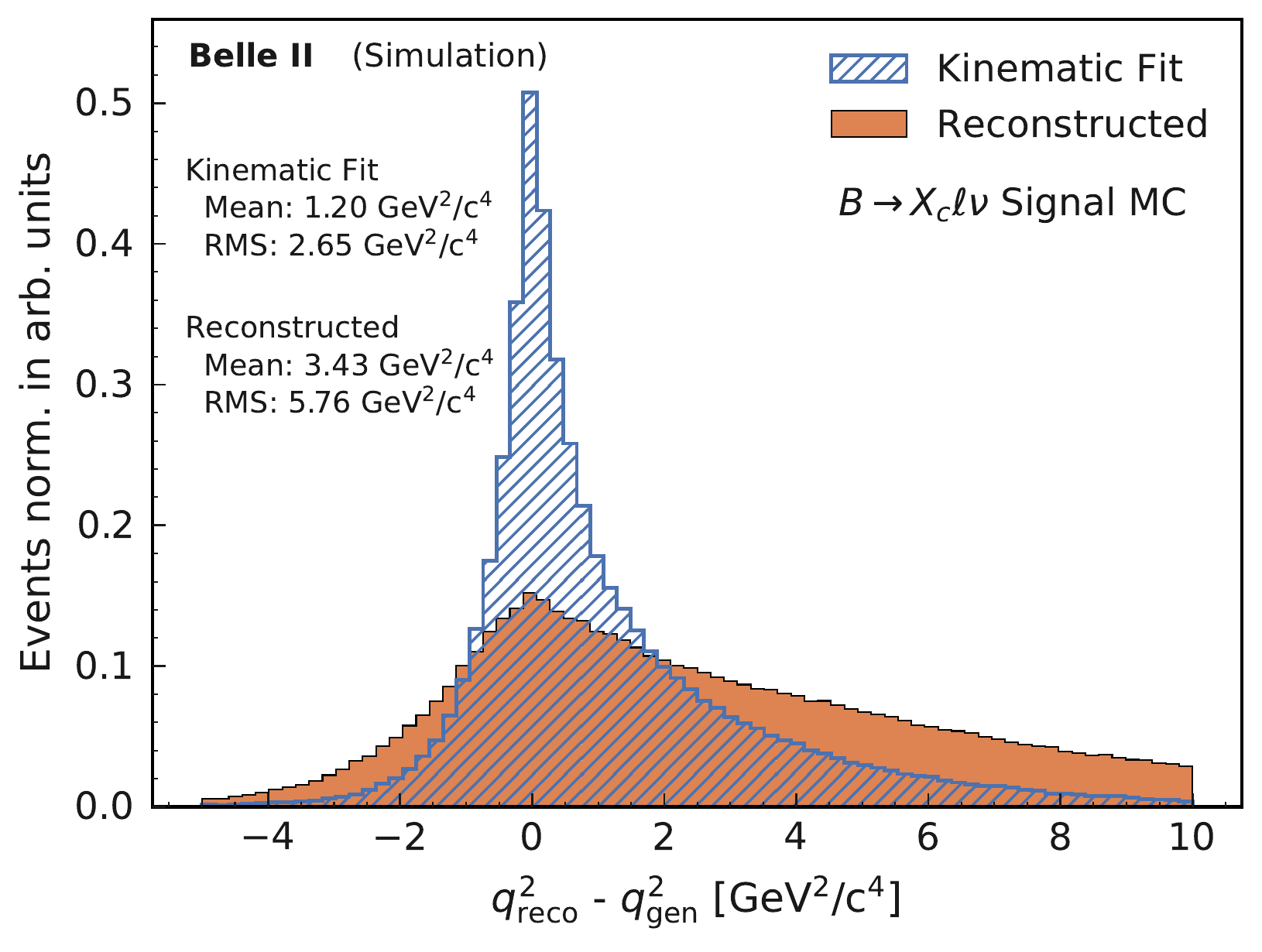}
  \caption{Comparison of reconstructed, fitted and generated \qsquared for \bclnu. The residuals are the difference of generated ('gen') and estimated ('reco') values. 
  }
  \label{fig:q2_mx_fit_vs_reco_resolution}
\end{figure}

To improve the resolution of \qsquared, we exploit the known kinematics of the $e^+\, e^-$ collision and fit for the four-momenta of \btag, $X$, $\ell$, and $\nu_\ell$.
We construct a $\chi^2$ function for each candidate of the form
\begin{linenomath}
\begin{align} \label{eq:chi2_kin}
 \chi^2 = \sum_{i \in \{ \btag, X, \ell \}} \left( \widehat{p}_i - p_i \right) C^{-1}_i  \left( \widehat{p}_i - p_i \right) \, ,
\end{align}
\end{linenomath}
where $ \widehat{p}_i$ is the fitted four-momentum, and $C_i$ is the covariance matrix of the four-momentum of a given final-state particle.
$C_\ell$ is given by the track fit result, while $C_{\btag}$ and $C_X$ are estimated using the corresponding four-momentum residuals.

Overall, we fit 14 parameters: The four-momenta components of the \btag and $X$ candidates and the momenta components of the signal lepton and neutrino.
The energies of the lepton and neutrino are calculated from the momenta assuming $p^2_\ell = m_\ell^2$ and $p^2_\nu = 0$. The kinematic fit is then performed by imposing the following constraints,
\begin{align}
  \widehat{p}_X^{\,2} > 0  \, ,  \,\, \,  \widehat{p}_{\btag}^{\,2}  = m_B^2 \, ,  \,\, \left(\widehat{p}_\ell + \widehat{p}_X + \widehat{p}_\nu \right)^2 = m_B^2 \, ,
\end{align}
and
\begin{align}
\left( \widehat{p}_{e^+ e^-} - \widehat{p}_{\btag} - \widehat{p}_\ell - \widehat{p}_X - \widehat{p}_\nu \right) = 0\,  
\end{align}
using Lagrange multipliers.
For each event the $\chi^2 $ function is numerically minimized with the constraints, following the algorithm described in Ref.~\cite{kraft1988software} implemented in $\mathtt{SciPy}$~\cite{2020SciPy-NMeth}.

Figure~\ref{fig:q2_mx_fit_vs_reco_resolution} show the distribution of the residuals of $q^2$ before and after the kinematic fit with simulated signal events.
Here the residual is calculated from the reconstructed and generated values.
The kinematic fit results in more symmetric residuals and a reduction in the tails of the residuals. The RMS improves from $\SI{5.76}{GeV^2/\clight^4}$ to $\SI{2.65}{GeV^2/\clight^4}$ and the bias reduces from $\SI{3.43}{GeV^2/\clight^4}$ to $\SI{1.20}{GeV^2/\clight^4}$. 


\section{Measurement of Lepton Mass Squared Moments}\label{sec:leptonic_inv_mass_moments}

To measure the lepton mass squared moments, background contributions from other processes must be subtracted from the \qsquared distribution.
Binned likelihood fits are applied to the $M_X$ distribution to determine the number of signal and background events.  
With this information and the shapes of backgrounds from simulation, an event-wise signal probability $w$ is constructed as a function of \qsquared.
We correct for acceptance and reconstruction effects by applying an event-wise calibration $q^2_\mathrm{reco} \to q^2_{\mathrm{calib}}$ and two additional calibration factors \ccalib and \ctrue, discussed in \cref{sec:calibration}. 
The background-subtracted \qsquared moment of order $n$ is calculated as a weighted mean 
\begin{align}\label{eq:moment_calculation}
  \qsqmoment{n} = \frac{\sum_i^{N_\mathrm{data}} w(\qsquared_i) \times \qsquarednevent{n}{calib}{i}}{\sum_j^{N_\mathrm{data}} w(\qsquared_j)} \times \ccalib \times \ctrue \, , \nonumber \\
\end{align}
with sums over all events. 
For each $q^2$ threshold, the binned likelihood fit to $M_X$ is repeated to update the event-wise signal probability weights. 
We use thresholds in the range $[1.5,8.5] \, \mathrm{GeV^2/\clight^4}$ in steps of $\SI{0.5}{GeV^2/\clight^4}$.

\subsection{Background Subtraction}\label{sec:bkg_subtraction}

The likelihood fit to the binned $M_X$ distribution is carried out separately in the  $B^+\ell^-$, $B^0\ell^-$, and $B^0\ell^+$ channels to account for efficiency differences in the FEI algorithm.
Electron and muon channels are not separated. Contributions from \bulnu decays are treated as background and have on average high \qsquared. We suppress this background by fitting the range with $M_X > \SI{0.5}{GeV/\clight^2}$. To determine the number of background events in each of these channels as well as for each $q^2$ threshold, we distinguish the following three event categories: 
\begin{enumerate}
  \item \bclnu signal (with yield $\eta_{\mathrm{sig}}$),
  \item  $e^+ e^- \to q \bar q$ continuum processes ($\eta_{q \bar q}$), and
  \item $B\Bbar$ background dominated by secondary leptons and hadronic $B$ decays misidentified as signal lepton candidates ($\eta_{B\Bbar}$).
\end{enumerate}
The likelihood is the product of Poisson likelihoods for each bin $i$ with $n_i$ observed events and $\nu_i$ expected events, with
\begin{linenomath}
\begin{align} \label{eq:fik}
\nu_i = \sum_{k}\eta_k \, f_{ki}  \, ,
\end{align}
\end{linenomath}
where $f_{ki}$ is the fraction of events of category $k$ reconstructed in bin $i$ as determined with simulated events.
The yield $\eta_{q \bar q}$ is constrained to its expectation as determined from off-resonance data. 
To reduce the dependence on the modeling of signal and backgrounds, the fit is carried out in five $M_X$ bins.
For each channel and \qsquared threshold, an adaptive binning is chosen. 
The likelihood is numerically maximized using the \texttt{Minuit} algorithm~\cite{JAMES1975343} in {\tt scikit\--hep/iminuit}~\cite{iminuit}.

\begin{figure}[t] 
  \centering
  \includegraphics[width=0.42\textwidth]{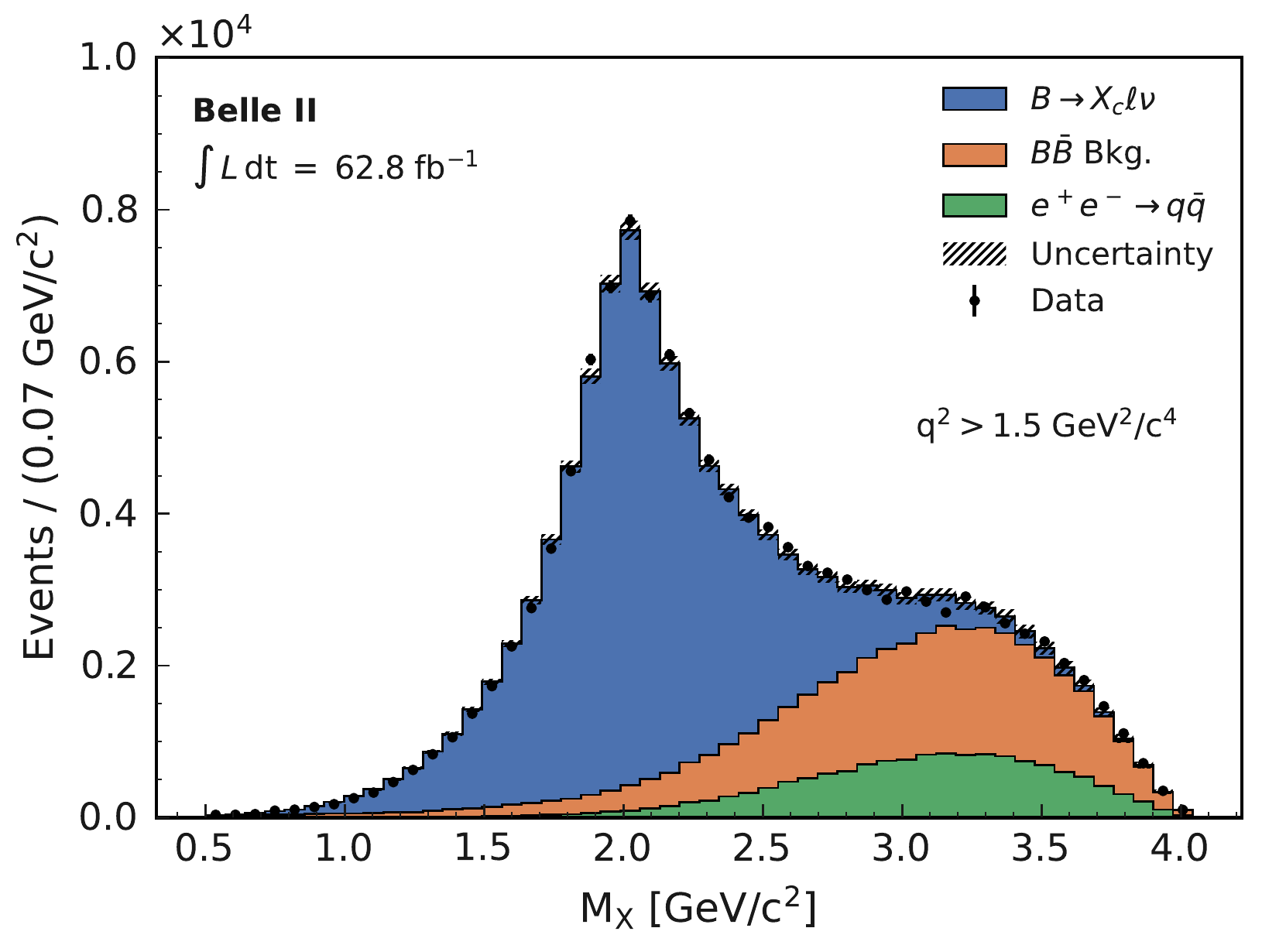}
  \includegraphics[width=0.42\textwidth]{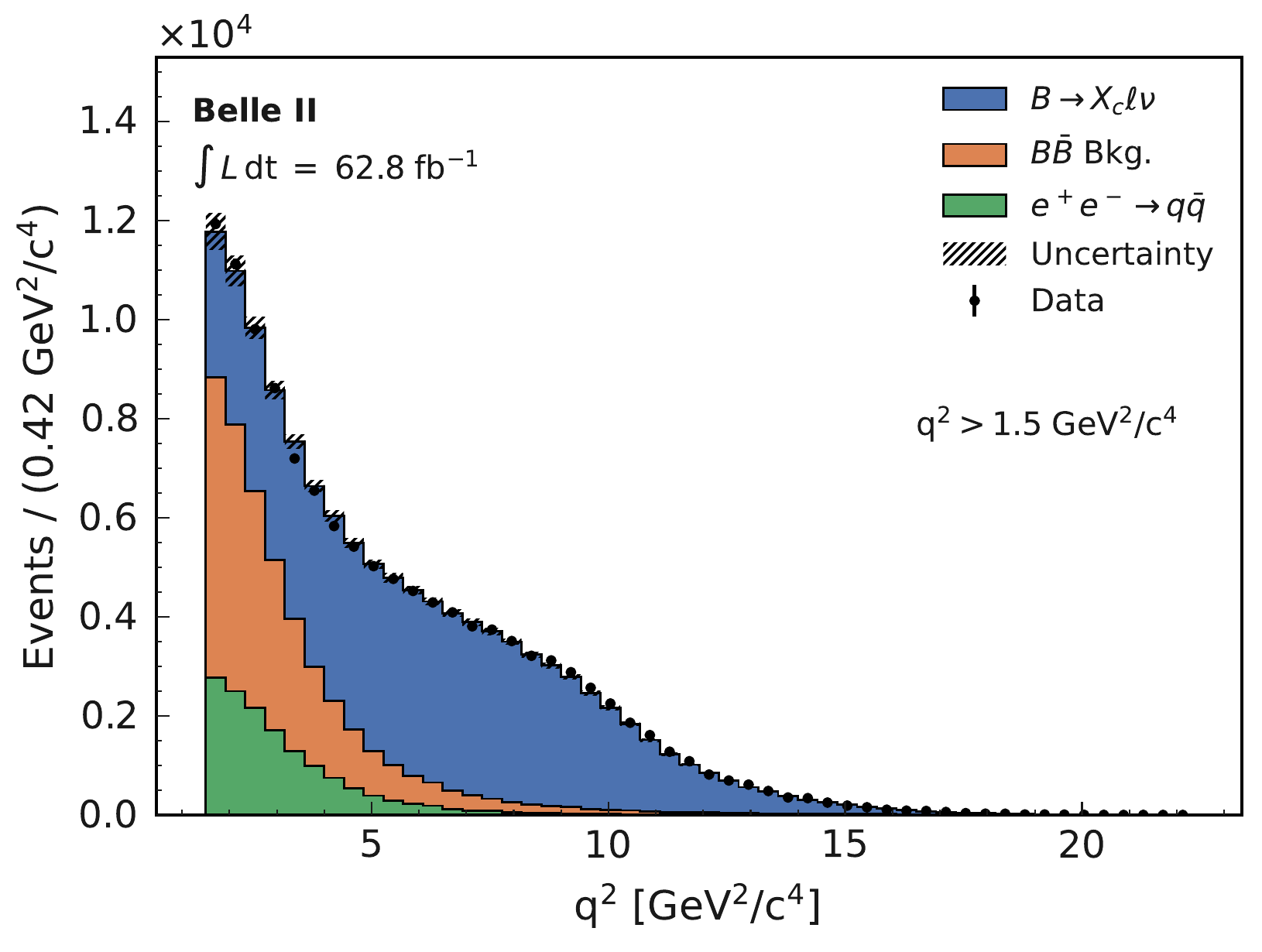}
  \caption{$M_X$ and \qsquared spectra with \bclnu and background components normalized to the results of the $M_X$ fits.}
  \label{fig:q2_and_mx_with_normalization}
\end{figure}

The  sample composition projections for $\qsquared > \SI{1.5}{GeV^2/\clight^4}$ are shown in \cref{app:example_fit}.
The $M_X$ and $q^2$ distributions with the fitted MC yields are shown in \cref{fig:q2_and_mx_with_normalization} for $\qsquared > \SI{1.5}{GeV^2/\clight^4}$ with finer granularity than used in the fit. 
The agreement is fair and the $p$ value from a $\chi^2$ test for the \qsquared distribution in the range of $1.5 - 15 \, \mathrm{GeV^2/\clight^4}$ is 30\%. 

\begin{figure}[t]
  \centering
  \includegraphics[width=0.45\textwidth]{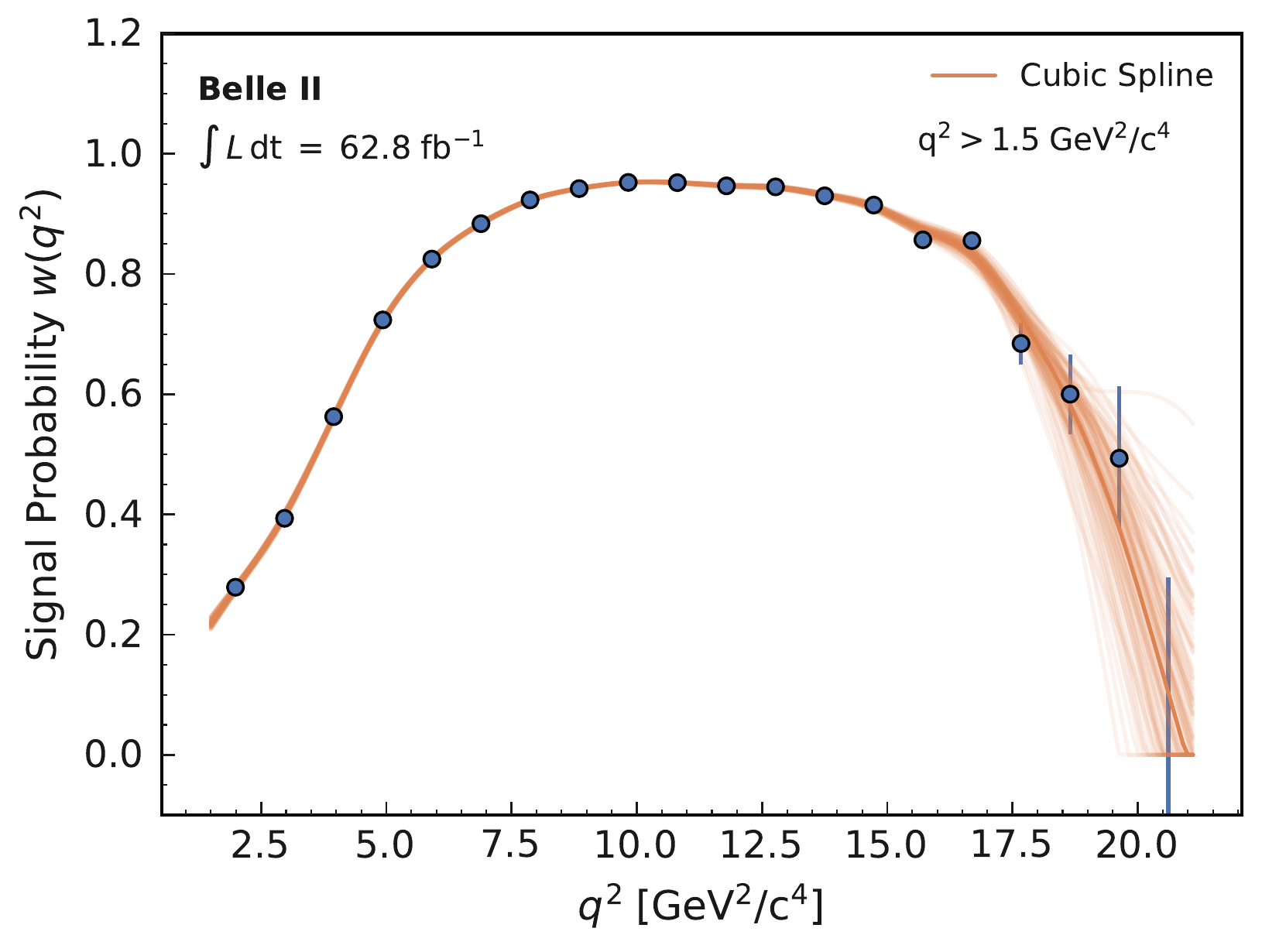}
  \caption{
  Binned signal probability $w_i$ together with a smoothed cubic-spline fit (dark red). In addition, variations of the signal spline fit (light red) determined with bootstrap replicas are shown.} 
  \label{fig:sig_prob_spline2}
\end{figure}

The event-wise signal probability $w$ is obtained by constructing a binned probability as a function of $q^2$ via
\begin{align}
w_i(\qsquared) = (n_i -  \tilde \eta_{B\Bbar} \, \tilde f_{i}^{B\Bbar} -  \tilde\eta_{q \bar q} \, \tilde f_{i}^{q \bar q}  )/n_i \, ,
\end{align}
where $ \tilde f_{i}$ is the estimated fraction of events reconstructed in bin $i$ of $q^2$ for a given background category estimated from the simulation and $\tilde \eta$ denote the sum of the estimated number of background events from the $M_X$ fits. 

We calculate a continuous  signal probability $w(q^2)$ by interpolating the binned distribution with smoothed cubic splines~\cite{csaps}. 
Negative probabilities are set to zero. 
The cubic-spline fit and statistical uncertainties of the signal probability are shown in \cref{fig:sig_prob_spline2}. 
The statistical uncertainty on \qsqmoment{n} is evaluated by a bootstrapping procedure~\cite{Hayes:1988xc} and a selection of spline fits from replicas is shown in \cref{fig:sig_prob_spline2}. 
The statistical uncertainty of $w(q^2)$ increases towards large $q^2$.

\subsection{\qsquared Calibration}
\label{sec:calibration}

\begin{figure}
  \includegraphics[width=0.45\textwidth]{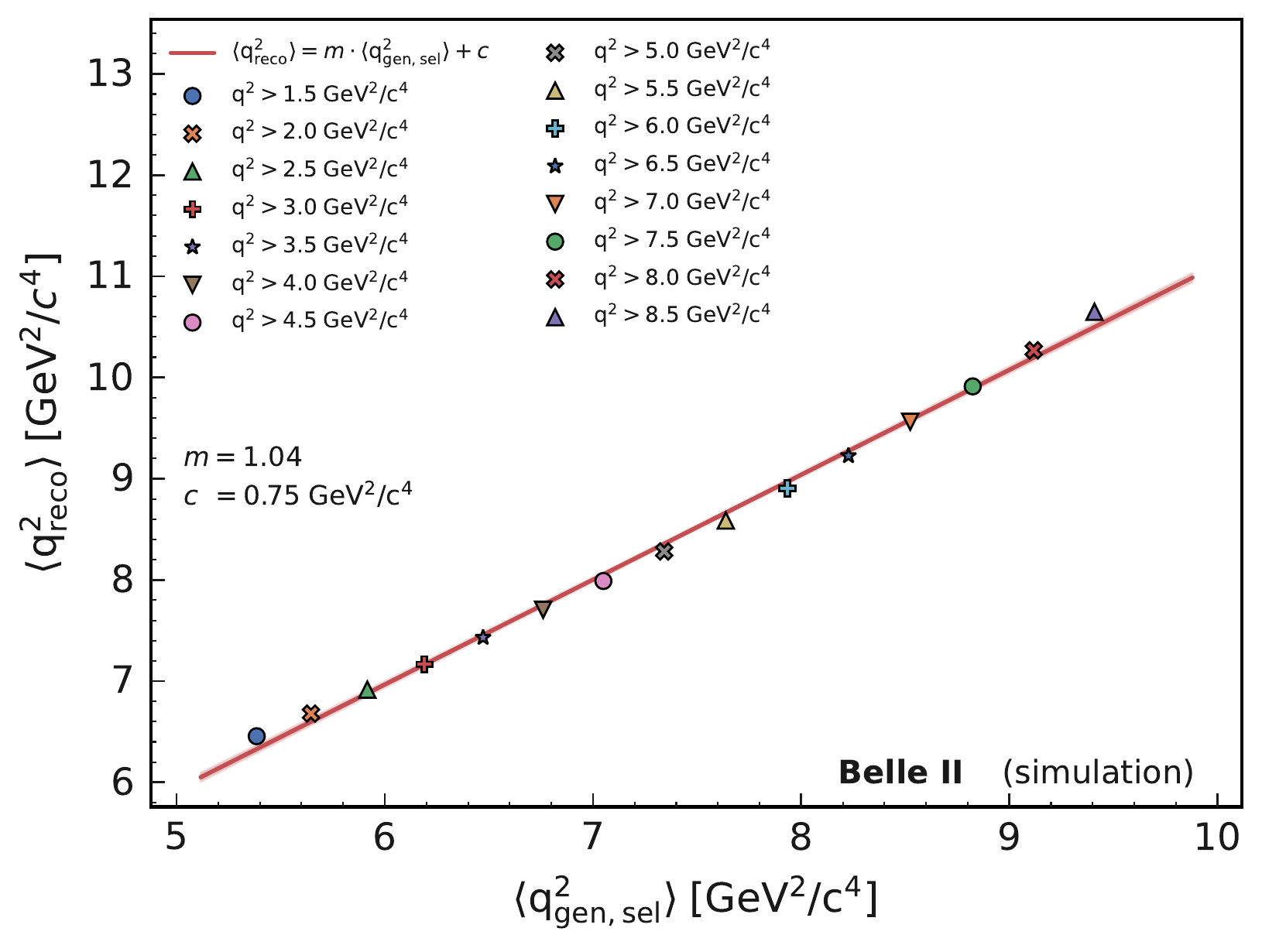}
  \caption{The linear calibration function for the first moment. The first moments are shown as a function of the minimum \qsquared requirement on the reconstructed and true underlying \qsquared distributions. }
  \label{fig:q2_calibration_curves}
\end{figure}

The \qsquared distribution from the kinematic fit is calibrated exploiting the linear relationship between reconstructed and generated moments. 
\Cref{fig:q2_calibration_curves} shows the linear relationship for simulated events for the first moment and as functions of \qsquared threshold between the reconstructed and true \qsquared distribution. 
We calibrate each event with
\begin{linenomath}
\begin{equation}
  \qsquaredncalib{n} = (\qsquarednreco{n} - c_n)/m_n,
  \label{eq:calibration_formula_inverted}
\end{equation}
\end{linenomath}
with $c_n$ and $m_n$ the intercept and slope of the linear relationship for a given moment of order $n$. More details on the linear calibration for the higher moments can be found in \cref{app:lin_calib_functions}.

Due to the linearity of the calibration, a small bias remains, which we corrected with an additional multiplicative calibration factor in \cref{eq:moment_calculation} calculated from simulated events by comparing the calibrated \qsqmomentcalib{n} and true generated \qsqmomenttrue{n} moments,
\
\begin{align}
 \ccalib =   \qsqmomenttrue{n} /  \qsqmomentcalib{n}  \, .
\end{align}
The \btag reconstruction and the \b2 detector acceptance and performance result in an additional bias.
To account for these effects we apply a second multiplicative calibration factor \ctrue\ by comparing the generated moments with all selection criteria applied (\qsqmomenttrue{n}) to their value without any selection applied (\qsqmomentgen{n}),
\begin{align}
 \ctrue =   \qsqmomentgen{n} / \qsqmomenttrue{n}  \, .
\end{align}
The \qsqmomentgen{n} are determined from an MC sample without \textsc{Photos} simulation and also corrects for FSR. 

Both \ccalib\ and \ctrue\ are determined for each \qsquared threshold and from independent samples from those used to determine the linear calibration function. 
The \ccalib factors range between $0.98$ and $1.02$ depending on the lower \qsquared threshold. 
The \ctrue factors vary between $0.90$ and $1.00$ with lower selection threshold values tending to have higher corrections. More details on the event-wise calibration can be found in \cref{app:bias_correction_factors}.

\subsection{Stability Checks}\label{sec:stab_checks}

The variables $M_X$ and $q^2$ are correlated.
We use simulated samples to test the robustness of the background subtraction. 
Tests are carried out with ensembles built from independent simulated samples. 
We observe small deviations of 0.01\% to 0.66\% caused by imperfections in the interpolation of $w(\qsquared)$. 

We also test the impact of systematically altered $q^2$ shapes for \bclnu. 
The altered shapes are obtained by completely removing the non-resonant \bclnu contributions or by applying scaling factors of $2$ or $0.5$ to the dominant \bdlnu or \bdslnu contributions. 
These variations are significantly outside of the quoted uncertainties of Table~\ref{tab:bfs}. 
The moments of the samples with the altered $q^2$ shapes are measured with the nominal \bclnu composition and the observed biases are well within the assigned uncertainties.

The consistency of the measurement for electron and muon final states is checked by separately determining the moments; we find good agreement.


\section{Systematic Uncertainties}\label{sec:systematics}

Several systematic uncertainties affect the \qsquared moments. Their sources can be grouped into two categories.
The first consists of systematic uncertainties originating from background subtraction.
The fit to the $M_X$ distribution assumes the composition of \bclnu and relies on data-driven corrections.
These and other uncertainties affect $w(q^2)$ and must be propagated to the moments.
The second category of uncertainties is related to assumptions when calibrating the moments.
Modeling of \bclnu and of the \b2 detector affects the calibration function and the calibration factors.
To assess the effect of each uncertainty source, we derive alternative sets of moments based on either a varied signal probability function or modified calibration.
The deviation from the nominal result is used to estimate the systematic uncertainty.

\subsection{$M_X$ Fit and Background Subtraction}
We include uncertainties from the signal and background compositions, MC statistics, and the data-driven correction factors directly into the likelihood function of the $M_X$ fit. 
This is achieved by introducing nuisance parameters $\theta_{ki}$ for event category $k$ and bin $i$, which are constrained with multivariate gaussians in the likelihood. The fraction of events is replaced  in \cref{eq:fik} by
\begin{equation}
  \label{eq:shape_nuisance_parameters}
  \frac{ f_{ki} + \sigma_{ki}\theta_{ki} }{\sum_{j}( f_{kj} + \sigma_{kj}\theta_{kj})} 
\end{equation}
and $\sigma_{ki}$ denotes the uncertainty on the fraction for event category $k$ and bin $i$.

The composition uncertainties of \mbox{\bclnu} are determined with the branching fraction uncertainties listed in \cref{tab:bfs}.
We evaluate the uncertainties of the BGL form-factor parameters for \bdlnu, \bdslnu using a set of orthogonal parameter variations for each decay. 
We include the uncertainty of the \bulnu branching fraction from Ref.~\cite{pdg:2020}. The efficiencies for identifying or misidentifying leptons and hadrons are estimated from ancillary measurements. 
We assign a track selection efficiency uncertainty of $0.69\%$ per track on the signal side. 

We propagate uncertainties on PID and tracking efficiencies, the \bulnu branching fraction, and the background yield obtained from the $M_X$ fit to $w_i(q^2)$ with all uncertainties varied according to a multivariate Gaussian distribution. We repeat the analysis with varied histograms and take the variation of the resulting moments as the systematic uncertainties due to these sources. 

We study the impact of the choice of the smoothing factor for the interpolation of the cubic splines used to derive $w(\qsquared)$ and find it to be negligible. 

\begin{figure}[hb!]
  \includegraphics[width=0.39\textwidth]{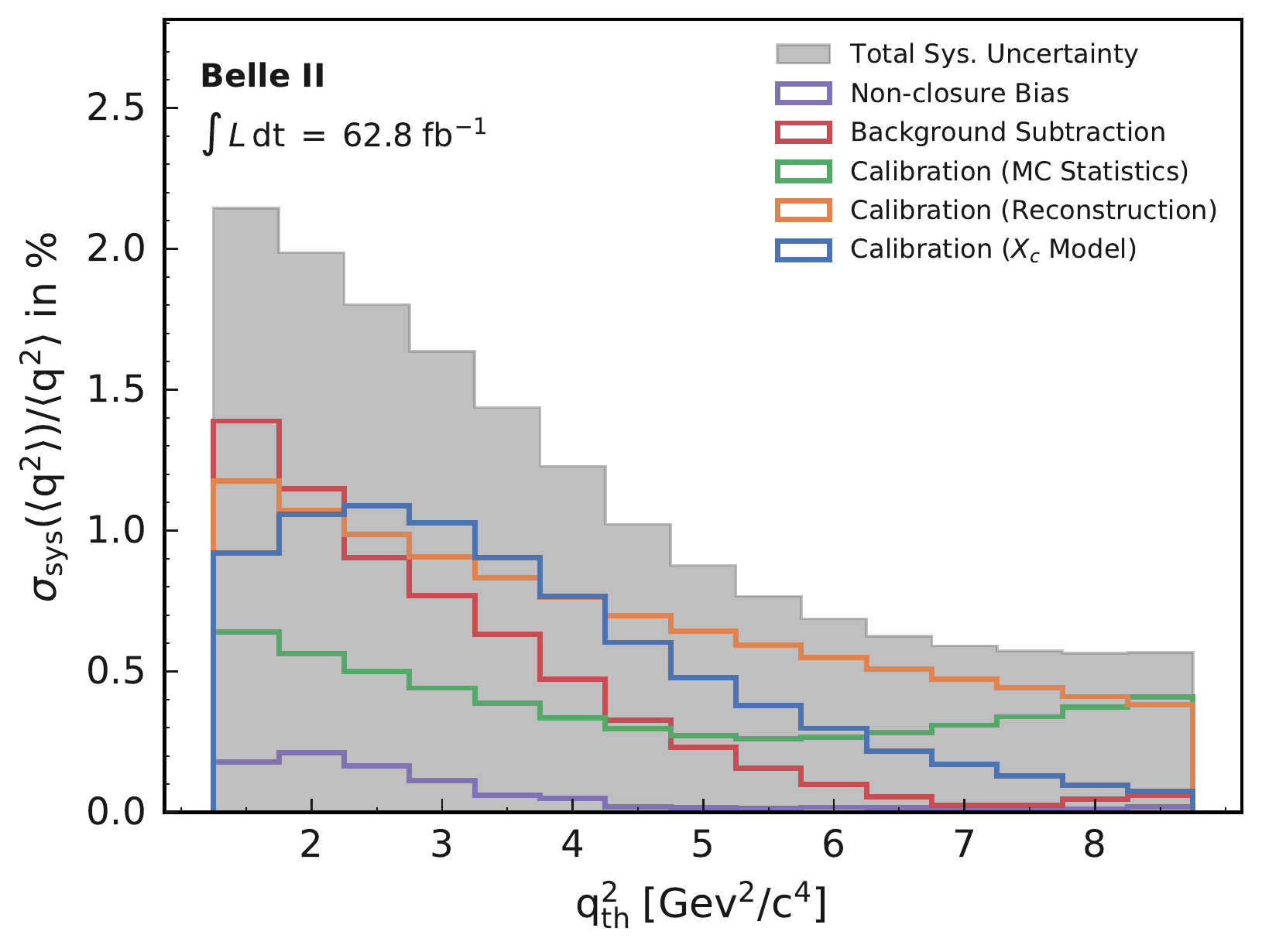}
  \includegraphics[width=0.39\textwidth]{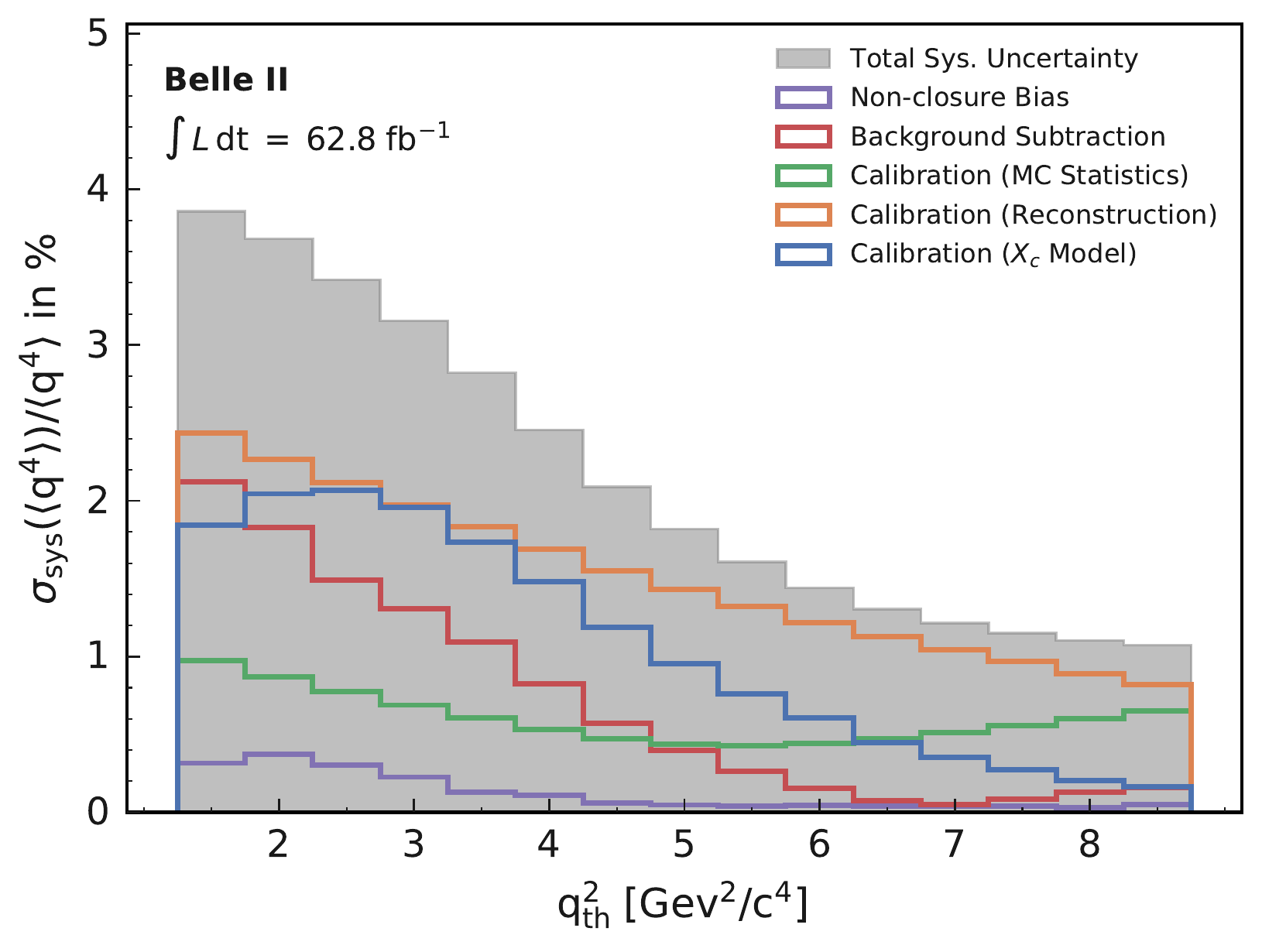}\\
  \includegraphics[width=0.39\textwidth]{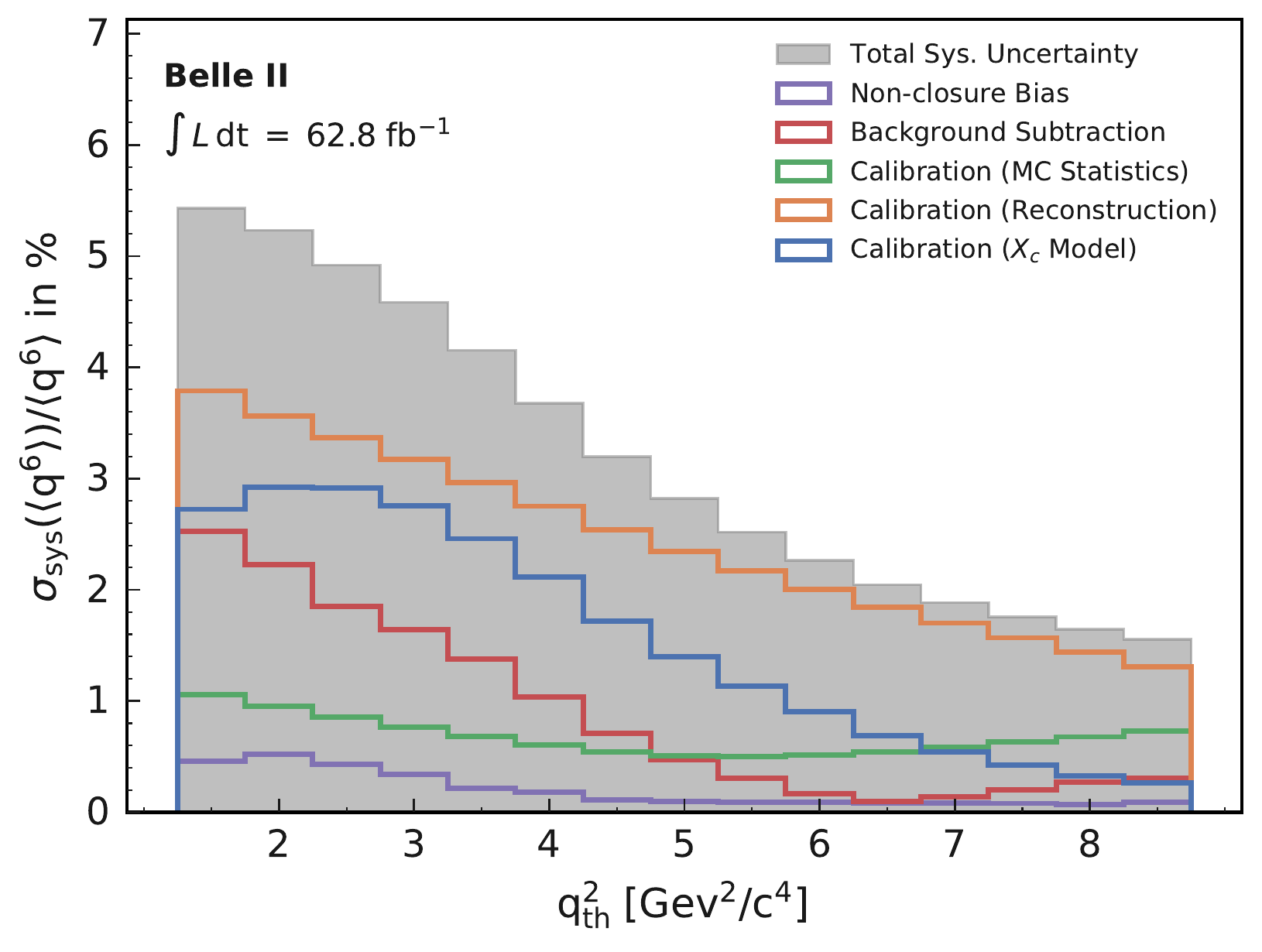}
  \includegraphics[width=0.39\textwidth]{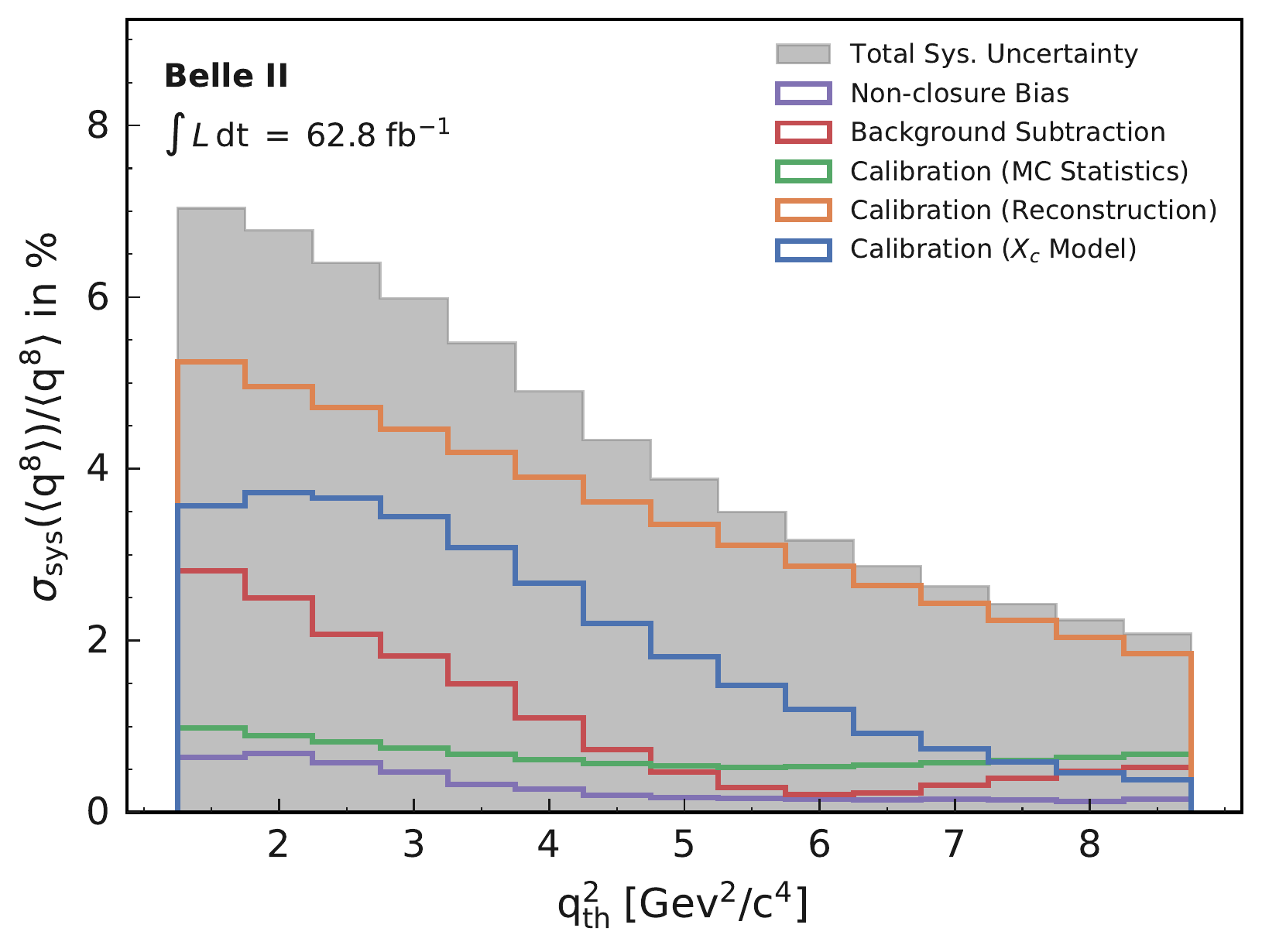}
  \caption{Total (gray) and grouped (colored histograms) relative systematic uncertainties of the raw \qsquared moments as functions of \qsquared threshold are shown.}
  \label{fig:systematics_breakdown}
\end{figure}

\subsection{Calibration of \qsquared Moments}

The calibration curves depend on the composition and modeling of \bclnu. 
We evaluate the impact of the branching fraction uncertainties in $B \to D \ell \bar\nu_\ell$, $B \to D^* \ell \bar\nu_\ell$, and $B \to D^{**} \ell \bar\nu_\ell$ by independently varying the branching fraction of each simulated component by one standard deviation and determining the corresponding variations of the calibration functions and calibration factors. 
To assess the effect of the poorly known non-resonant and gap modes, calibration procedures from two different approaches are compared. 
The first model removes contributions from $B \to D^{(*)} \pi \pi \ell \bar \nu_\ell$ and $B \to D^{(*)} \eta \ell \bar\nu_\ell$ decays. 
The second model replaces them with decays to $D^{**}$ states ($D_0^*$ and $D_1^\prime$). 
Although there is no experimental evidence for additional decays of charm $1P$ states into other final states or the existence of an additional broad state in semileptonic transitions, this provides an alternative kinematic description of the three-body decay, $B \to D^{**}_{\mathrm{gap}} \, \ell \bar \nu_\ell$.
We also evaluate the sensitivity of the calibration functions and factors to the \bdlnu and \bdslnu BGL form-factor parameters.
For each orthogonal variation of the BGL parameters we repeat the calibration.

Modeling of the photon and charged-particle multiplicities directly affects the resolution on $q^2$ and contributes a systematic uncertainty caused by differences between data and MC in how final-state particles are assigned to the signal and tag side.
We select a signal-enriched region by requiring $M_X < \SI[]{3.0}{GeV/\clight^2}$ and $p_\ell^* > \SI{1}{GeV/\clight}$ and calculate correction factors for both multiplicities independently.

We observe differences between data and MC in $E_{\mathrm{miss}} - |\bold{p}_{\mathrm{miss}}|$.
We parameterize the differences using a smoothed cubic spline and correct MC events to evaluate the impact on the calibration.

We evaluate the uncertainty from the track finding efficiency and of PID efficiency on the calibration curves. 

We propagate the statistical uncertainty on the parameters of the calibration function by varying the calibration curve parameters by one standard deviation.
For the calibration factors, we vary the statistical uncertainty on $\ccalib \times \ctrue$ within one standard deviation and repeat the calculation of the \qsquared moments.

The deviation from the closure for the measurement method discussed in \cref{sec:stab_checks} is assigned as an uncertainty.
Its size is subdominant for all moments. 

\subsection{Breakdown of the systematic uncertainties}
\Cref{fig:systematics_breakdown} shows the relative systematic uncertainty for the raw moments.
A more detailed breakdown of the relative systematic uncertainties is given in \cref{sec:app_mom}.
For each moment, the total systematic uncertainty decreases with increasing \qsquared threshold, whereas the statistical uncertainty increases. 
At low $q^2$ thresholds and for the first and second moments, the $q^2$ resolution from mismodeling of the number of charged particles in the $X$ system, the \bclnu\ modeling, and the uncertainty from the background subtraction are of similar size. 

The branching fraction and BGL parameter uncertainties of the resonant decays \bdlnu and \bdslnu are smaller than the uncertainty due to the composition of the higher mass states of the $X_c$ spectrum. 

At high $q^2$ thresholds, MC simulation statistics also can be sizeable sources of uncertainty for the first and second moments. 
For the third and fourth moments, the dominant uncertainty at high $q^2$ thresholds is from the mismodeling of the number of charged particles in the $X$ system, followed by MC simulation statistics, and \bclnu modeling.

\section{Results}\label{sec:results}

\begin{figure*}[ht!]
  \includegraphics[width=0.42\textwidth]{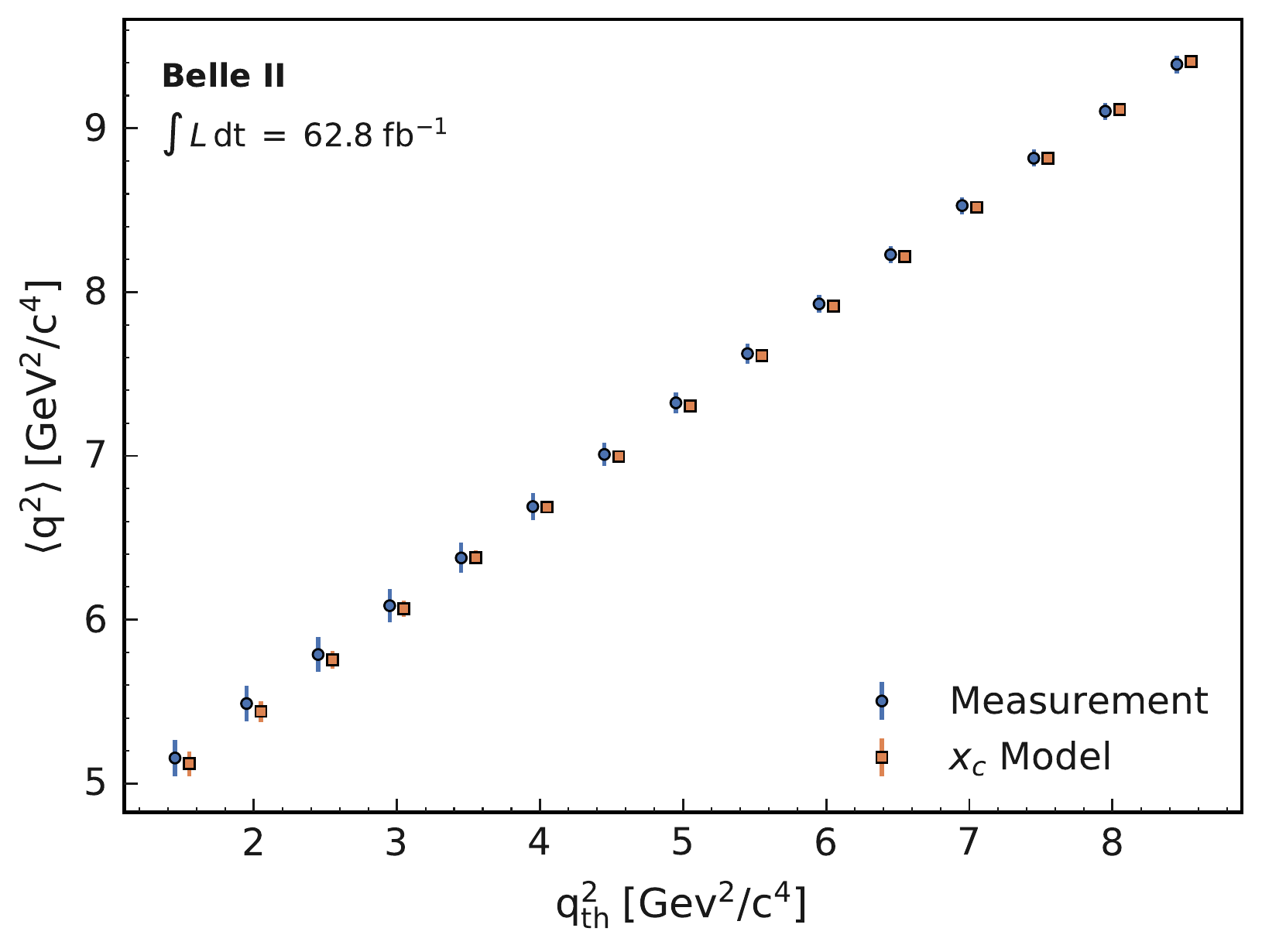}
  \includegraphics[width=0.42\textwidth]{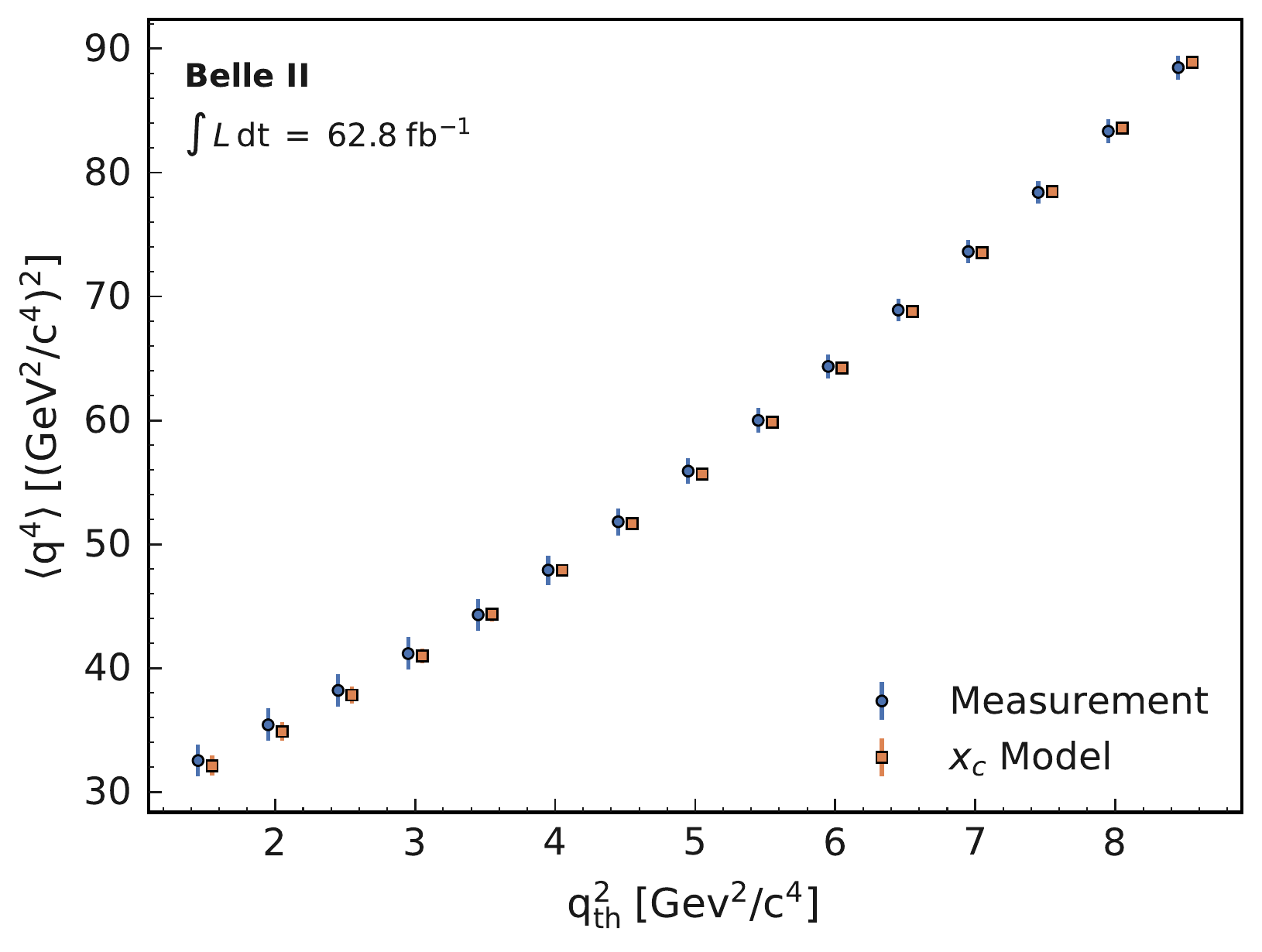}\\
  \includegraphics[width=0.42\textwidth]{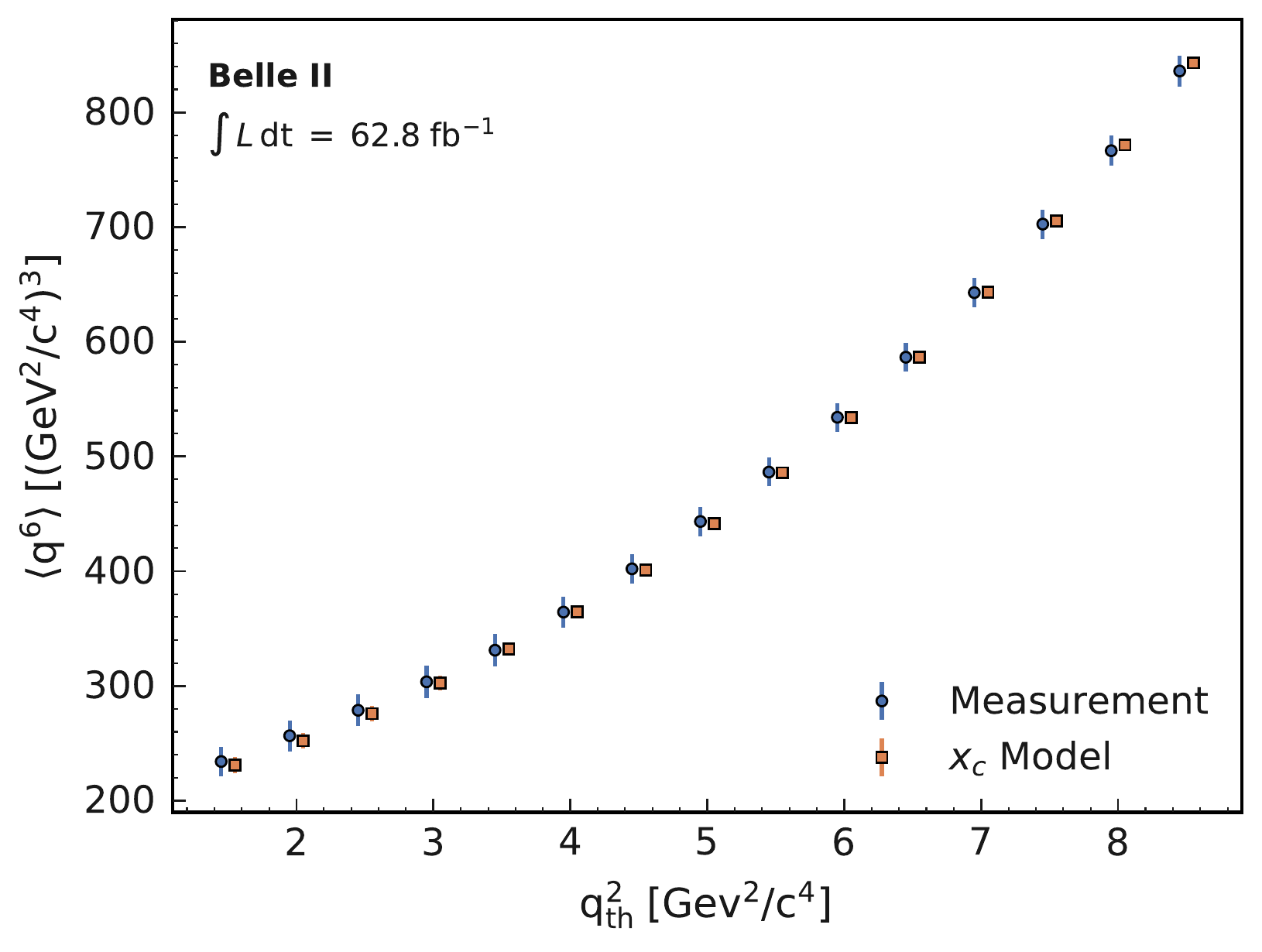}
  \includegraphics[width=0.42\textwidth]{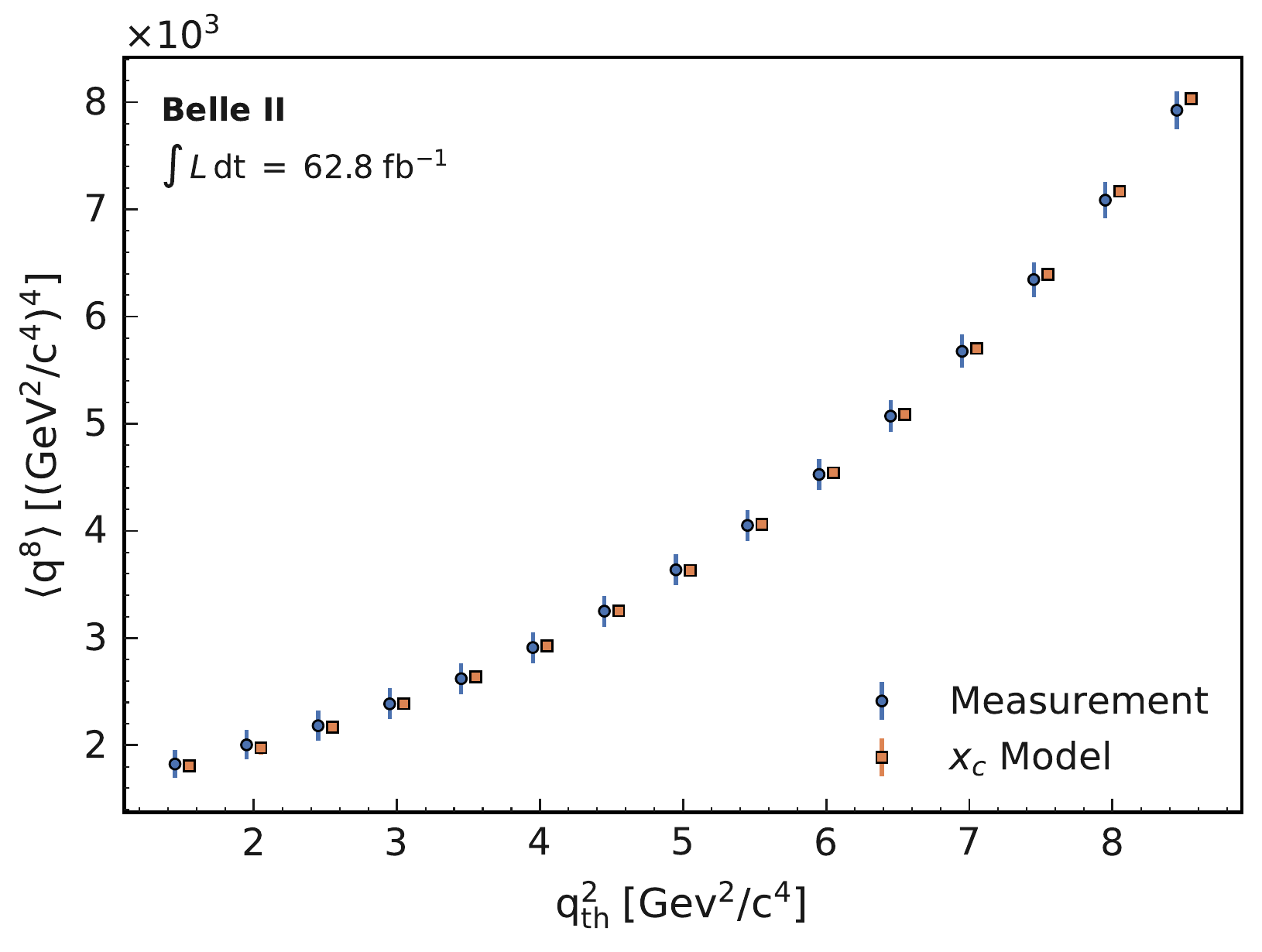}
  \caption{ \qsquared moments (blue) as functions of \qsquared threshold with full uncertainties. The simulated moments (orange) are shown for comparison. }
  \label{fig:results_on_data_q2_moments}
\end{figure*}

The $\qsqmoment{n}$ moments for \mbox{$n=1$--4} are shown in \cref{fig:results_on_data_q2_moments} for \qsquared thresholds ranging from $\SI{1.5}{GeV^2/\clight^4}$ to $\SI{8.5}{GeV^2/\clight^4}$ in $\SI{0.5}{GeV^2/\clight^4}$ increments. 
Numerical values are given in \cref{sec:app_mom} in \cref{tab:results_uncertainties_q2_1,tab:results_uncertainties_q2_2,tab:results_uncertainties_q2_3,tab:results_uncertainties_q2_4}.
 Moments with similar \qsquared thresholds are strongly correlated. 
 The estimated correlation coefficients are given in \cref{app:raw_moment_correlations}. 

\Cref{fig:results_on_data_q2_moments} also shows the moments calculated from the simulated \bclnu\ sample. 
The simulated moments include uncertainties from the \bclnu composition and \bdbsblnu BGL-form-factor parameters.
We observe a fair agreement between measured and simulated moments. 
We compare the raw moments for each order with the simulated moments using $\chi^2$ tests.
To obtain numerically stable results, each test only includes measurements with correlation below $95\%$.
The resulting $p$ values range from 27\% to 94\%.

We calculate values for the central \qsquared moments by expanding the binomial relation
\begin{align}
  \langle (\qsquared - \langle \qsquared \rangle)^n\rangle = \sum_{j=0}^n \binom{n}{j} (-1)^{n-j} \qsqmoment{j} {\qsqmoment{}}^{n-j} \nonumber \\
\end{align}
and applying the following non-linear transformation
\begin{linenomath}
\begin{equation}
  \label{eq:central_eqn}
      \begin{pmatrix}
          \qsqmoment{}\\
          \langle q^4 \rangle\\
          \langle q^6 \rangle \\
          \langle q^8 \rangle
      \end{pmatrix}
      \qquad
      \rightarrow 
      \qquad
      \begin{pmatrix}
          \langle q^{2} \rangle \\
          \langle ( q^{2}- \langle q^{2} \rangle)^{2}\rangle\\
          \langle ( q^{2}- \langle q^{2} \rangle)^{3}\rangle\\
          \langle ( q^{2}- \langle q^{2} \rangle)^{4}\rangle\\
      \end{pmatrix} \, .
\end{equation}
\end{linenomath}
The covariance matrix of the central moments $C^\prime$ is calculated using Gaussian uncertainty propagation \mbox{$C^\prime=J \, C \, J^\intercal$}.
Here, $J$ is the Jacobian matrix for the transformation in \cref{eq:central_eqn}. 

\begin{figure*}[th!]
  \includegraphics[width=0.32\textwidth]{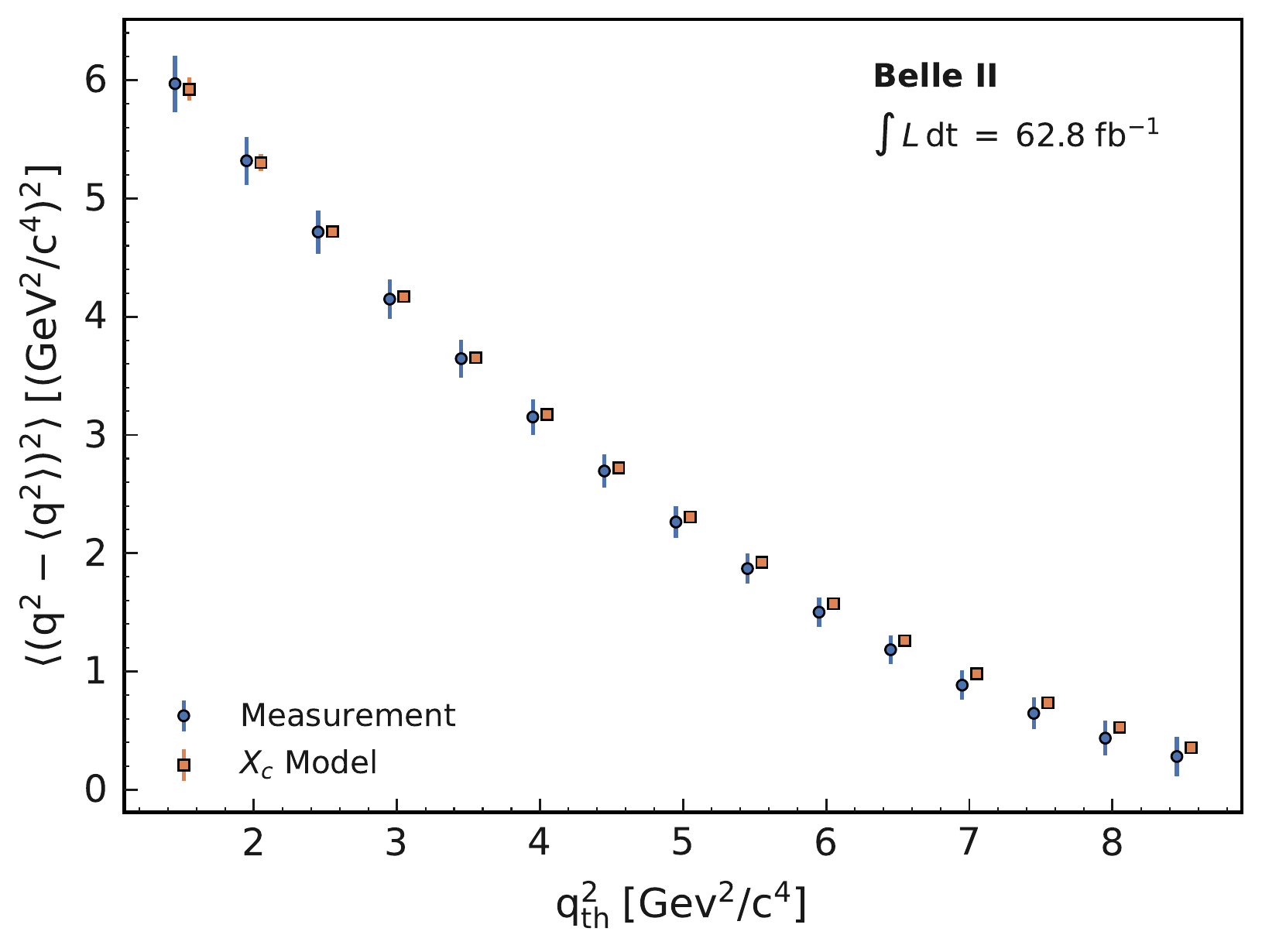}
  \includegraphics[width=0.32\textwidth]{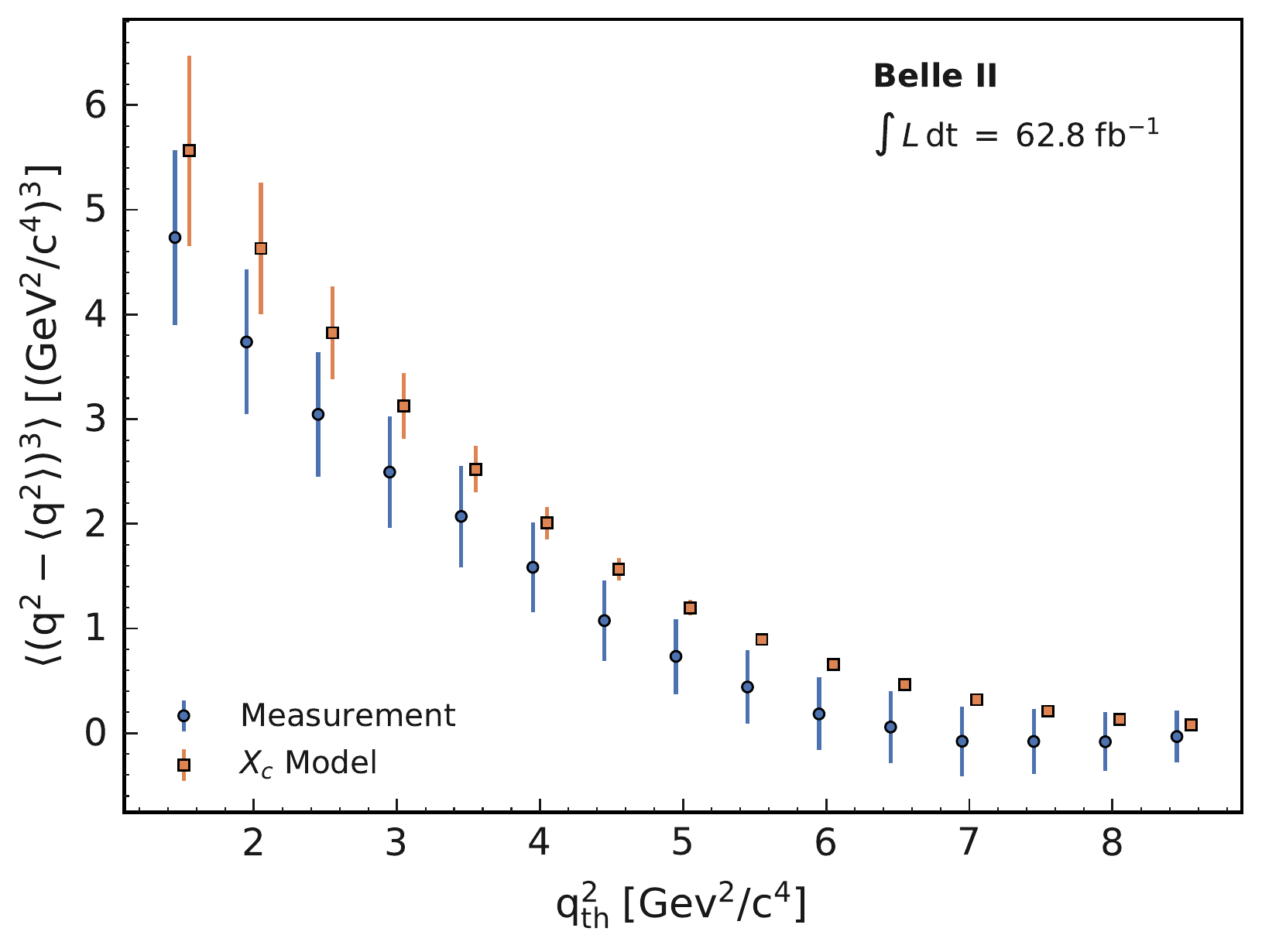}
  \includegraphics[width=0.32\textwidth]{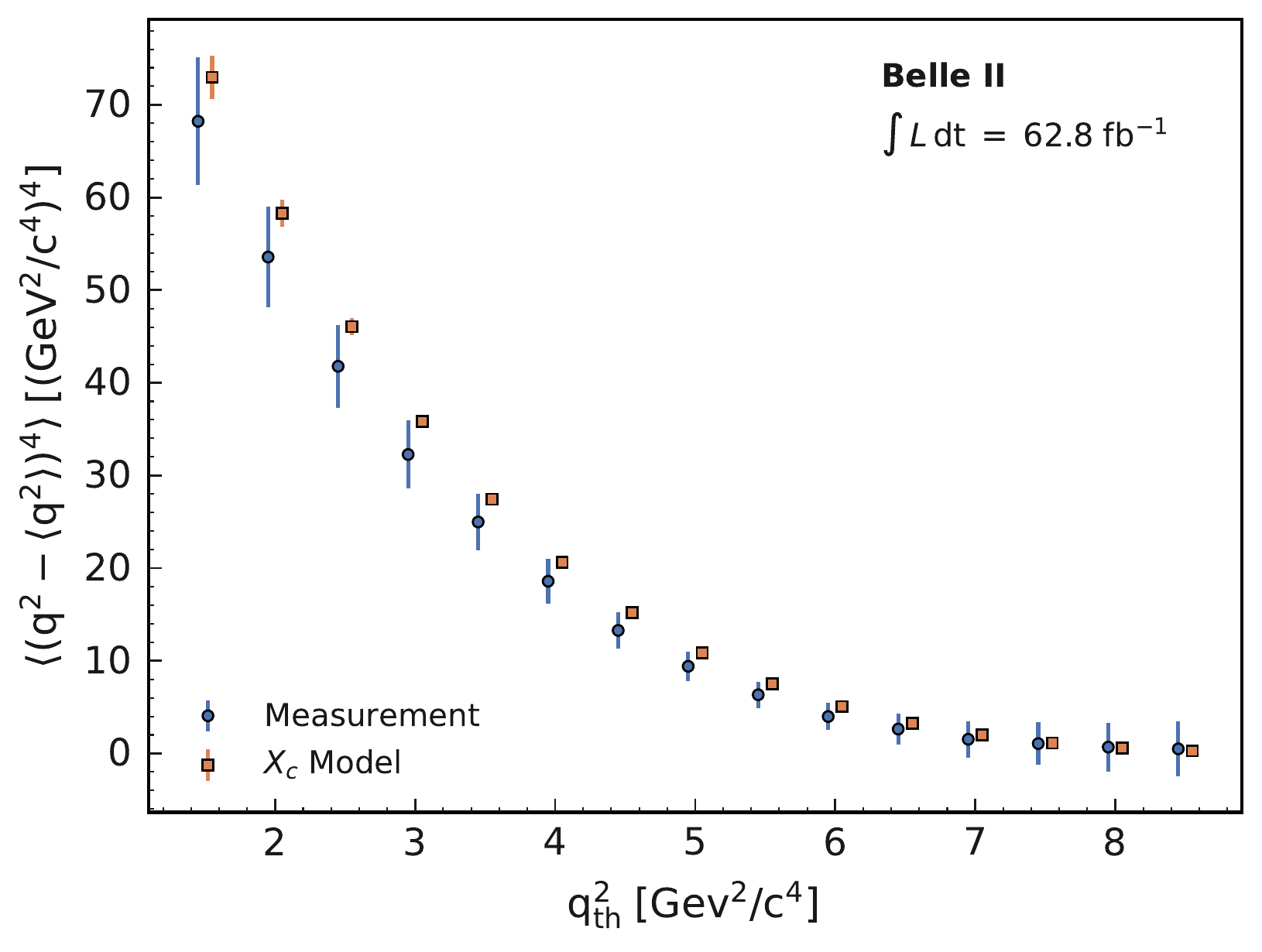}
  \caption{ Central \qsquared moments as functions of \qsquared threshold with full uncertainties. The simulated moments (orange) are shown for comparison. }
  \label{fig:results_on_data_q2_central_moments}
\end{figure*}

\Cref{fig:results_on_data_q2_central_moments} shows the second, third, and fourth central moments as functions of \qsquared threshold.
The central moments are less correlated with each other than the raw moments, but have larger variances. 
We observe negative correlations between some of the central moments. 
The full correlation matrix is given in \cref{app:central_moment_correlations}.
Comparisons of the measured and simulated moments using $\chi^2$ tests show $p$ values greater than $98\%$.

The Belle Collaboration recently presented a measurement similar to this one~\cite{Belle:2021idw}.
This work provides additional new measurements of the raw and central \qsquared moments with comparable precision.
We present measurements starting at lower \qsquared thresholds of $1.5$, $2.0$, and $\SI{2.5}{GeV^2/\clight^4}$, which retain more information about the inclusive $X_c$ spectrum and allow for reductions of the uncertainty on \AbsVcb.
We compare the overlapping measurements of the raw moments from both analyses for \qsquared thresholds between $3.0$ and $\SI{8.5}{GeV^2/\clight^4}$ using a $\chi^2$ test including again only measurements with different lower \qsquared selections having an observed correlation below $95\%$.
The tests yield $p$ values between 5\% and 72\%.
Here, we assumed the systematic uncertainties for the simulation of the $X_c$ spectrum are fully correlated between the Belle and \b2 measurements.

\section{Summary and Conclusion}\label{sec:summary}

We measure the first to fourth moments of the \qsquared spectrum of \mbox{\bclnu} from $1.5$ to $\SI{8.5}{GeV^2/\clight^4}$.
The precise determinations of these moments are a crucial experimental input for determinations of \AbsVcb and HQE parameters, proposed by the authors of Ref.~\cite{Fael:2018vsp}.
This analysis probes up to 77\% of the accessible \bclnu phase space, improving on the measurement of Ref.~\cite{Belle:2021idw}, and includes the experimentally challenging \qsquared region of $ [ 1.5, 2.5 ] \, \mathrm{GeV^2/\clight^4}$. 
The measured moments are also transformed into central moments, which are less correlated, but have larger variances than the raw moments.

The uncertainty for the \qsquared moments is dominantly systematic, with the uncertainties from the background yield and shape, composition of the $X_c$ system, and the simulated detector resolution dominating.
A better understanding of the detector and backgrounds will lead to a more precise determination of the \qsquared moments in the future and will allow measurements with a \qsquared threshold below $\SI{1.5}{GeV^2/\clight^4}$.

We provide numerical results and covariance matricess on HEPData (\href{https://www.hepdata.net}{https://www.hepdata.net}).

\acknowledgments
We thank Keri Vos, Kevin Olschewsky, and Matteo Fael for useful discussions.
We thank the SuperKEKB group for the excellent operation of the
accelerator; the KEK cryogenics group for the efficient
operation of the solenoid; the KEK computer group for
on-site computing support.

\bibliographystyle{apsrev4-1}
\bibliography{BtoXclnu}


\clearpage

\onecolumngrid
\begin{appendix}

\section{$M_X$ Fit Results for $q^2 > \SI{1.5}{GeV^2/\clight^4}$ }\label{app:example_fit}

\begin{figure}[h]
  \includegraphics[width=0.45\textwidth]{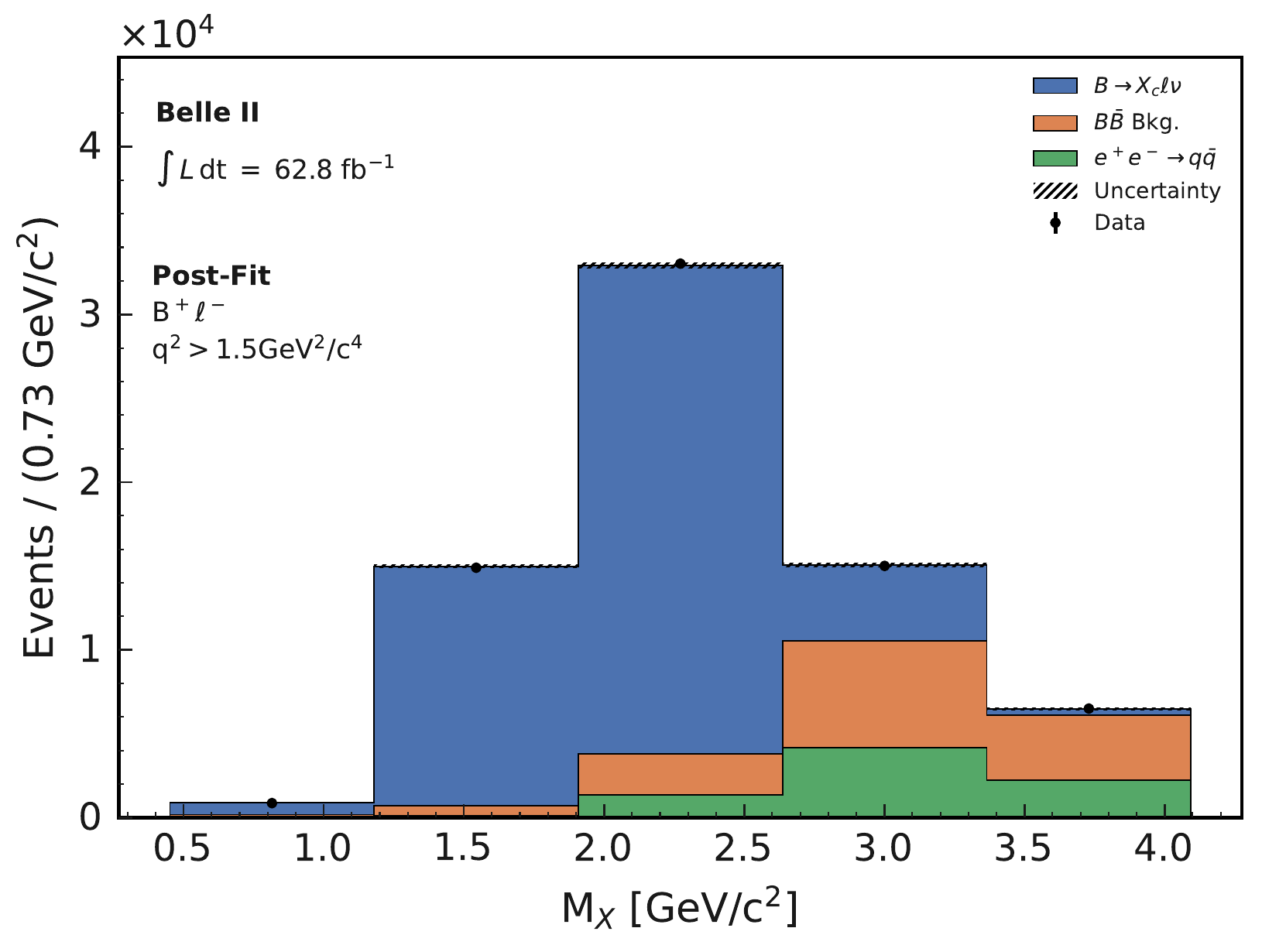} 
  \includegraphics[width=0.45\textwidth]{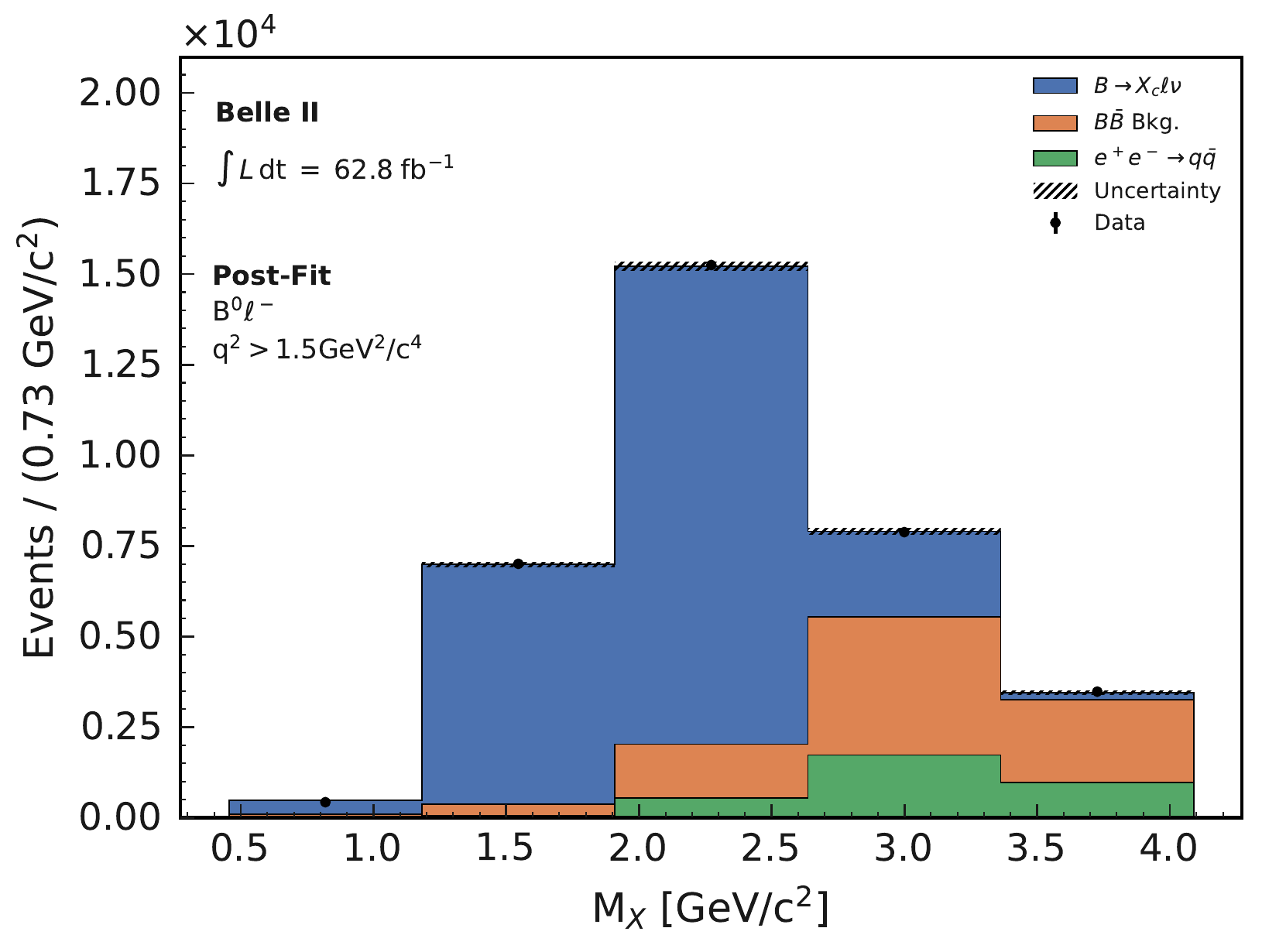} \\
  \includegraphics[width=0.45\textwidth]{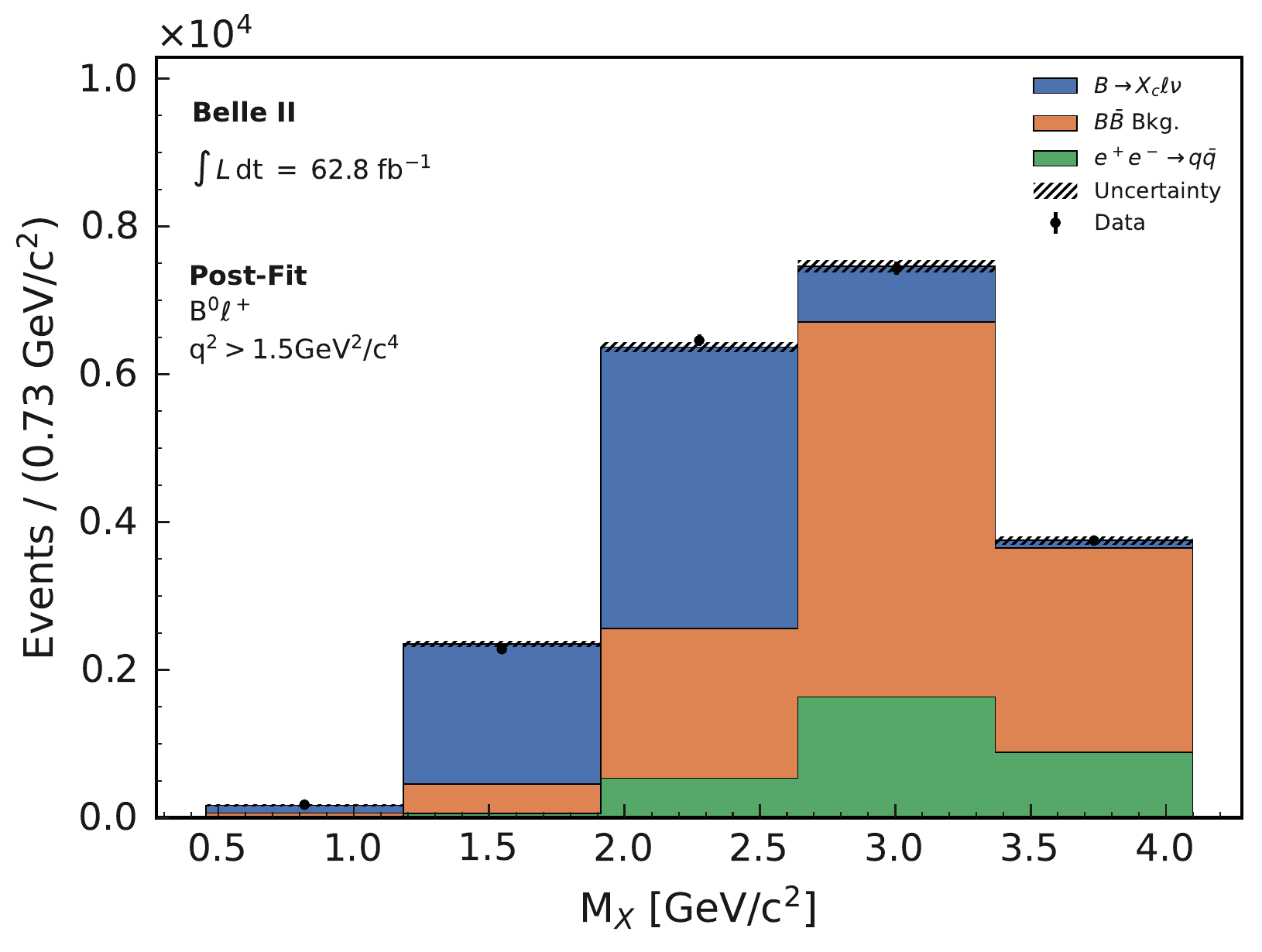}
  \caption{
    Fits to $M_X$ for $B^+ \ell^-$, $B^0 \ell^-$, $B^+ \ell^+$ tag candidates for $q^2 > \SI{1.5}{GeV^2/\clight^4}$.
  }
  \label{fig:mX_fit_first_cut}
\end{figure}

Figure~\ref{fig:mX_fit_first_cut} shows the binned likelihood fits of $M_X$ for $B^+ \ell^-$, $B^0 \ell^-$, and $B^+ \ell^+$ tag candidates for $q^2 > \SI{1.5}{GeV^2/\clight^4}$. 
The fit uses a coarse binning in $M_X$ to reduce the dependence of the composition and modeling of the \bclnu transition (blue). 
The background contribution from continuum (green) is constrained to its expectation, whereas background contributions from $B$ meson decays (due to secondary and fake leptons, orange) are allowed to float.
The fits incorporate nuisance parameters for all templates to account for systematic uncertainties. 
The total uncertainty on the sum of the post-fit templates is shown as hatched histograms.

\clearpage

\section{Linear Calibration Functions}\label{app:lin_calib_functions}

\begin{figure}[h]
  \includegraphics[width=0.45\textwidth]{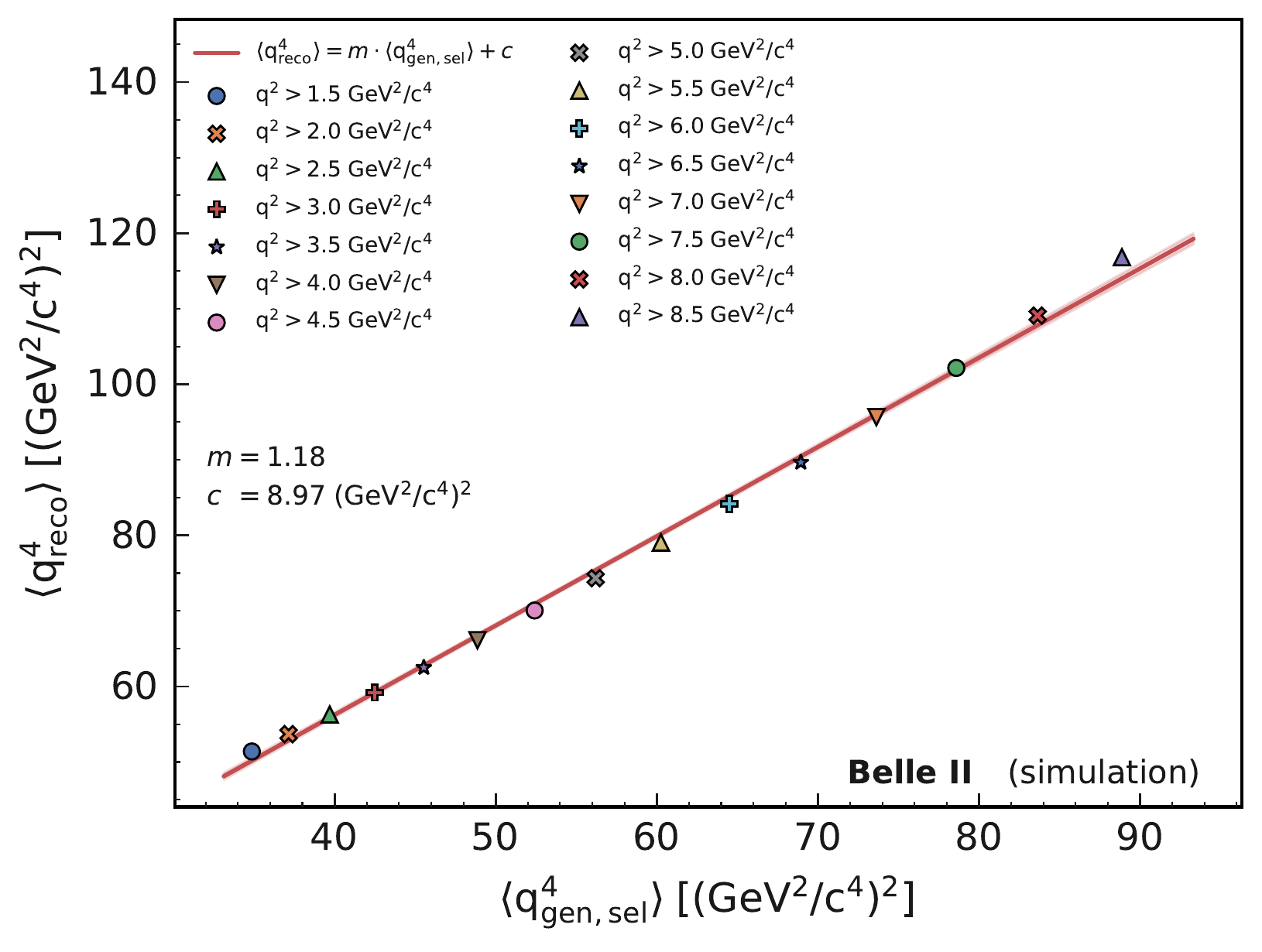} 
  \includegraphics[width=0.45\textwidth]{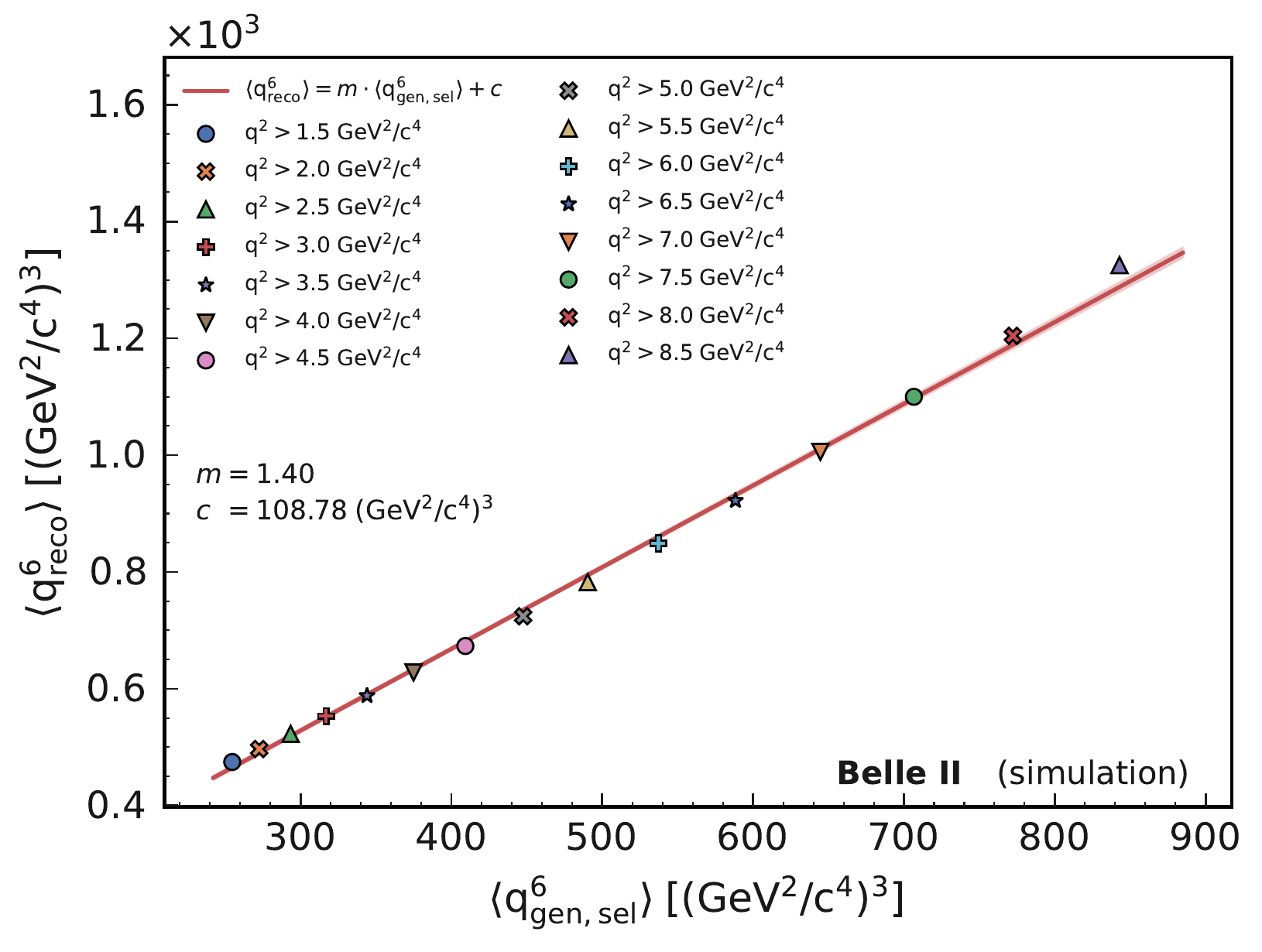} \\
  \includegraphics[width=0.45\textwidth]{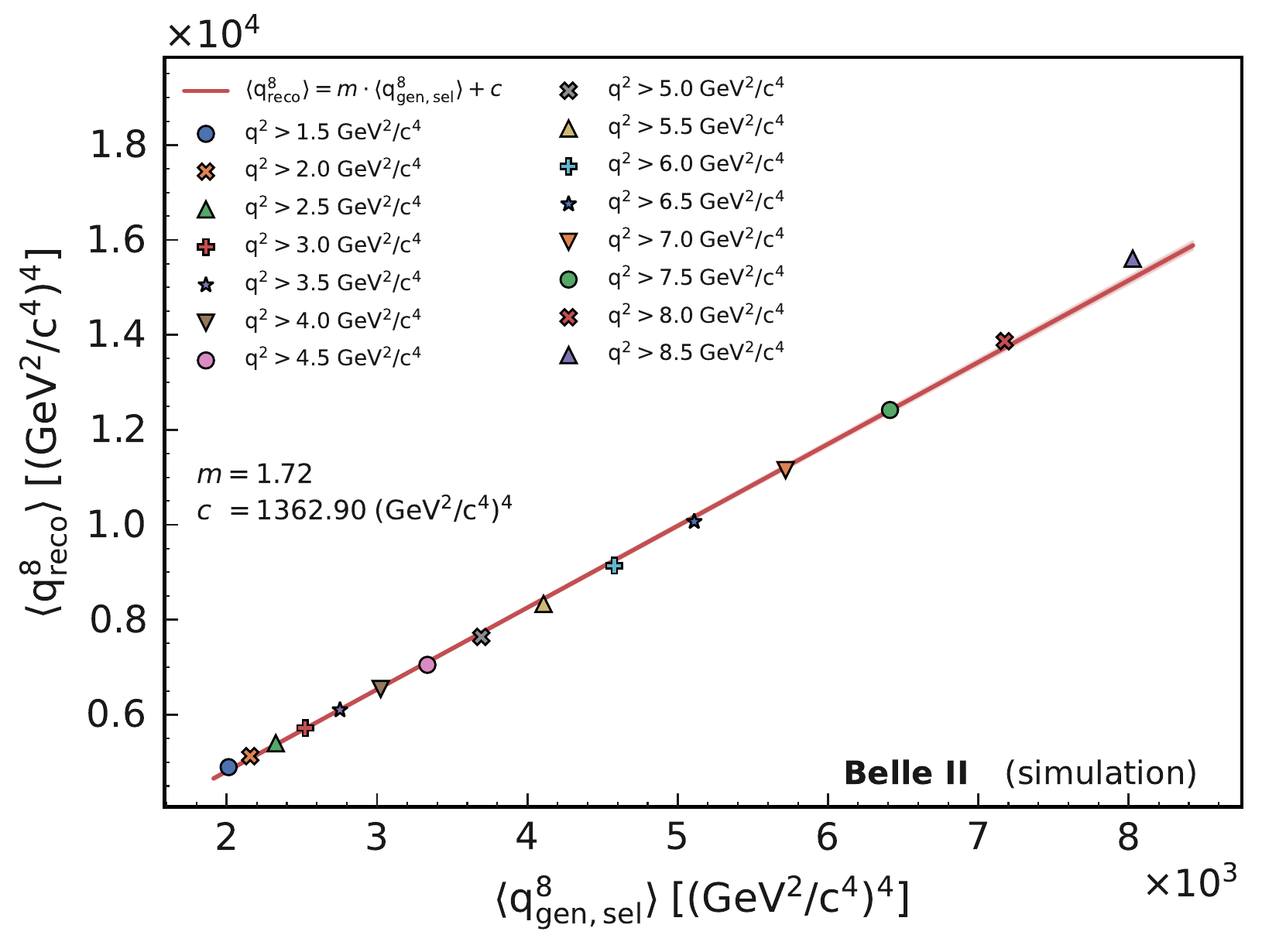} 
  \caption{ Values of the calibration curves (line) for the second to the fourth moment. }
  \label{fig:q2_calibration_curves_app}
\end{figure}

\cref{fig:q2_calibration_curves_app} shows the linear relationships for the second to fourth moments, which are used to derive the linear calibration functions $  \qsquaredncalib{n} = (\qsquarednreco{n} - c_n)/m_n$. 
The moments are shown as functions of \qsquared threshold on the reconstructed and true underlying \qsquared distributions. 
The obtained values for $m_n$ range from $1.18$ to $1.72$. 
The values of $c_n$, parameterizing the overall shift between reconstructed and generated moments, range from $\SI{8.97}{(\mathrm{GeV^2/\clight^4})^2}$ to $\SI{1362.9}{(\mathrm{GeV^2/\clight^4})^2}$.

g

\clearpage

\section{Calibration Factors \ccalib and \ctrue}\label{app:bias_correction_factors}

\begin{figure}[h]
  \includegraphics[width=0.45\textwidth]{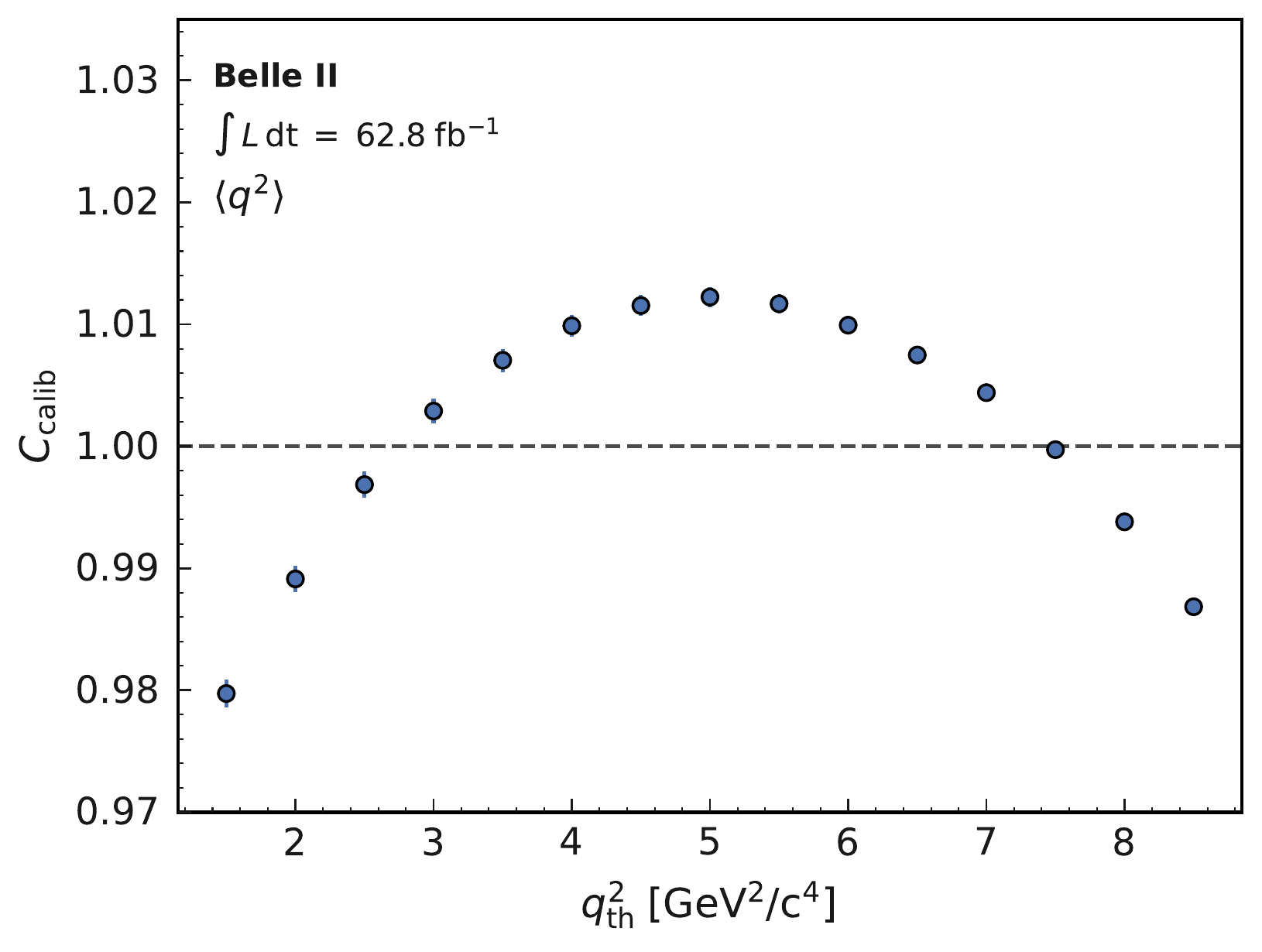}
  \includegraphics[width=0.45\textwidth]{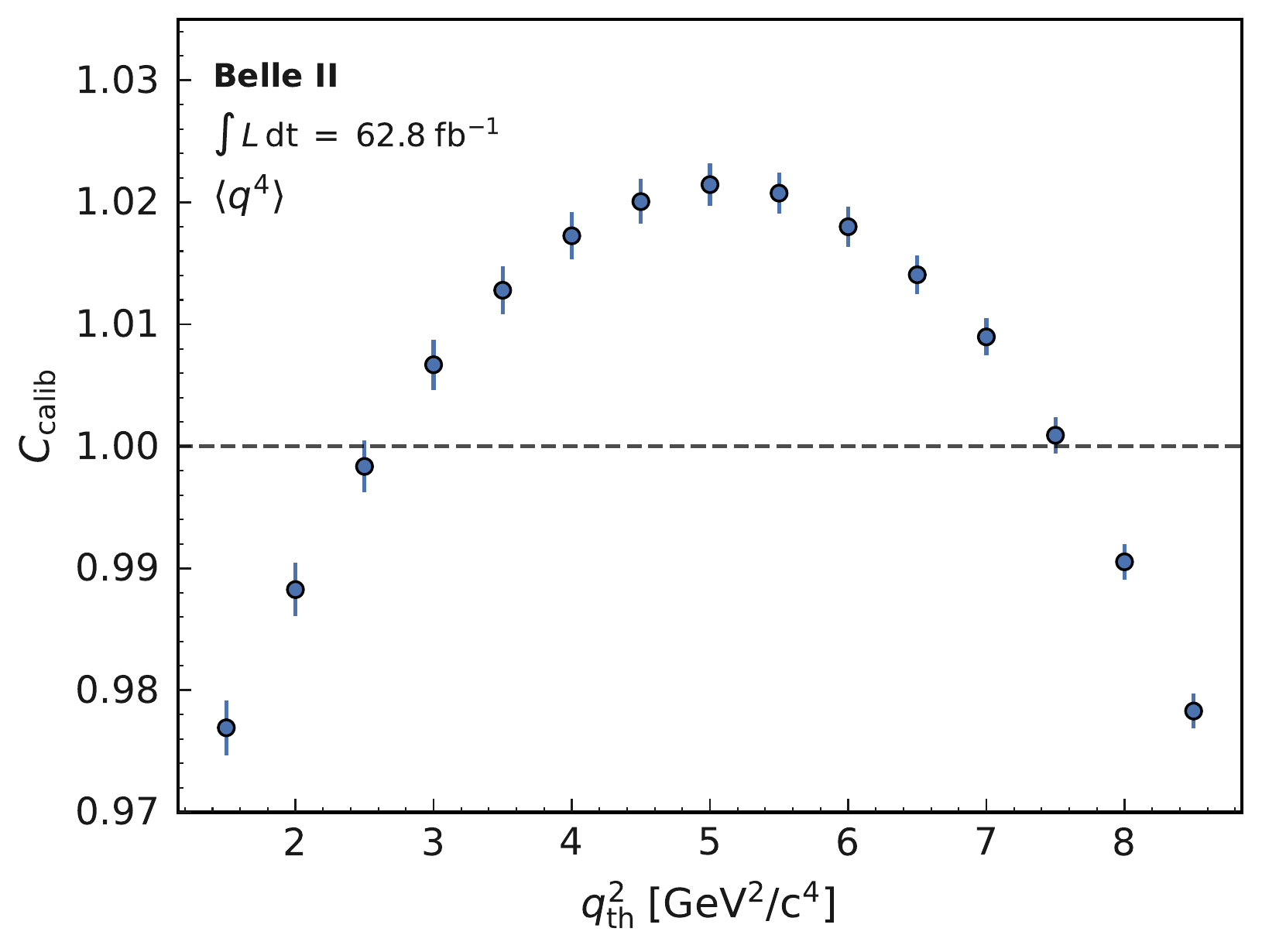}\\
  \includegraphics[width=0.45\textwidth]{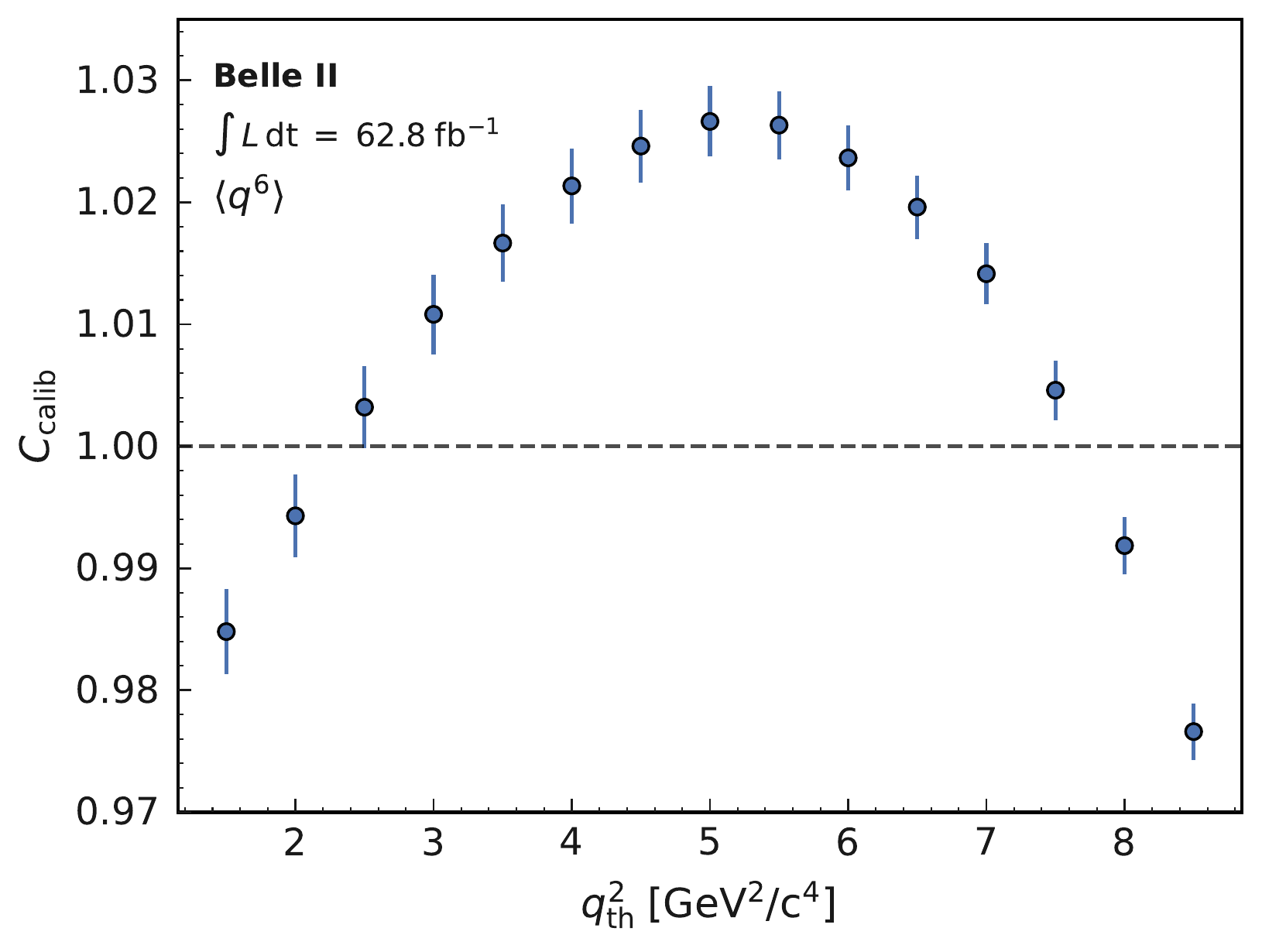}
  \includegraphics[width=0.45\textwidth]{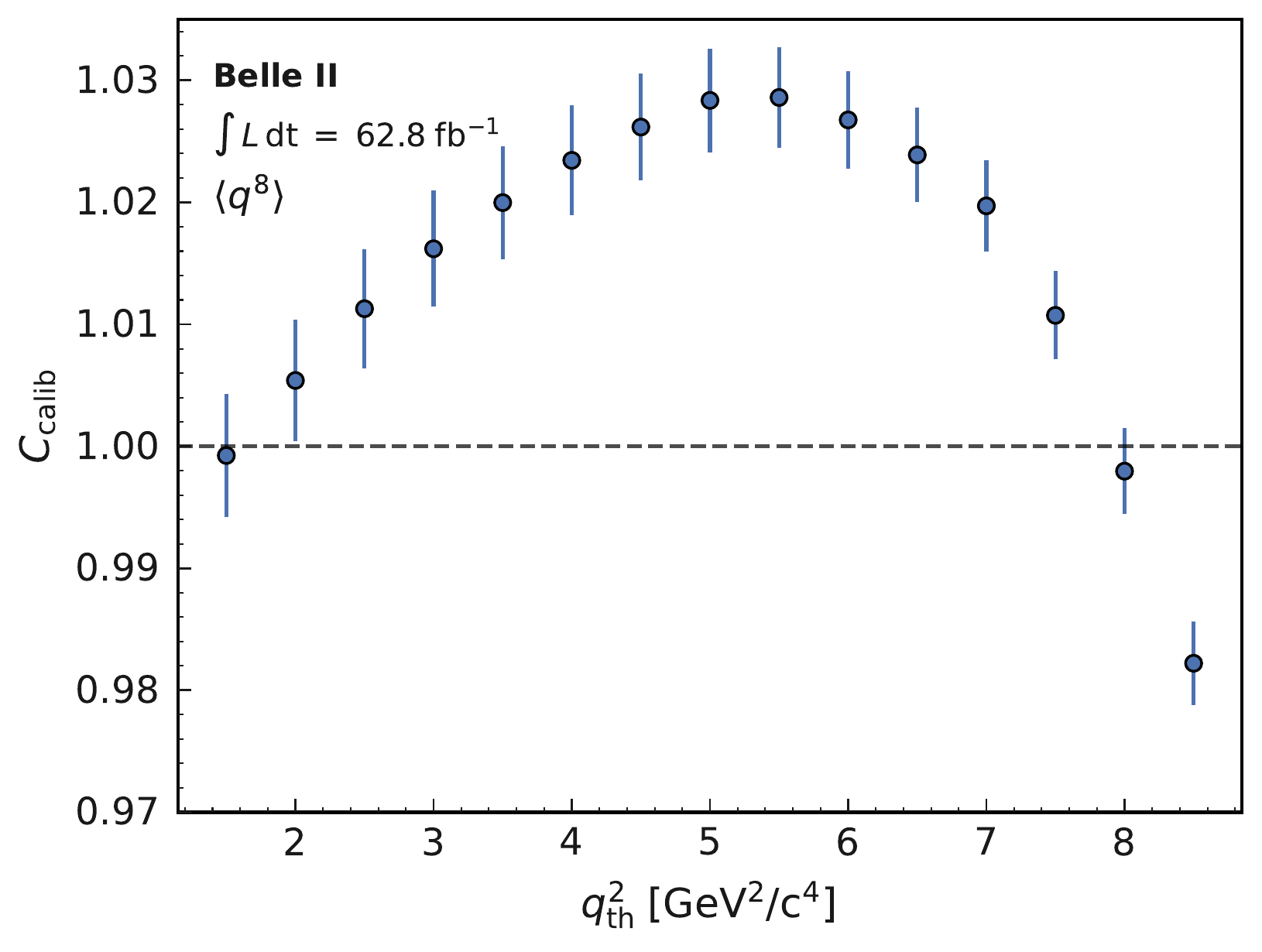}
  \caption{Calibration factors \ccalib applied in the calculation of the first to fourth \qsquared moment.}
  \label{fig:calib_bias_correction_factors}
\end{figure}

\begin{figure}[t]
  \includegraphics[width=0.45\textwidth]{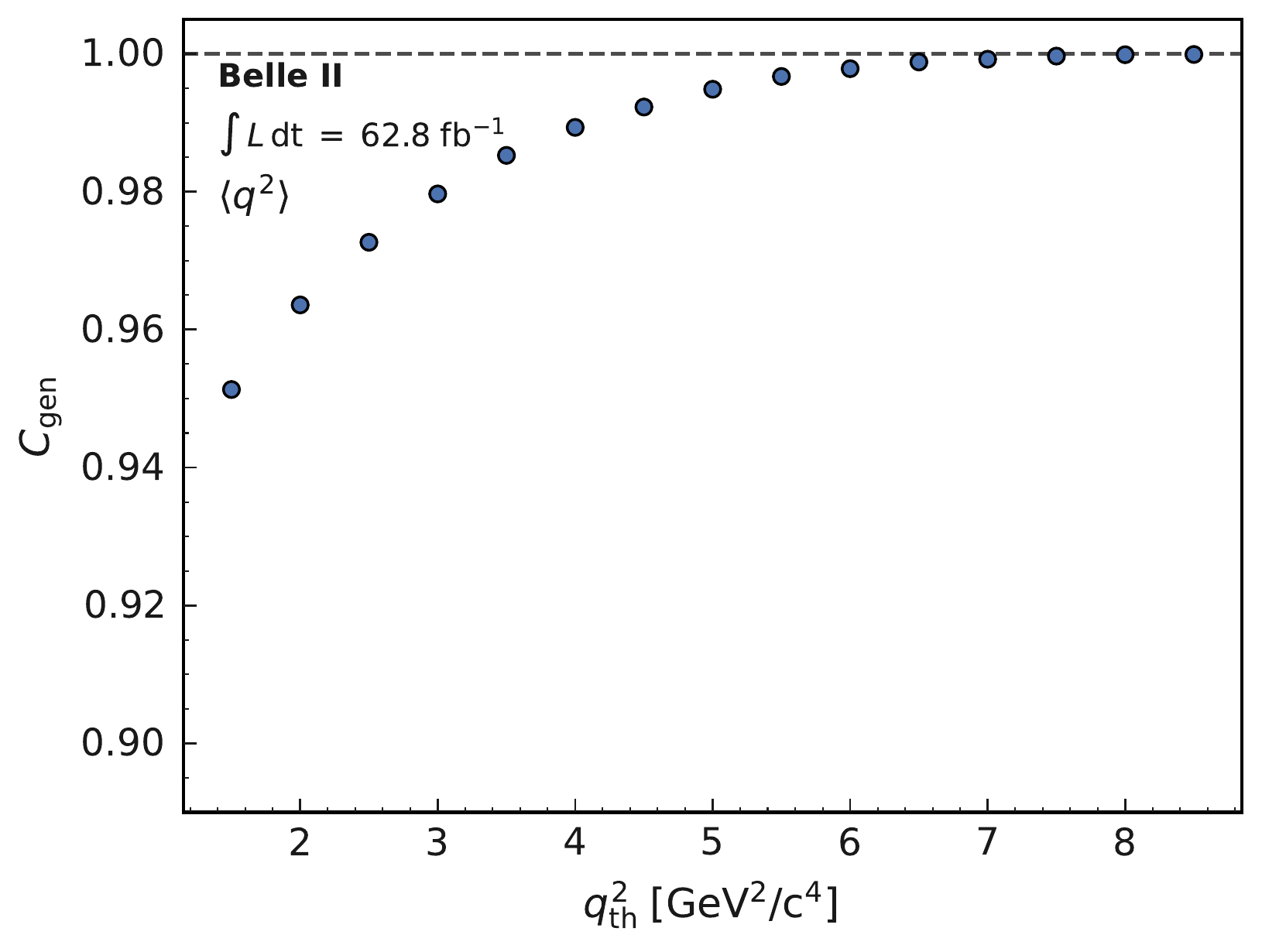}
  \includegraphics[width=0.45\textwidth]{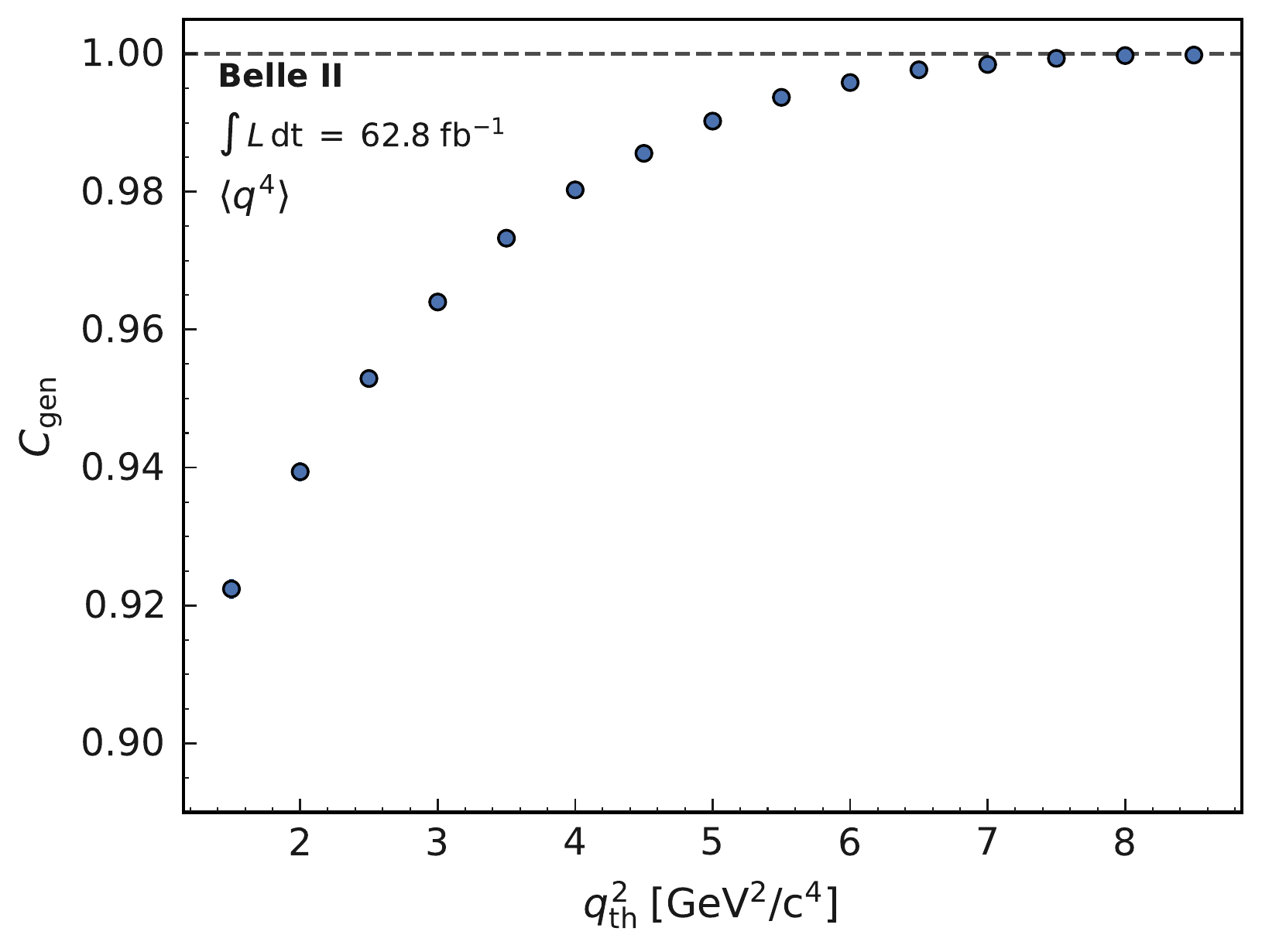}\\
  \includegraphics[width=0.45\textwidth]{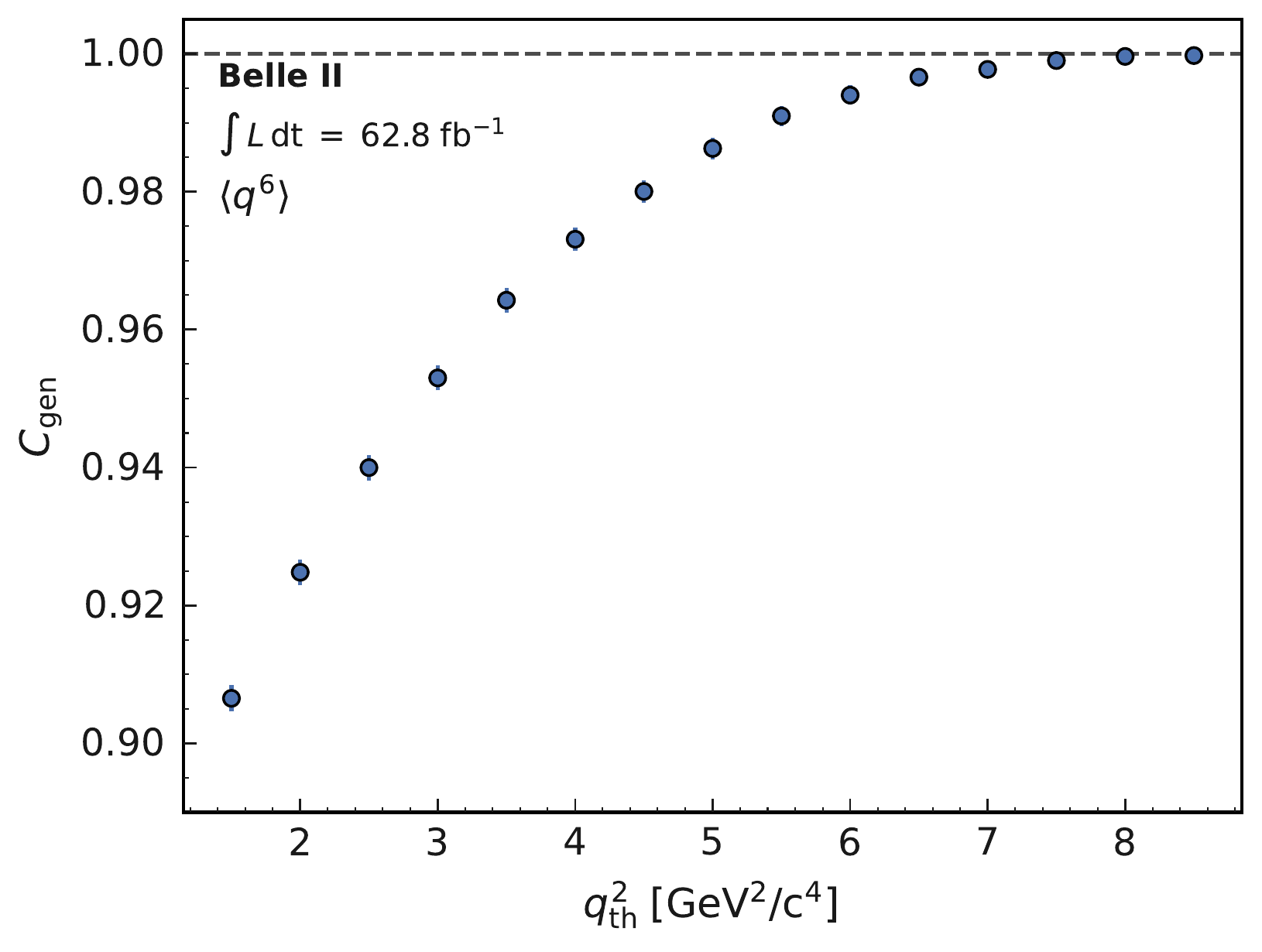}
  \includegraphics[width=0.45\textwidth]{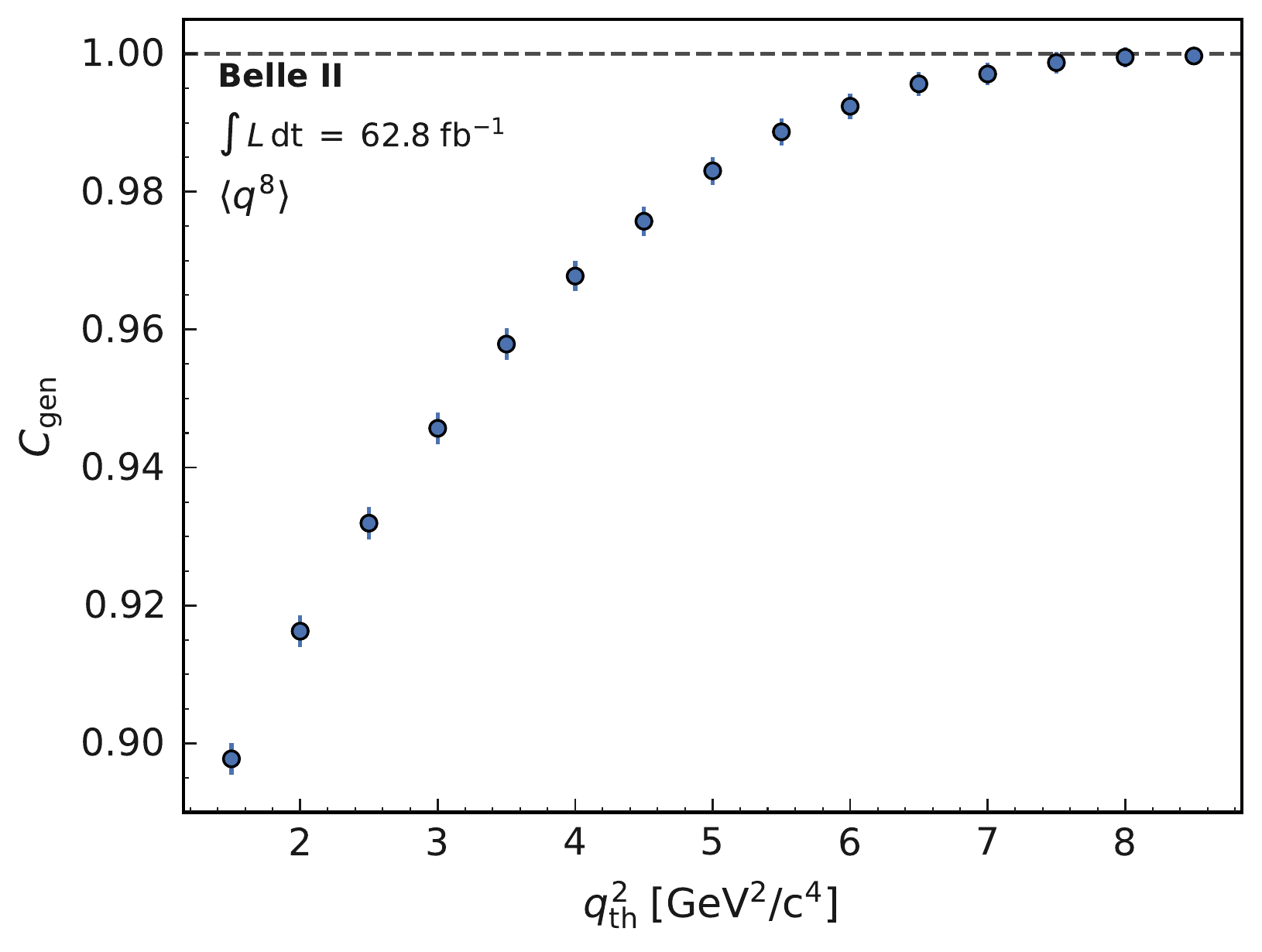}
  \caption{Calibration factors \ctrue applied in the calculation of the first to fourth \qsquared moment.}
  \label{fig:true_bias_correction_factors}
\end{figure}

\cref{fig:calib_bias_correction_factors,fig:true_bias_correction_factors} show the calibration factors \ccalib and \ctrue as functions of $q^2$ threshold.
The factors are determined using independent simulated samples of signal \bclnu decays. 
The corrections from \ccalib are small, typically below 2\%, and correct deviations from the linear relationships between reconstructed and generated moments. 
The corrections from \ctrue decrease with the $q^2$ threshold. 

\clearpage

\section{Numerical Values for the Raw $q^2$ Moments}\label{sec:app_mom}

\cref{tab:results_uncertainties_q2_1,tab:results_uncertainties_q2_2,tab:results_uncertainties_q2_3,tab:results_uncertainties_q2_4} summarize the $q^2$ moments and the systematic uncertainties.
The uncertainties are grouped into uncertainties from the background subtraction and calibration.
At low $q^2$ thresholds the uncertainty on the background shape limits the precision.
At high $q^2$ thresholds the uncertainties on the simulation of the \b2 detector are the largest systematic uncertainties.

\begin{table}[htb]
  \caption{Central values and uncertainties for the measurement of \qsqxmoment{2}. All uncertainties are given as relative uncertainties in \%.}
    \resizebox{\textwidth}{!}{%
  \begin{tabular}{llrrrrrrrrrrrrrrr}
\toprule
      & $q^2_\mathrm{th} \; [\mathrm{Gev^2/\clight^4}]$ &   1.5 &   2.0 &   2.5 &   3.0 &   3.5 &   4.0 &   4.5 &   5.0 &   5.5 &   6.0 &   6.5 &   7.0 &   7.5 &   8.0 &   8.5 \\
\midrule
{} & $\langle q^2 \rangle \; [\mathrm{Gev^2/\clight^4}]$ &  5.16 &  5.49 &  5.79 &  6.09 &  6.38 &  6.69 &  7.01 &  7.32 &  7.62 &  7.93 &  8.23 &  8.53 &  8.82 &  9.10 &  9.39 \\
\midrule
Calibration (MC Statistics) & Calib. Curve (Stat. Unc.) &  0.63 &  0.56 &  0.49 &  0.43 &  0.38 &  0.33 &  0.29 &  0.26 &  0.25 &  0.26 &  0.28 &  0.30 &  0.33 &  0.37 &  0.40 \\
      & Bias Corr. (Stat. Unc.) &  0.10 &  0.09 &  0.09 &  0.08 &  0.08 &  0.08 &  0.07 &  0.07 &  0.07 &  0.07 &  0.06 &  0.06 &  0.06 &  0.06 &  0.06 \\
\cmidrule{2-17}
Calibration ($X_c$ Model) & $\mathcal{B}(B\rightarrow D  \ell \nu)$ &  0.10 &  0.09 &  0.08 &  0.07 &  0.06 &  0.05 &  0.04 &  0.04 &  0.03 &  0.02 &  0.02 &  0.01 &  0.01 &  0.00 &  0.00 \\
      & $\mathcal{B}(B\rightarrow D^\ast \ell \nu)$ &  0.33 &  0.29 &  0.24 &  0.21 &  0.17 &  0.14 &  0.11 &  0.09 &  0.07 &  0.05 &  0.04 &  0.03 &  0.02 &  0.01 &  0.00 \\
      & $\mathcal{B}(B\rightarrow D^{\ast\ast} \ell \nu)$ &  0.71 &  0.63 &  0.55 &  0.48 &  0.40 &  0.34 &  0.28 &  0.23 &  0.18 &  0.13 &  0.10 &  0.07 &  0.05 &  0.03 &  0.02 \\
      & Non-Res. $X_c$ Dropped &  0.31 &  0.63 &  0.75 &  0.76 &  0.69 &  0.60 &  0.48 &  0.39 &  0.32 &  0.25 &  0.18 &  0.14 &  0.11 &  0.08 &  0.06 \\
      & Non-Res. $X_c$ Repl. w/ {$D_1^\prime$, $D_0^*$} &  0.34 &  0.49 &  0.51 &  0.45 &  0.37 &  0.29 &  0.18 &  0.10 &  0.04 &  0.02 &  0.00 &  0.03 &  0.03 &  0.03 &  0.01 \\
      & $B\rightarrow D  \ell \nu$ Form Factor &  0.01 &  0.01 &  0.01 &  0.01 &  0.01 &  0.01 &  0.01 &  0.01 &  0.00 &  0.00 &  0.00 &  0.00 &  0.00 &  0.00 &  0.00 \\
      & $B\rightarrow D^\ast \ell \nu$ Form Factor &  0.08 &  0.07 &  0.07 &  0.07 &  0.06 &  0.06 &  0.06 &  0.05 &  0.05 &  0.05 &  0.04 &  0.04 &  0.04 &  0.04 &  0.03 \\
\cmidrule{2-17}
Calibration (Reconstruction) & PID Uncertainty &  0.14 &  0.12 &  0.11 &  0.09 &  0.08 &  0.07 &  0.05 &  0.04 &  0.03 &  0.02 &  0.02 &  0.01 &  0.01 &  0.01 &  0.01 \\
      & $N_\gamma$ Reweighted &  0.30 &  0.27 &  0.24 &  0.22 &  0.20 &  0.18 &  0.16 &  0.14 &  0.14 &  0.13 &  0.13 &  0.12 &  0.11 &  0.10 &  0.10 \\
      & $N_\mathrm{tracks}$ Reweighted &  1.09 &  1.00 &  0.92 &  0.85 &  0.78 &  0.72 &  0.65 &  0.60 &  0.55 &  0.51 &  0.47 &  0.44 &  0.41 &  0.38 &  0.35 \\
      & $E_\mathrm{miss} - p_\mathrm{miss}$ Reweighted &  0.26 &  0.22 &  0.21 &  0.19 &  0.18 &  0.17 &  0.15 &  0.15 &  0.14 &  0.14 &  0.13 &  0.12 &  0.12 &  0.11 &  0.09 \\
      & Tracking Efficiency &  0.13 &  0.12 &  0.11 &  0.10 &  0.09 &  0.09 &  0.08 &  0.07 &  0.06 &  0.06 &  0.05 &  0.05 &  0.05 &  0.04 &  0.04 \\
\cmidrule{2-17}
Background Subtraction & Spline Smooth. Factor &  0.00 &  0.00 &  0.00 &  0.00 &  0.00 &  0.00 &  0.00 &  0.00 &  0.00 &  0.00 &  0.00 &  0.00 &  0.00 &  0.00 &  0.00 \\
      & Bkg. Yield \& Shape &  1.39 &  1.15 &  0.90 &  0.77 &  0.63 &  0.47 &  0.33 &  0.23 &  0.16 &  0.10 &  0.06 &  0.03 &  0.02 &  0.05 &  0.06 \\
\cmidrule{2-17}
Other & Non-Closure Bias &  0.18 &  0.21 &  0.16 &  0.11 &  0.06 &  0.05 &  0.02 &  0.02 &  0.01 &  0.02 &  0.02 &  0.02 &  0.01 &  0.01 &  0.02 \\
\midrule
      & Stat. Uncertainty &  0.27 &  0.24 &  0.21 &  0.20 &  0.18 &  0.16 &  0.16 &  0.15 &  0.14 &  0.14 &  0.13 &  0.13 &  0.13 &  0.13 &  0.13 \\
      & Syst. Uncertainty &  2.14 &  1.99 &  1.80 &  1.64 &  1.44 &  1.23 &  1.02 &  0.88 &  0.77 &  0.69 &  0.62 &  0.59 &  0.57 &  0.56 &  0.57 \\
      & Total Uncertainty &  2.16 &  2.00 &  1.81 &  1.65 &  1.45 &  1.24 &  1.03 &  0.89 &  0.78 &  0.70 &  0.64 &  0.61 &  0.59 &  0.58 &  0.58 \\
\bottomrule
\end{tabular}
  }
  \label{tab:results_uncertainties_q2_1}
\end{table}

\begin{table}[htb]
  \caption{Central values and uncertainties for the measurement of \qsqxmoment{4}.  All uncertainties are given as relative uncertainties in \%.}
  \resizebox{\textwidth}{!}{%
  \begin{tabular}{llrrrrrrrrrrrrrrr}
\toprule
      & $q^2_\mathrm{th} \; [\mathrm{Gev^2/\clight^4}]$ &    1.5 &    2.0 &    2.5 &    3.0 &    3.5 &    4.0 &    4.5 &    5.0 &    5.5 &    6.0 &    6.5 &    7.0 &    7.5 &    8.0 &    8.5 \\
\midrule
{} & $\langle q^4 \rangle \; [(\mathrm{Gev^2/\clight^4})^2] $ &  32.55 &  35.44 &  38.21 &  41.18 &  44.31 &  47.92 &  51.82 &  55.90 &  60.00 &  64.35 &  68.90 &  73.62 &  78.40 &  83.33 &  88.47 \\
\midrule
Calibration (MC Statistics) & Calib. Curve (Stat. Unc.) &   0.96 &   0.85 &   0.75 &   0.67 &   0.58 &   0.50 &   0.44 &   0.41 &   0.40 &   0.42 &   0.45 &   0.49 &   0.54 &   0.59 &   0.64 \\
      & Bias Corr. (Stat. Unc.) &   0.20 &   0.19 &   0.18 &   0.18 &   0.17 &   0.16 &   0.16 &   0.15 &   0.15 &   0.14 &   0.14 &   0.14 &   0.14 &   0.14 &   0.14 \\
\cmidrule{2-17}
Calibration ($X_c$ Model) & $\mathcal{B}(B\rightarrow D  \ell \nu)$ &   0.18 &   0.16 &   0.15 &   0.13 &   0.12 &   0.10 &   0.08 &   0.07 &   0.06 &   0.04 &   0.03 &   0.02 &   0.01 &   0.01 &   0.01 \\
      & $\mathcal{B}(B\rightarrow D^\ast \ell \nu)$ &   0.60 &   0.52 &   0.45 &   0.39 &   0.33 &   0.27 &   0.22 &   0.18 &   0.14 &   0.10 &   0.07 &   0.05 &   0.03 &   0.02 &   0.01 \\
      & $\mathcal{B}(B\rightarrow D^{\ast\ast} \ell \nu)$ &   1.30 &   1.17 &   1.04 &   0.91 &   0.79 &   0.67 &   0.56 &   0.45 &   0.36 &   0.27 &   0.20 &   0.14 &   0.09 &   0.06 &   0.05 \\
      & Non-Res. $X_c$ Dropped &   0.91 &   1.31 &   1.47 &   1.47 &   1.35 &   1.18 &   0.96 &   0.79 &   0.64 &   0.52 &   0.38 &   0.30 &   0.23 &   0.16 &   0.13 \\
      & Non-Res. $X_c$ Repl. w/ {$D_1^\prime$, $D_0^*$} &   0.69 &   0.87 &   0.89 &   0.79 &   0.66 &   0.51 &   0.31 &   0.17 &   0.07 &   0.03 &   0.02 &   0.06 &   0.07 &   0.06 &   0.03 \\
      & $B\rightarrow D  \ell \nu$ Form Factor &   0.02 &   0.02 &   0.01 &   0.01 &   0.01 &   0.01 &   0.01 &   0.01 &   0.01 &   0.01 &   0.01 &   0.01 &   0.01 &   0.01 &   0.01 \\
      & $B\rightarrow D^\ast \ell \nu$ Form Factor &   0.17 &   0.16 &   0.15 &   0.15 &   0.14 &   0.13 &   0.12 &   0.12 &   0.11 &   0.10 &   0.10 &   0.09 &   0.09 &   0.08 &   0.08 \\
\cmidrule{2-17}
Calibration (Reconstruction) & PID Uncertainty &   0.25 &   0.23 &   0.20 &   0.17 &   0.15 &   0.13 &   0.10 &   0.08 &   0.06 &   0.05 &   0.04 &   0.03 &   0.02 &   0.02 &   0.02 \\
      & $N_\gamma$ Reweighted &   0.61 &   0.57 &   0.52 &   0.49 &   0.45 &   0.40 &   0.36 &   0.33 &   0.32 &   0.30 &   0.28 &   0.26 &   0.25 &   0.23 &   0.22 \\
      & $N_\mathrm{tracks}$ Reweighted &   2.27 &   2.11 &   1.98 &   1.85 &   1.72 &   1.58 &   1.46 &   1.34 &   1.24 &   1.14 &   1.05 &   0.97 &   0.90 &   0.83 &   0.76 \\
      & $E_\mathrm{miss} - p_\mathrm{miss}$ Reweighted &   0.53 &   0.48 &   0.45 &   0.42 &   0.39 &   0.37 &   0.34 &   0.32 &   0.31 &   0.30 &   0.28 &   0.26 &   0.24 &   0.21 &   0.18 \\
      & Tracking Efficiency &   0.28 &   0.26 &   0.24 &   0.22 &   0.20 &   0.19 &   0.17 &   0.16 &   0.14 &   0.13 &   0.12 &   0.11 &   0.10 &   0.09 &   0.09 \\
\cmidrule{2-17}
Background Subtraction & Spline Smooth. Factor &   0.00 &   0.00 &   0.00 &   0.00 &   0.00 &   0.00 &   0.00 &   0.00 &   0.00 &   0.00 &   0.00 &   0.00 &   0.01 &   0.01 &   0.01 \\
      & Bkg. Yield \& Shape &   2.12 &   1.83 &   1.49 &   1.31 &   1.10 &   0.83 &   0.57 &   0.40 &   0.27 &   0.16 &   0.08 &   0.05 &   0.08 &   0.13 &   0.16 \\
\cmidrule{2-17}
Other & Non-Closure Bias &   0.32 &   0.37 &   0.30 &   0.23 &   0.13 &   0.11 &   0.06 &   0.05 &   0.04 &   0.05 &   0.04 &   0.04 &   0.04 &   0.03 &   0.05 \\
\midrule
      & Stat. Uncertainty &   0.49 &   0.46 &   0.43 &   0.40 &   0.37 &   0.35 &   0.34 &   0.33 &   0.31 &   0.31 &   0.30 &   0.29 &   0.29 &   0.29 &   0.29 \\
      & Syst. Uncertainty &   3.86 &   3.68 &   3.42 &   3.16 &   2.82 &   2.46 &   2.09 &   1.82 &   1.61 &   1.44 &   1.30 &   1.21 &   1.15 &   1.10 &   1.07 \\
      & Total Uncertainty &   3.89 &   3.71 &   3.45 &   3.18 &   2.85 &   2.48 &   2.12 &   1.85 &   1.64 &   1.47 &   1.34 &   1.25 &   1.19 &   1.14 &   1.11 \\
\bottomrule
\end{tabular}
  }
  \label{tab:results_uncertainties_q2_2}
\end{table}

\begin{table}[htb]
  \caption{Central values and uncertainties for the measurement of \qsqxmoment{6}. All uncertainties are given as relative uncertainties in \%. }
  \resizebox{\textwidth}{!}{%
  \begin{tabular}{llrrrrrrrrrrrrrrr}
\toprule
      & $q^2_\mathrm{th} \; [\mathrm{Gev^2/\clight^4}]$ &     1.5 &     2.0 &     2.5 &     3.0 &     3.5 &     4.0 &     4.5 &     5.0 &     5.5 &     6.0 &     6.5 &     7.0 &     7.5 &     8.0 &     8.5 \\
\midrule
{} & $\langle q^6 \rangle \; [(\mathrm{Gev^2/\clight^4})^3] $ &  234.11 &  256.58 &  278.78 &  303.60 &  331.14 &  364.36 &  402.07 &  443.33 &  486.42 &  534.18 &  586.53 &  642.87 &  702.59 &  766.54 &  836.02 \\
\midrule
Calibration (MC Statistics) & Calib. Curve (Stat. Unc.) &    1.01 &    0.90 &    0.80 &    0.71 &    0.62 &    0.54 &    0.48 &    0.44 &    0.44 &    0.46 &    0.49 &    0.54 &    0.59 &    0.64 &    0.69 \\
      & Bias Corr. (Stat. Unc.) &    0.31 &    0.31 &    0.30 &    0.29 &    0.28 &    0.27 &    0.26 &    0.25 &    0.25 &    0.24 &    0.23 &    0.23 &    0.23 &    0.22 &    0.22 \\
\cmidrule{2-17}
Calibration ($X_c$ Model) & $\mathcal{B}(B\rightarrow D  \ell \nu)$ &    0.24 &    0.22 &    0.20 &    0.18 &    0.16 &    0.14 &    0.12 &    0.10 &    0.08 &    0.06 &    0.04 &    0.02 &    0.01 &    0.01 &    0.02 \\
      & $\mathcal{B}(B\rightarrow D^\ast \ell \nu)$ &    0.79 &    0.70 &    0.62 &    0.54 &    0.46 &    0.39 &    0.32 &    0.26 &    0.20 &    0.15 &    0.11 &    0.07 &    0.05 &    0.02 &    0.02 \\
      & $\mathcal{B}(B\rightarrow D^{\ast\ast} \ell \nu)$ &    1.75 &    1.60 &    1.44 &    1.29 &    1.12 &    0.96 &    0.81 &    0.67 &    0.53 &    0.40 &    0.30 &    0.22 &    0.15 &    0.10 &    0.08 \\
      & Non-Res. $X_c$ Dropped &    1.66 &    2.03 &    2.16 &    2.12 &    1.94 &    1.71 &    1.41 &    1.17 &    0.96 &    0.78 &    0.58 &    0.46 &    0.35 &    0.25 &    0.20 \\
      & Non-Res. $X_c$ Repl. w/ {$D_1^\prime$, $D_0^*$} &    0.93 &    1.12 &    1.14 &    1.02 &    0.85 &    0.66 &    0.40 &    0.21 &    0.08 &    0.03 &    0.04 &    0.09 &    0.11 &    0.10 &    0.05 \\
      & $B\rightarrow D  \ell \nu$ Form Factor &    0.03 &    0.02 &    0.02 &    0.02 &    0.02 &    0.02 &    0.02 &    0.02 &    0.02 &    0.02 &    0.02 &    0.02 &    0.02 &    0.02 &    0.01 \\
      & $B\rightarrow D^\ast \ell \nu$ Form Factor &    0.27 &    0.26 &    0.25 &    0.24 &    0.23 &    0.21 &    0.20 &    0.19 &    0.18 &    0.17 &    0.17 &    0.16 &    0.15 &    0.14 &    0.14 \\
\cmidrule{2-17}
Calibration (Reconstruction) & PID Uncertainty &    0.34 &    0.31 &    0.28 &    0.24 &    0.21 &    0.18 &    0.15 &    0.12 &    0.10 &    0.08 &    0.06 &    0.05 &    0.04 &    0.04 &    0.04 \\
      & $N_\gamma$ Reweighted &    0.96 &    0.90 &    0.84 &    0.79 &    0.73 &    0.67 &    0.61 &    0.56 &    0.53 &    0.50 &    0.48 &    0.44 &    0.42 &    0.38 &    0.36 \\
      & $N_\mathrm{tracks}$ Reweighted &    3.53 &    3.33 &    3.15 &    2.97 &    2.78 &    2.58 &    2.38 &    2.20 &    2.03 &    1.87 &    1.72 &    1.58 &    1.46 &    1.34 &    1.23 \\
      & $E_\mathrm{miss} - p_\mathrm{miss}$ Reweighted &    0.81 &    0.74 &    0.70 &    0.67 &    0.62 &    0.58 &    0.54 &    0.50 &    0.48 &    0.46 &    0.42 &    0.39 &    0.35 &    0.30 &    0.25 \\
      & Tracking Efficiency &    0.43 &    0.40 &    0.38 &    0.35 &    0.33 &    0.30 &    0.28 &    0.26 &    0.23 &    0.22 &    0.20 &    0.18 &    0.17 &    0.15 &    0.14 \\
\cmidrule{2-17}
Background Subtraction & Spline Smooth. Factor &    0.00 &    0.01 &    0.01 &    0.01 &    0.01 &    0.00 &    0.00 &    0.01 &    0.01 &    0.01 &    0.01 &    0.00 &    0.01 &    0.01 &    0.01 \\
      & Bkg. Yield \& Shape &    2.53 &    2.23 &    1.85 &    1.64 &    1.38 &    1.03 &    0.71 &    0.48 &    0.30 &    0.17 &    0.10 &    0.14 &    0.20 &    0.27 &    0.31 \\
\cmidrule{2-17}
Other & Non-Closure Bias &    0.46 &    0.52 &    0.43 &    0.34 &    0.22 &    0.18 &    0.11 &    0.10 &    0.09 &    0.09 &    0.08 &    0.08 &    0.08 &    0.07 &    0.09 \\
\midrule
      & Stat. Uncertainty &    0.74 &    0.71 &    0.67 &    0.65 &    0.61 &    0.59 &    0.57 &    0.55 &    0.53 &    0.52 &    0.50 &    0.50 &    0.49 &    0.49 &    0.49 \\
      & Syst. Uncertainty &    5.43 &    5.23 &    4.92 &    4.59 &    4.15 &    3.68 &    3.20 &    2.82 &    2.52 &    2.27 &    2.05 &    1.89 &    1.76 &    1.65 &    1.56 \\
      & Total Uncertainty &    5.48 &    5.28 &    4.97 &    4.63 &    4.20 &    3.72 &    3.25 &    2.87 &    2.57 &    2.32 &    2.11 &    1.95 &    1.82 &    1.72 &    1.63 \\
\bottomrule
\end{tabular}
  }
  \label{tab:results_uncertainties_q2_3}
\end{table}

\begin{table}[htb]
  \caption{Central values and uncertainties for the measurement of \qsqxmoment{8}. All uncertainties are given as relative uncertainties in \%. }
  \resizebox{\textwidth}{!}{%
  \begin{tabular}{llrrrrrrrrrrrrrrr}
\toprule
      & $q^2_\mathrm{th} \; [\mathrm{Gev^2/\clight^4}]$ &      1.5 &      2.0 &      2.5 &      3.0 &      3.5 &      4.0 &      4.5 &      5.0 &      5.5 &      6.0 &      6.5 &      7.0 &      7.5 &      8.0 &      8.5 \\
\midrule
{} & $\langle q^8 \rangle \; [(\mathrm{Gev^2/\clight^4})^4] $ &  1824.48 &  2003.76 &  2182.03 &  2386.22 &  2621.05 &  2911.47 &  3251.23 &  3636.73 &  4051.07 &  4526.33 &  5071.04 &  5675.12 &  6344.75 &  7085.85 &  7924.67 \\
\midrule
Calibration (MC Statistics) & Calib. Curve (Stat. Unc.) &     0.87 &     0.77 &     0.69 &     0.61 &     0.54 &     0.46 &     0.41 &     0.38 &     0.38 &     0.39 &     0.42 &     0.46 &     0.50 &     0.55 &     0.59 \\
      & Bias Corr. (Stat. Unc.) &     0.46 &     0.45 &     0.44 &     0.43 &     0.42 &     0.41 &     0.39 &     0.38 &     0.37 &     0.36 &     0.35 &     0.34 &     0.34 &     0.33 &     0.33 \\
\cmidrule{2-17}
Calibration ($X_c$ Model) & $\mathcal{B}(B\rightarrow D  \ell \nu)$ &     0.28 &     0.26 &     0.24 &     0.22 &     0.19 &     0.17 &     0.14 &     0.12 &     0.09 &     0.07 &     0.04 &     0.02 &     0.02 &     0.01 &     0.03 \\
      & $\mathcal{B}(B\rightarrow D^\ast \ell \nu)$ &     0.93 &     0.84 &     0.74 &     0.65 &     0.57 &     0.48 &     0.40 &     0.32 &     0.25 &     0.19 &     0.14 &     0.09 &     0.06 &     0.03 &     0.02 \\
      & $\mathcal{B}(B\rightarrow D^{\ast\ast} \ell \nu)$ &     2.10 &     1.94 &     1.77 &     1.59 &     1.40 &     1.22 &     1.03 &     0.86 &     0.69 &     0.53 &     0.40 &     0.29 &     0.20 &     0.14 &     0.11 \\
      & Non-Res. $X_c$ Dropped &     2.47 &     2.76 &     2.82 &     2.72 &     2.48 &     2.18 &     1.81 &     1.51 &     1.25 &     1.02 &     0.78 &     0.62 &     0.47 &     0.35 &     0.28 \\
      & Non-Res. $X_c$ Repl. w/ {$D_1^\prime$, $D_0^*$} &     1.06 &     1.26 &     1.28 &     1.15 &     0.96 &     0.75 &     0.46 &     0.24 &     0.08 &     0.02 &     0.07 &     0.14 &     0.16 &     0.14 &     0.08 \\
      & $B\rightarrow D  \ell \nu$ Form Factor &     0.04 &     0.03 &     0.03 &     0.03 &     0.03 &     0.03 &     0.03 &     0.03 &     0.03 &     0.03 &     0.03 &     0.03 &     0.03 &     0.02 &     0.02 \\
      & $B\rightarrow D^\ast \ell \nu$ Form Factor &     0.38 &     0.37 &     0.35 &     0.34 &     0.33 &     0.31 &     0.30 &     0.28 &     0.27 &     0.26 &     0.25 &     0.24 &     0.23 &     0.22 &     0.21 \\
\cmidrule{2-17}
Calibration (Reconstruction) & PID Uncertainty &     0.41 &     0.38 &     0.34 &     0.30 &     0.26 &     0.23 &     0.19 &     0.16 &     0.13 &     0.11 &     0.09 &     0.07 &     0.06 &     0.06 &     0.05 \\
      & $N_\gamma$ Reweighted &     1.35 &     1.27 &     1.20 &     1.13 &     1.06 &     0.98 &     0.90 &     0.83 &     0.79 &     0.74 &     0.70 &     0.65 &     0.61 &     0.56 &     0.52 \\
      & $N_\mathrm{tracks}$ Reweighted &     4.90 &     4.64 &     4.41 &     4.18 &     3.93 &     3.67 &     3.40 &     3.15 &     2.91 &     2.69 &     2.47 &     2.27 &     2.09 &     1.91 &     1.74 \\
      & $E_\mathrm{miss} - p_\mathrm{miss}$ Reweighted &     1.09 &     1.00 &     0.95 &     0.90 &     0.85 &     0.79 &     0.73 &     0.68 &     0.64 &     0.60 &     0.55 &     0.50 &     0.44 &     0.38 &     0.31 \\
      & Tracking Efficiency &     0.58 &     0.55 &     0.52 &     0.49 &     0.46 &     0.43 &     0.40 &     0.37 &     0.34 &     0.31 &     0.28 &     0.26 &     0.24 &     0.21 &     0.19 \\
\cmidrule{2-17}
Background Subtraction & Spline Smooth. Factor &     0.01 &     0.01 &     0.01 &     0.01 &     0.01 &     0.01 &     0.01 &     0.01 &     0.01 &     0.01 &     0.01 &     0.00 &     0.02 &     0.03 &     0.02 \\
      & Bkg. Yield \& Shape &     2.81 &     2.49 &     2.08 &     1.82 &     1.50 &     1.10 &     0.73 &     0.47 &     0.29 &     0.21 &     0.22 &     0.32 &     0.39 &     0.48 &     0.52 \\
\cmidrule{2-17}
Other & Non-Closure Bias &     0.64 &     0.69 &     0.58 &     0.47 &     0.32 &     0.27 &     0.20 &     0.17 &     0.16 &     0.15 &     0.14 &     0.15 &     0.14 &     0.13 &     0.15 \\
\midrule
      & Stat. Uncertainty &     1.08 &     1.04 &     1.00 &     0.96 &     0.92 &     0.89 &     0.86 &     0.83 &     0.81 &     0.79 &     0.77 &     0.76 &     0.74 &     0.74 &     0.74 \\
      & Syst. Uncertainty &     7.04 &     6.78 &     6.40 &     5.99 &     5.47 &     4.90 &     4.34 &     3.88 &     3.50 &     3.17 &     2.86 &     2.63 &     2.43 &     2.24 &     2.08 \\
      & Total Uncertainty &     7.12 &     6.86 &     6.48 &     6.06 &     5.54 &     4.98 &     4.42 &     3.97 &     3.59 &     3.26 &     2.97 &     2.74 &     2.54 &     2.36 &     2.20 \\
\bottomrule
\end{tabular}
  }
  \label{tab:results_uncertainties_q2_4}
\end{table}

\clearpage
\section{Correlation Coefficients of the Raw Moments}
\label{app:raw_moment_correlations}

\begin{figure}[h]
  \includegraphics[width=1\textwidth]{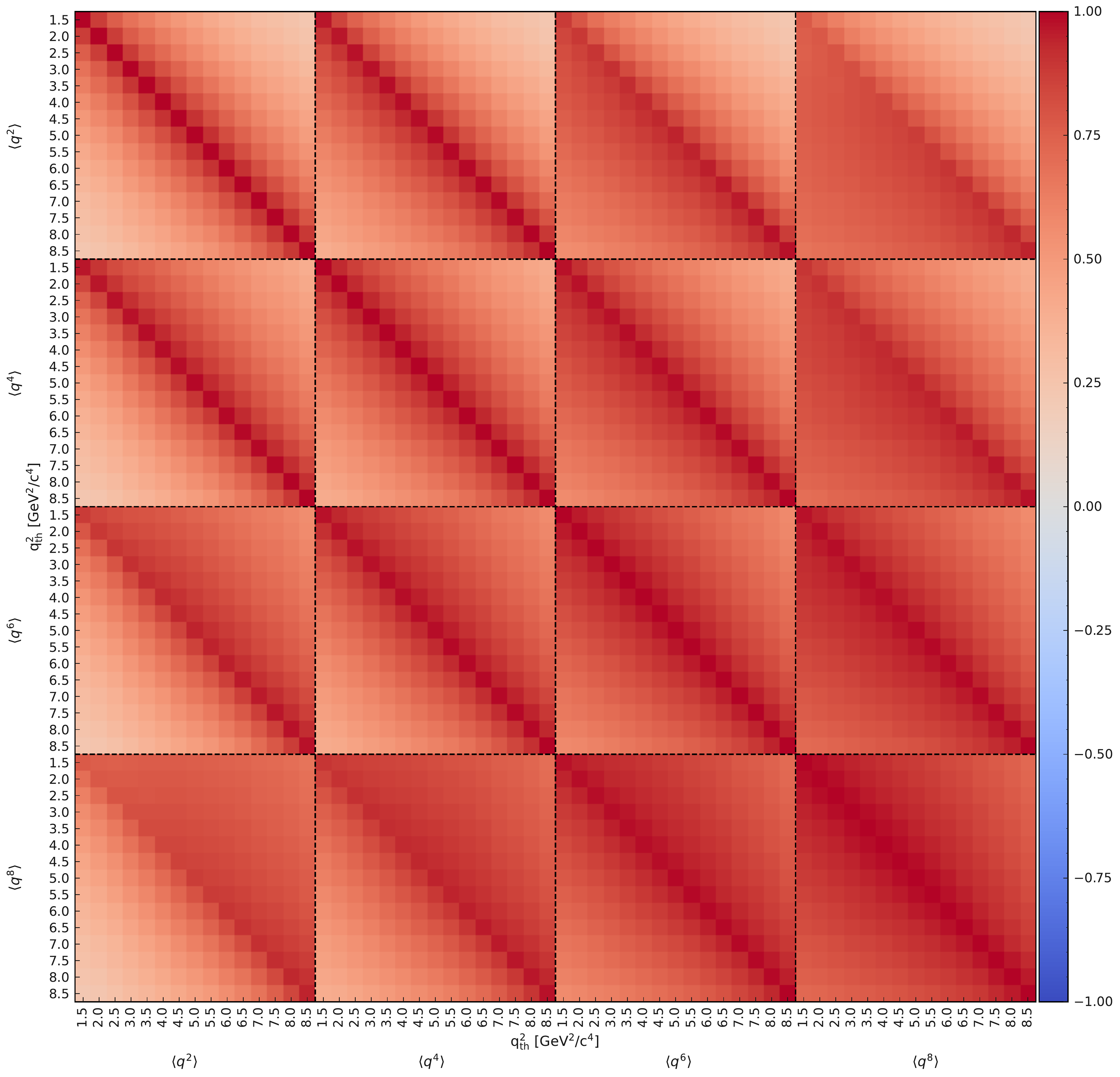}
  \caption{Statistical correlations between \qsqmoment{} and \qsqmoment{n} for \mbox{$n=1$--4}.}
  \label{fig:results_on_data_q2_moments_stat_correlation_q2_all}
\end{figure}

\begin{figure}[h]
  \includegraphics[width=1\textwidth]{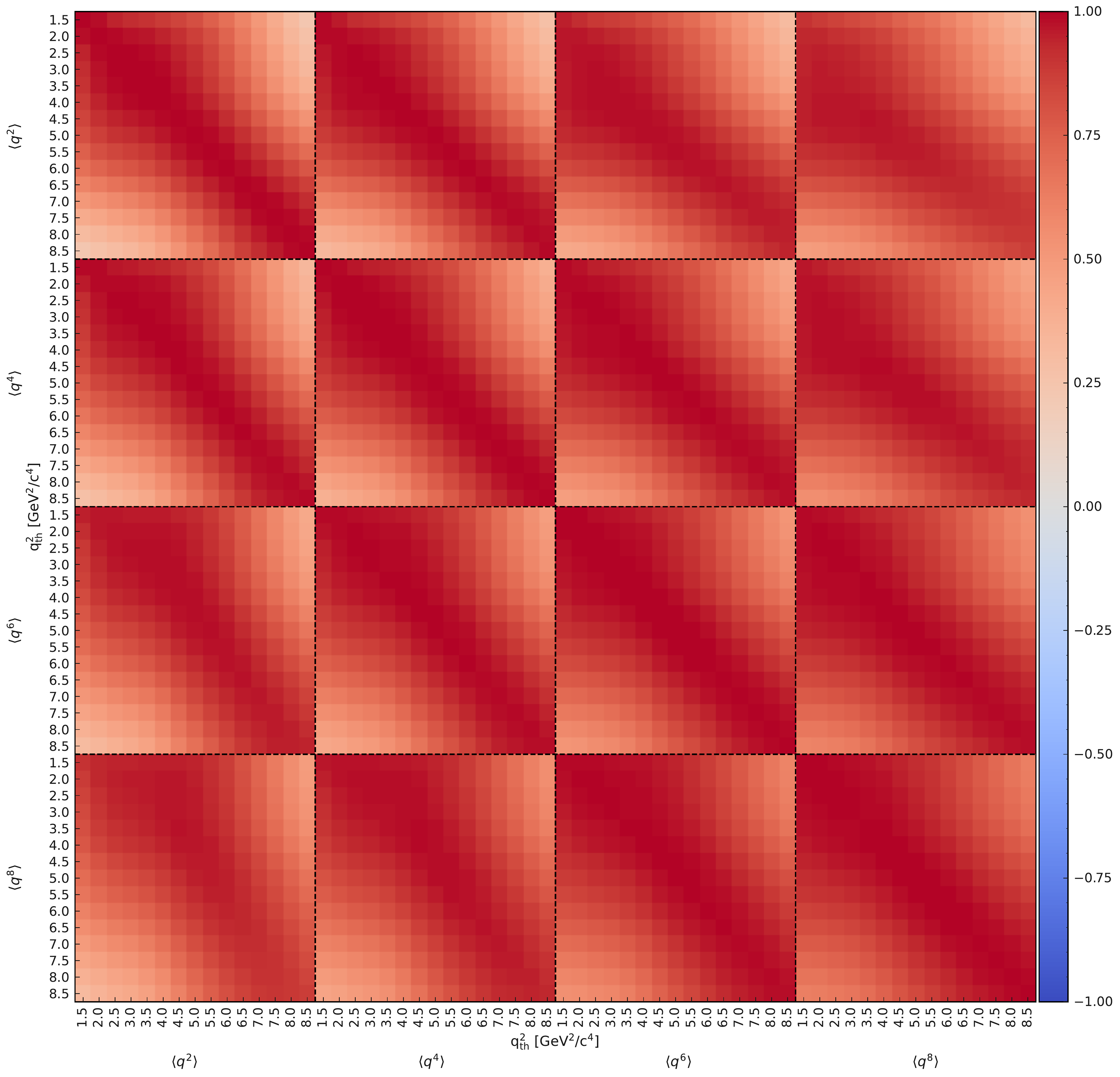}
  \caption{Experimental correlations between \qsqmoment{} and \qsqmoment{n} for \mbox{$n=1$--4}.}
  \label{fig:results_on_data_q2_moments_total_correlation_q2_all}
\end{figure}

The statistical correlation coefficients for the raw moments are shown in \cref{fig:results_on_data_q2_moments_stat_correlation_q2_all}. 
Moments with similar $q^2$ thresholds are strongly correlated.
\cref{fig:results_on_data_q2_moments_total_correlation_q2_all} shows the full experimental correlations taking systematic uncertainties into account.
Systematic uncertainties further increase the correlations of neighboring thresholds. 

\clearpage

\section{Correlation Coefficients of the Central Moments}\label{app:central_moment_correlations}

\begin{figure}[h]
  \includegraphics[width=1\textwidth]{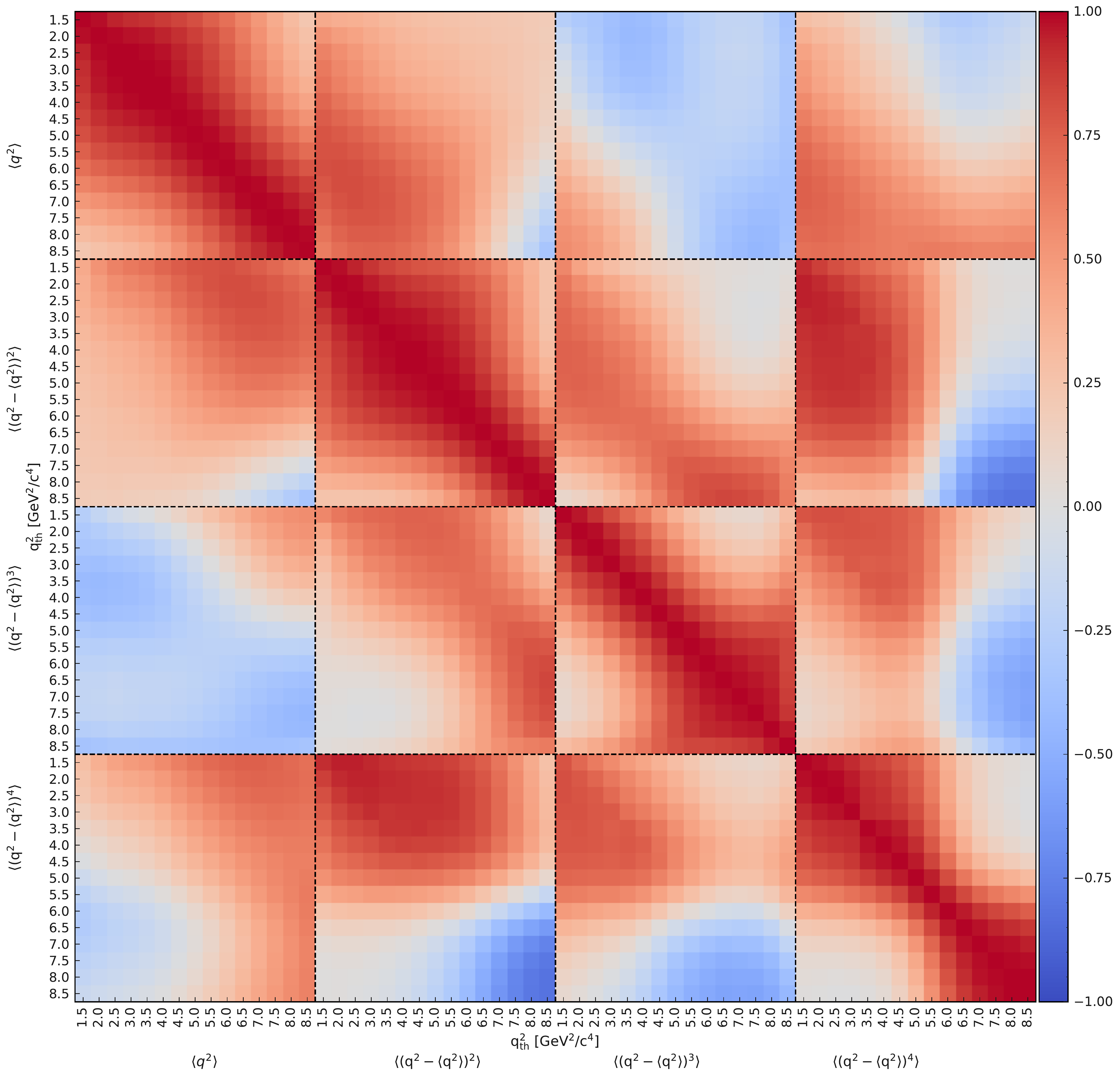}
  \caption{Correlations between \qsqmoment{} and $\langle ( q^{2}- \langle q^{2} \rangle)^{n}\rangle$  for \mbox{$n=2$--4} and for central moments of different order.}
  \label{fig:results_on_data_q2_moments_correlation_central_q2_all}
\end{figure}

The experimental correlation coefficients between the first raw moment and central moments and for the central moments of different order are shown in \cref{fig:results_on_data_q2_moments_correlation_central_q2_all}. 
The central moments are less correlated and some moments show anti-correlations.

\end{appendix}


\end{document}